\documentclass[aps,twocolumn,showpacs,amsmath,amssymb,superscriptaddress,longbibliography,prx,10pt]{revtex4-2}

\def\XXint#1#2#3{{\setbox0=\hbox{$#1{#2#3}{\int}$}
     \vcenter{\hbox{$#2#3$}}\kern-.5\wd0}}

\makeatletter
\def\@bibdataout@aps{%
 \immediate\write\@bibdataout{%
  @CONTROL{%
   apsrev42Control%
   \longbibliography@sw{%
    ,author="48",editor="1",pages="0",title="0",year="1"%
   }{%
    ,author="48",editor="1",pages="0",title="",year="1"%
   }%
  }%
 }%
 \if@filesw
  \immediate\write\@auxout{\string\citation{apsrev42Control}}%
 \fi
}%
\makeatother

\usepackage[colorlinks=true,allcolors=blue,hypertexnames=false]{hyperref}
\usepackage{amsmath}
\usepackage{dsfont}
\usepackage{hyperref}
\usepackage{textcomp}
\usepackage{tabularx}
\usepackage{graphicx}
\usepackage{soul}
\usepackage{feynmp-auto}
\usepackage{footmisc}
\usepackage[export]{adjustbox}
\usepackage{wrapfig}
\usepackage{tikz}

\newcommand{\bk}{{\bf k}}

\newcommand{\bq}{{\bf q}}

\newcommand{\bj}{{\bf j}}

\newcommand{\bn}{{\bf n}}

\newcommand{\bQ}{{\bf Q}}

\newcommand{\br}{{\bf r}}

\newcommand{\bK}{{\bf K}}

\newcommand{\vF}{v_F}

\newcommand{\w}{\omega}

\newcommand{\bp}{{\bf p}}

\newcommand{\beg}{\begin{equation}}
\newcommand{\en}{\end{equation}}

\newcommand \bel  {\begin{align}}
\newcommand \enl  {\end{align}}

\newcommand{\veps}{\varepsilon}
\newcommand{\eps}{\epsilon}

\newcommand{\up}{\uparrow}
\newcommand{\dn}{\downarrow}

\newcommand{\Tr}{\mathrm{Tr}\,}

\def \be{\begin{equation}}
\def \ee{\end{equation}}
\def \bea{\begin{eqnarray}}
\def \eea{\end{eqnarray}}

\begin{document}

\title{Light-induced magnetization in $d$-wave superconductors}

\author{Maxim Dzero}
\affiliation{Department of Physics, Kent State University, Kent, Ohio 44242, USA}

\author{Vladyslav Kozii}
\affiliation{Department of Physics, Carnegie Mellon University, Pittsburgh, Pennsylvania 15213, USA}

\begin{abstract}
We develop a microscopic theory of the inverse Faraday effect in weakly
disordered $s$- and $d$-wave superconductors. The Keldysh--Nambu quasiclassical
formalism is extended to retain the particle-hole asymmetric terms, and the dc
current induced by monochromatic radiation is computed to second order in the
field and to first order in the wave vector, with elastic impurity scattering
treated in the Born approximation. At finite wave vector the radiation induces an oscillation of the
order parameter amplitude -- the Schmid--Higgs amplitude mode -- that is linear in the
field. It is driven by the component of the electric field along the photon
wave vector and appears for both $s$- and $d$-wave pairing.  The amplitude mode contribution to the static magnetization sets in at $\omega=2\Delta$ and diverges as $(\omega-2\Delta)^{-1}$ for
an isotropic gap, whereas for $d$-wave pairing it is finite at all frequencies
and changes sign near $\omega\approx 2\sqrt{2}\Delta$. This contribution is of the first order in the particle-hole asymmetry, and it
is not enhanced by the disorder scattering rate. The orbital magnetization that exists
already in the normal state and the contribution of the charge imbalance are
not evaluated.
\end{abstract}

\maketitle

\section{Introduction}
Studies of photogalvanic phenomena in superconductors have a rich history dating back to the 1970s \cite{Artemenko1974,Aronov1976,Artemenko1979}. Initially it was  argued that it would be impossible to generate a constant electric field deep in the bulk of a superconductor, i.e. on length scales greatly exceeding the London penetration depth. The argument is based on the equation of motion for the condensate $\dot{\bp}_s=e{\mathbf E}+\mbox{\boldmath $\nabla$}\mu$, where $\bp_s$ is the momentum of the condensate, $\mu$ is electrochemical potential and ${\mathbf E}$ is the static electric field. One expects the gradient of the electrochemical potential to be small and, if one neglects it, it follows that the condensate will be accelerated by the electric field continuously. However, experiments showed that the static electric fields can indeed be generated in the bulk of a superconductor without accelerating the condensate \cite{Artemenko1979}. It was then immediately realized that in the steady state the equality $e{\mathbf E}=-\mbox{\boldmath $\nabla$}\mu$ must hold  \cite{Aronov1976}. This condition implies that under the action of external electromagnetic field there is a redistribution of electronic charge between the electron-like and hole-like branches which leads to the appearance of nonzero gradient of the electrochemical potential, $\mbox{\boldmath $\nabla$}\mu\not =0$.

Subsequent works have demonstrated that indeed the effect turns out to be small, i.e. of the order of  $T_c/\veps_F$ where $T_c$ is the superconducting critical temperature and $\veps_F$ is the Fermi energy \cite{Aronov1976,Zaitsev1986}. It must be mentioned that later studies have demonstrated that in the presence of magnetic impurities (or spin-orbit coupling) the magnitude of the effect will be defined by the corresponding scattering rates rather than by the ratio $T_c/\veps_F$ \cite{ZaiZai1982}. In particular, this means that in unconventional superconductors, irrespective of whether they are nodal or nodeless, the effect is expected to be small, since even a small amount of nonmagnetic impurities is sufficient to strongly suppress superconductivity \cite{MineevSamokhin1999}. Whether the magnitude of the effect will be of the order $T_c/\veps_F$ or will become of the order of $1/(\tau_uT_c)$, where $\tau_u^{-1}$ is the relaxation rate for the scattering on potential impurities, remains an open problem. 

In conventional superconductors the branch population imbalance induced by an external field leads to a number of remarkable physical effects, such as thermoelectric effects \cite{Elesin1981}, photoelectric effect in superconducting-normal hybrid structures \cite{Zaikin2015}, pairing instabilities \cite{Galperin1981} and inverse Faraday effect \cite{Mironov2021-IFESC}. The inverse Faraday effect (IFE) consists in the static magnetization induced by circularly polarized light \cite{Pershan1966,Battiato2014,yang2022inverse,mou2023reversed,mou2023chiral}. In the existing theories the IFE of the superconducting condensate originates from the particle-hole asymmetry, which makes the relaxation constant of the time-dependent GL equation complex, and, as for the branch imbalance discussed above, the induced magnetization is proportional to a dimensionless constant $\sim T_c/\veps_F$ \cite{Mironov2021-IFESC,Parafilo2022Fl,Croitoru2022,Buzdin2023,Putilov2023-IFESC,Dzero-IFE2024}. 

Recent theories of the IFE in conventional superconductors have employed the Ginzburg-Landau (GL) formalism \cite{Mironov2021-IFESC,Putilov2023-IFESC}. It is well known that GL theory is accurate at temperatures close to the critical temperature  \cite{GL}. Given that the IFE is intrinsically nonequilibrium phenomenon, the applicability of the GL formalism is further limited by the condition that the time scale on which the order parameter varies $\tau_\Delta(T)=\hbar/\Delta(T)$ represents the longest time scale in the problem, i.e. quasiparticle excitations must relax much faster than the superconducting order parameter $\Delta(t)$ \cite{Elihu1966,GorkovEliashberg1968a,GorkovEliashberg1968b,Kopnin2001}. Naively, in the absence of paramagnetic impurities the applicability of the GL approach is limited by the condition $\Delta^2/T_c\ll T_c^2/\veps_F$. As it was shown by Gor'kov and Eliashberg \cite{GorkovEliashberg1968a} conceptually the problem lies in the fact that the temporal fluctuations of the order parameter $\Delta(t)$ are generally non-local and, as a result, the GL expansion of the anomalous propagator in powers of the Fourier components  $\Delta_\omega$ of the pairing amplitude will contain integrals with singular kernels at frequencies which correspond to the single-particle excitation threshold $\omega_{\mathrm{th}}=2\Delta$. This type of singularities renders such an expansion uncontrolled. 

 These restrictions can be circumvented when pair breaking processes become dominant. Such a situation is realized when either the temperature is very close to the critical temperature or when paramagnetic impurities are present in a superconductor such that the order parameter is vanishingly small, so that single particles equilibrate on a pair breaking time scale $\tau_s\ll\tau_\Delta(T)$. As a result, in the gapless state the expansion in powers of the Fourier components of $\Delta(t)$ produces non-singular integrals which renders the GL theory suitable to describe the non-equilibrium dynamics. Note that in this case the magnitude of the IFE will be determined by the ratio $1/(T_c\tau_s)$ rather than $T_c/\veps_F$. 

A consistent microscopic theory of the IFE that goes
beyond the GL paradigm is thus lacking even for conventional
superconductors. Combined with the recent theoretical and experimental
interest in the nonlinear optical response of unconventional superconductors
\cite{Katsumi2018,Katsumi2020,Gedik2026,Awelewa2025,Kazi2026}, this defines the void the present work is aiming to fill. At first glance the quasiclassical approach to superconductivity should be an ideal tool for the description of the photogalvanic phenomena in superconductors. It has been noted by several authors, however, that the quasiclassical description in its canonical form assumes the particle-hole symmetry from the outset and for this reason is not suitable for the description of the effects which originate from the branch population imbalance \cite{Belzig1999,Kalenkov2014,Zaikin2015,Masaki2018}. In order to capture such effects with the quasiclassical formalism, one necessarily needs to extend it by including the terms of order $\eps/\veps_F$ (here $\eps$ is some characteristic energy scale), which are absent from the canonical theory.

A separate motivation to study this problem is provided by the physics of the collective amplitude (Schmid-Higgs)
mode. Since the amplitude mode is charge neutral, it couples
to the electromagnetic field only quadratically at zero photon momentum. It is
therefore invisible in the linear response, and its observation typically relies on
intense terahertz driving and higher-harmonic generation --- an approach
recently extended to the $d$-wave cuprates \cite{Katsumi2018,Katsumi2020},
where the dynamics of the corresponding Schmid--Higgs mode exhibits a number of
peculiar features \cite{Awelewa2025,Kazi2026}. As we propose below, the rectified
response offers an alternative, dc route to the same physics: at finite photon
momentum and to first order in the particle-hole asymmetry the radiation
induces \emph{a linear-in-field} correction to the order parameter amplitude. This
correction is fully symmetric under the point group of the crystal, exists for
any singlet pairing, and contributes to the dc current with a weight
proportional to the pair susceptibility. Furthermore, it can be resonantly enhanced
near the pair-breaking threshold, $\omega=2\Delta$ for an $s$-wave pairing gap and
the maximal-gap threshold $\omega=2\sqrt2\Delta$ for $d$-wave pairing. The inverse Faraday effect thus
carries a fingerprint of the Higgs mode in a static observable, with the
pairing symmetry encoded in the lineshape of this feature rather than in a
selection rule.

The rectified response has one more ingredient which can be easily overlooked:
potential disorder. A dc current at second order in the field is produced only
if the radiation is absorbed, and a quasiparticle in a state with momentum
$\bp$ and energy $E_\bp$ can absorb a photon of energy $\omega$ only if its
momentum changes during the absorption, $\bp\to\bp'$: in a single-band metal
there is no state with energy $E_\bp+\omega$ at the same momentum. Keeping the photon momentum finite
does not help, since the corresponding energy transfer,
$v_Fq=(v_F/c)\,\omega\ll \omega$, is too small. Elastic scattering on
potential impurities provides the momentum relaxation channel as it unpins the
quasiparticles from the condensate, which for isotropic pairing remains immune
to potential scattering. The same obstruction is well known for the nonlinear
response at vanishing photon momentum: a spatially uniform drive can be
removed from a clean, Galilean-invariant condensate by a boost, so that the
coupling linear in the vector potential excites neither the Higgs mode nor a
third-harmonic signal unless impurity scattering is present
\cite{Silaev2019-Disorder}, and a spatially uniform drive of a clean $d$-wave
superconductor produces pair breaking but no absorption, and hence no
Eliashberg effect \cite{SteinemanDzero-Eliashberg}. At finite photon momentum
the particle-hole asymmetric vertices provide an excitation channel for the amplitude mode at linear
order in the field, without any help from disorder, but the
conversion of this oscillation into a dc current again requires absorption.
In a strictly clean superconductor this channel of the rectified current
vanishes, and the limit of vanishing scattering rate is singular: the
impurity-assisted absorption amplitude is proportional to the scattering rate
$1/\tau$, while the current-carrying, anisotropic part of the nonequilibrium
distribution relaxes by the same elastic scattering over the time $\tau$. The
two factors cancel each other, so that for any finite amount of disorder
the result is finite and, in the clean limit, independent of $\tau$ --- just as
the photon-drag current of a normal metal at $\omega\tau\gg1$. In
contrast with the Eliashberg effect \cite{SteinemanDzero-Eliashberg}, inelastic
relaxation does not enter the result, since the isotropic part of the
nonequilibrium distribution, whose steady state it would control, does not
carry current at this order. One also has to keep in mind that for $d$-wave pairing potential disorder is also a
pair breaker \cite{MineevSamokhin1999}, which restricts our treatment to the
ballistic regime, $\tau T_c\gg 1$, only.

In this paper we formulate a fully microscopic theory of the inverse Faraday
effect in conventional ($s$-wave) and unconventional ($d$-wave)
superconductors. Our theory
is based on an extension of the quasiclassical formalism that retains
the particle-hole asymmetric vertices responsible for the rectified
response and is applicable in a wide range of temperatures. We
compute self-consistently the two collective fields induced at linear order
--- the electrochemical potential describing the imbalance and the amplitude
(Schmid--Higgs) oscillation of the order parameter --- and then evaluate the
rectified current density, which is quadratic in the field amplitude and
linear in the photon momentum. The terms of order $\eps/\veps_F$ which control the
order parameter response are obtained from the exact finite-momentum
expansion of the Dyson equation rather than from a gradient expansion.  As it turns out, the result can be organized into a small number of
additive channels with sharply different physical content: a photon-drag
channel, a diamagnetic channel, an imbalance channel controlled by the
functions $\zeta_\w$ and $\Pi_\mu$, and an order parameter channel through
which the light-induced Higgs mode imprints the pair susceptibility
$\chi_{SH}(\w)$ directly onto the static magnetization. We evaluate two leading channels -- the
order parameter and diamagnetic ones -- while the imbalance channel is identified
but not evaluated. The order parameter channel is driven by
the longitudinal part of the in-plane electric field and it can be distinguished from the imbalance channel by its
frequency dependence. We show that it is
of first order in the particle-hole asymmetry, as anticipated by the naive
estimate $\sim T_c/\varepsilon_F$, and in the clean limit it does not depend
on the disorder scattering rate, so that it contains no relaxation enhancement. In what follows 
we will use the units $\hbar=c=1$ throughout
this paper and we adopt the following notations: $(\mathbf{a}\mathbf{b})\equiv\mathbf
a\cdot\mathbf b$ denotes the scalar product, $[\mathbf a\times\mathbf b]$ is the
vector product, $\langle\dots\rangle_\bn$ is the average over the direction
of the Fermi velocity, and $[A,B]_\pm=AB\pm BA$.

\section{Model and formalism}
We consider a one-band model of 2D fermions with an attractive interaction in the $d$-wave Cooper channel in the presence of potential disorder. Our model Hamiltonian reads: 
\beg\label{Eq1}
\begin{aligned}
\hat{\cal H}&=\sum_{\bk,\sigma} \xi_\bk \hat{c}^\dagger_{\bk\sigma}\hat{c}_{\bk\sigma}\\&+
    \sum_{\bk,\bp,\bK} V_{\bk\bp}^{({\mathrm{d}})} \hat{c}^\dagger_{\bk+\bK/2\uparrow} \hat{c}^\dagger_{-\bk+\bK/2\downarrow}\hat{c}_{-\bp+\bK/2\downarrow}\hat{c}_{\bp+\bK/2\uparrow}\\&+
    \sum_{\bk,\bp,\sigma} U_{\bk-\bp}\,\hat{c}^\dagger_{\bk\sigma}\hat{c}_{\bp\sigma},
\end{aligned}
\en
 where 
 $\hat{c}_{\bk\sigma}^\dagger$ ($\hat{c}_{\bk\sigma}$) are the creation (annihilation) fermionic operators, $\bk$ is momentum, $\sigma=\up,\dn$ is a spin,  
 $V_{\bk\bp}^{({\mathrm{d}})} $ is the pairing interaction,  $\xi_\bk= k^2/2m - \veps_F$ is the single particle dispersion, $\veps_F$ is the Fermi energy, and $U_{\bk-\bp}$ is the Fourier component of the potential $U(\br)=\sum_ju(\br-\br_j)$ of nonmagnetic impurities randomly distributed with concentration $n_{\rm imp}$. 
 We project $V_{\bk\bp}^{({\mathrm{d}})} $ into the $d$-wave channel  and within the weak-coupling limit approximate it as  
\beg\label{Vkp}
V_{\bk\bp}^{(\mathrm{d})}=- g{\cal Y}(\phi_\bk)\, {\cal Y}(\phi_\bp),
\en
where $g>0$ is the coupling constant, ${\cal Y}(\phi_\bk)=\sqrt{2}\cos2\phi_\bk$ is the normalized $d$-wave form factor and $\phi_\bk$ defines the direction of the momentum on the Fermi surface. As in any weak coupling theory the interaction acts within the energy shell $|\xi_\bk|,|\xi_\bp|\leq\omega_D$. The $s$-wave case is obtained by setting ${\cal Y}=1$.

We take the impurities to be point-like, $u(\br)=u_0\,\delta(\br)$, so that the scattering is isotropic, and account for it in the Born approximation through the impurity self-energy of the quasiclassical equations \cite{Kopnin2001,SteinemanDzero-Eliashberg},  with the elastic scattering rate 
\beg\label{ElasticRate}
\frac{1}{\tau}=2\pi\nu_Fn_{\rm imp}|u_0|^2, 
\en
where $\nu_F$ is the density of states per spin at the Fermi level. We will consider the clean limit $1/\tau\ll{\rm min}(\Delta,\omega)$.     For $d$-wave pairing, where potential scattering is pair breaking \cite{MineevSamokhin1999}, the clean limit also requires that the scattering rate satisfies $1/\tau\ll T_c$. The impurity scattering is essential for the driven problem: as explained in the Introduction, it provides the momentum relaxation without which the quasiparticles cannot absorb the radiation.

We consider the situation when a superconductor is subjected to an external electromagnetic radiation with the vector potential 
\beg\label{VectorPotential}
{\mathbf A}(\br,t)=\left(\frac{{\mathbf E}}{i\omega}\right)e^{i(\bq\br-\omega t)}+\left(-\frac{{\mathbf E}^*}{i\omega}\right)e^{-i(\bq\br-\omega t)},
\en
where $\bq$ is the in-plane component of the photon momentum, $\omega$ is the frequency and ${\mathbf E}$ is the complex amplitude of the in-plane electric field.
In order to describe the non-equilibrium state of our system, we introduce the Green's function $\check{G}$ which is a four-by-four matrix defined in Keldysh and Nambu spaces. We use the Keldysh formulation rather than the Matsubara technique with a subsequent analytic continuation for the following reason: the dc current at second order in the field is carried by the light-induced change of the quasiparticle populations, whose creation requires the impurity-assisted absorption of photons (Sec.~IV), and by the branch-imbalance potential, which is fixed by a kinetic constraint. Both are naturally handled in the kinetic-equation language. For the linear response of the order parameter, where the two techniques are equally convenient, we use the Matsubara form and continue analytically (Appendix \ref{AppendixA}), which also provides an independent check of the structure of the Keldysh kernels. The function $\check G$ satisfies the Dyson equation \cite{LO}:
\beg\label{DysonEq}
\left(\hat{G}_0^{-1}-\check{\Sigma}\right)\circ\check{G}=1, \quad \check{G}=\left(\begin{matrix} \hat{G}^R & \hat{G}^K \\ 0 & \hat{G}^A \end{matrix}\right).
\en
Here $\hat{G}_0^{-1}=i\partial_t-\hat{H}_0$, where $\hat{H}_0$ is the non-interacting part of the model Hamiltonian \eqref{Eq1}. It includes the external electromagnetic field, which enters the single-particle energy through the substitution $\bk\to\bk-e{\mathbf A}$, but not the impurity potential, which is accounted for by the self-energy. The self-energy $\check{\Sigma}$ is due to the pairing interaction and to impurity scattering, and $(\hat{A}\circ\hat{B})(x,x')=\int dx''\hat{A}(x,x'')\hat{B}(x'',x')$ defines the convolution of two matrix functions, with $x=(\br,t)$. In passing we note that $\check{G}$ also satisfies the Dyson equation with respect to its second coordinate, $\check{G}\circ(\hat{G}_0^{-1}-\check{\Sigma})=1$.  
We proceed by applying the Wigner transformation:
\beg\label{Wigner}
\check{G}(x_1,x_2)=\int\frac{d\eps}{2\pi}\int\frac{d^2\bp}{(2\pi)^2}\check{G}_{\bp\eps}(x)e^{i(\bp\delta\br-\eps\delta t)},
\en
where $\delta\br=\br_1-\br_2$, $\delta t=t_1-t_2$ and $x=(x_1+x_2)/2$. We combine \eqref{DysonEq} with the second Dyson equation which acts on the second argument of the Green's function. This results in the following equation for $\check{G}_{\bp\eps}(\br,t)$ \cite{Zaikin2015}:
\beg\label{EilenMainG}
\begin{aligned}
&\left[\eps\check{\tau}_3+\check{\Sigma}_\bn\stackrel{\circ},\check{G}_{\bp\eps}\right]_{-}+\frac{i}{2}\left[\check{\tau}_3,\partial_t\check{G}_{\bp\eps}\right]_{+}\\&+\frac{e}{m}\left[\bp{\mathbf A}(\br,t)\check{\tau}_3\stackrel{\circ},\check{G}_{\bp\eps}\right]_{-}=-\frac{i}{m}(\bp\mbox{\boldmath $\nabla$}_\br)\check{G}_{\bp\eps},
\end{aligned}
\en
where $[A,B]_{\pm}=AB\pm BA$, $\check{\Sigma}_\bn(\br,t)=\check\Delta_\bn(\br,t)+\mu(\br,t)\check{\tau}_0-\check\sigma_{\rm dis}(\br,t)$,
$\mu(\br,t)$ is the electrochemical potential which appears due to induced population imbalance between the electron-like ($\xi_\bk>0$) and hole-like ($\xi_\bk<0$) branches, and $\check\sigma_{\rm dis}$ is the self-energy due to the scattering on impurities, Eq.~\eqref{sigma-dis} below.  Here and below $\check G_{\bp\eps}$ denotes $\check\tau_3$ times the Wigner transform of the function in Eq.~\eqref{DysonEq}. In this representation the pairing field and the impurity self-energy enter through $\check\Sigma_\bn-\mu\check\tau_0=-\check\Sigma\check\tau_3$, while $\mu$ is the self-consistent potential determined in Sec.~III. Superconducting pairing field is
\beg\label{SigmaPairing}
\check{\Delta}_\bn(\br,t)={\cal Y}_\bn\left[(i\check{\tau}_2)\Delta(\br,t)+(i\check{\tau}_1){\Delta}'(\br,t)\right],
\en
where ${\cal Y}_\bn\equiv{\cal Y}(\phi_\bn)$, $\Delta$ and $\Delta'$ are the real and imaginary parts of the order parameter, matrices $\check{\tau}_a$ act in Nambu space and are all diagonal in Keldysh space. The convolution is given by the Groenewold-Moyal product rule:
\beg\label{Convo}
\begin{aligned}
&\left[\check{A}(\br,t)\stackrel{\circ},\check{B}(\br,t)\right]_{-}\\&=\check{A}(\br,t)e^{\frac{i}{2}\left(\stackrel{\leftarrow}\partial_\br\stackrel{\rightarrow}\partial_\bp-\stackrel{\leftarrow}\partial_t\stackrel{\rightarrow}\partial_\eps-\stackrel{\leftarrow}\partial_\bp\stackrel{\rightarrow}\partial_\br+\stackrel{\leftarrow}\partial_\eps\stackrel{\rightarrow}\partial_t\right)}\check{B}(\br,t)\\&-
\check{B}(\br,t)e^{\frac{i}{2}\left(\stackrel{\leftarrow}\partial_\br\stackrel{\rightarrow}\partial_\bp-\stackrel{\leftarrow}\partial_t\stackrel{\rightarrow}\partial_\eps-\stackrel{\leftarrow}\partial_\bp\stackrel{\rightarrow}\partial_\br+\stackrel{\leftarrow}\partial_\eps\stackrel{\rightarrow}\partial_t\right)}\check{A}(\br,t).
\end{aligned}
\en
In the convolution with the pairing field we keep the time dependence exactly and neglect the spatial gradients. The order parameter oscillates at the frequency $\omega\sim 2\Delta$, so that an expansion in its time derivatives is not allowed. Its spatial gradients can be neglected because the order parameter is uniform in equilibrium and its field-induced correction is already of order $\eps/\veps_F$, so that the gradient terms it generates are of second order in the particle-hole asymmetry. Throughout we also assume $v_Fq\ll\Delta$. Thus,
   \beg\label{AppConvo}
   \left[\check{\Delta}_\bn\stackrel{\circ},\check{G}_{\bp\eps}\right]_{-}\approx\check{\Delta}_\bn e^{-\frac{i}{2}\stackrel{\leftarrow}\partial_t\stackrel{\rightarrow}\partial_\eps}\check{G}_{\bp\eps}-\check{G}_{\bp\eps}e^{\frac{i}{2}\stackrel{\leftarrow}\partial_\eps\stackrel{\rightarrow}\partial_t}\check{\Delta}_\bn .
   \en
At the same time, as we have mentioned above, the gradient term with respect to $\br$ involving the electrochemical potential turns out to be very important for the observation of various nonlinear transport phenomena such as thermoelectric and photogalvanic effects since the external electromagnetic field causes an inhomogeneous charge redistribution. Since we expect the effect to be small, we will make the following approximation:
\beg\label{PhiConvo}
\begin{aligned}
&\left[{\mu}(x)\check{\tau}_0\stackrel{\circ},\check{G}_{\bp\eps}(x)\right]_{-}\approx
{i}{\mbox{\boldmath $\nabla$}}_\br{\mu}\cdot{\mbox{\boldmath $\nabla$}}_\bp\check{G}_{\bp\eps}\\&+{\mu}(x)e^{-\frac{i}{2}\stackrel{\leftarrow}\partial_t\stackrel{\rightarrow}\partial_\eps}\check{G}_{\bp\eps}(x)-\check{G}_{\bp\eps}(x)e^{\frac{i}{2}\stackrel{\leftarrow}\partial_\eps\stackrel{\rightarrow}\partial_t}{\mu}(x).
\end{aligned}
\en
It also proves convenient to re-write the momentum gradient of the Green's function as
\beg\label{NablapGpeps}
\begin{aligned}
{\mbox{\boldmath $\nabla$}}_\bp\check{G}_{\bp\eps}(\br,t)&=(v_F\mathbf n)\,\frac{p(\xi_{\mathbf p})}{p_F}\,
\frac{\partial\check G_{\mathbf p\epsilon}}{\partial\xi_{\mathbf p}}
\\&+
\frac{v_F\hat{\phi}_{\mathbf n}}{2\varepsilon_F}
\left(1+\frac{\xi_{\mathbf p}}{\varepsilon_F}\right)^{-1/2}
\frac{\partial\check G_{\mathbf p\epsilon}}{\partial\phi_{\mathbf n}},
\end{aligned}
\en
where $p(\xi_\bp)=p_F\sqrt{1+\xi_\bp/\veps_F}$ and $\hat{\phi}_{\mathbf n}=(-\sin\phi_{\mathbf n},\cos\phi_{\mathbf n})$ is a unit vector. Acting on the equilibrium function, the second term is nonzero only for anisotropic pairing. Acting on the field-induced corrections, which depend on the direction $\bn$, it is nonzero for any pairing. Once the gradient acts on a function of $\br$ with the wave vector $\bq$, both terms are of the same order, $O(v_Fq/\veps_F)$, and both must be retained. In Appendix \ref{AppendixA} the linear response of the order parameter is obtained from the exact solution of the Dyson equation for a single Fourier component of the field, which contains the radial and the angular parts of the momentum gradient, as well as the momentum dependence of the velocity in the vertices, without any expansion. 
With these provisions, we can now develop a perturbative solution of the kinetic equation \eqref{EilenMainG} up to the second order in powers of ${\mathbf E}$, Eq. \eqref{VectorPotential}. 

We also need the gradient terms of first order in $q$ in the last term on the left hand side of Eq.~\eqref{EilenMainG}. Taking into account the spatial and momentum gradients only we have
\beg\label{ExpandLinearA}
\begin{aligned}
&\left[\bp{\mathbf A}(\br,t)\check{\tau}_3\stackrel{\circ},\check{G}_{\bp\eps}\right]_{-}\approx p_F\left[\bn{\mathbf A}(\br,t)\check{\tau}_3\stackrel{\circ},\check{G}_{\bp\eps}\right]_{-}\\&-\frac{i}{2}\left[\check{\tau}_3,({\mathbf A}{\mbox{\boldmath $\nabla$}}_\br)\check{G}_{\bp\eps}\right]_{+}+\frac{i}{2}{\mbox{\boldmath $\nabla$}}_\br(\bp{\mathbf A})\left[\check{\tau}_3,{\mbox{\boldmath $\nabla$}}_\bp\check{G}_{\bp\eps}\right]_{+}.
\end{aligned}
\en
 In the first term we have set $\bp=p_F\bn$. The correction $[p(\xi_\bp)-p_F](\bn{\mathbf A})$ is of order $\xi_\bp/\veps_F$ and contributes to the dc current at first order in $q$ only together with the longitudinal $O(q)$ part of $\check g_1$. These contributions will not be evaluated in this work. As we will see below, the terms of order $\eps/\veps_F$ produce a linear correction to the order parameter order parameter amplitude which then feeds back into the dc component of the current density at second order in the electric field.  We emphasize that in the linear response of the order parameter three and only three sources of particle-hole asymmetry appear: the momentum dependence of the quasiparticle velocity in the paramagnetic vertex, the same dependence in the drift term, and the shift of the direction of the momentum, which changes the gap function $\Delta_\bn$ for anisotropic pairing. The density of states of a two-dimensional parabolic band carries no energy dependence. We use the gradient-expanded kinetic equation for the particle-hole symmetric part of the problem and for the particle-hole asymmetric sources at second order. The $O(\eps/\veps_F)$ contribution to the linear order parameter response is obtained from the exact expansion of Appendix~\ref{AppendixA}.

\section{Perturbative solution of the Eilenberger equation}
The Eilenberger equation can be derived from equation \eqref{EilenMainG} by integrating it over $\xi_\bp$. The quasiclassical function is defined according to \cite{Eilenberger1968,LO-Vortex}
\beg\label{Quasi}
\check{g}(\bn,\eps;\br,t)=\frac{i}{\pi}\int\limits_{-\infty}^{\infty}d\xi_\bp \check{G}_{\bp\eps}(\br,t)
\en
and it satisfies the normalization condition \cite{Belzig1999}
\beg\label{norm}
\check{g}\circ\check{g}=\check{1}.
\en

The equation for the function $\check{g}$ reads
\begin{widetext}
\beg\label{EilenMain}
\begin{aligned}
&[\eps\check{\tau}_3+\check\Delta_\bn(\br,t)\stackrel{\circ},\check{g}]_{-}+\left[{\mu}(\br,t)\check{\tau}_0\stackrel{\circ},\check{g}\right]_{-}+i{v}_F(\bn{\mbox{\boldmath $\nabla$}_\br})\check{g}+\frac{i}{2}\left[\check{\tau}_3,\partial_t\check{g}\right]_{+}+{ev_F}[\bn{\mathbf A}(\br,t)\check{\tau}_3\stackrel{\circ},\check{g}]_{-}-\big[\check\sigma_{\rm dis}\stackrel{\circ},\check g\big]_{-}\\&=-i(v_F\bn)\left({\mbox{\boldmath $\nabla$}}_\br\mu(\br,t)\right)\check{\Lambda}_{\bn\eps}(\br,t)-\frac{iv_F}{2\veps_F}\big((\hat\phi_\bn\!\cdot\!{\mbox{\boldmath $\nabla$}_\br})\mu(\br,t)\big)\,\partial_{\phi_\bn}\check g-\;\frac{i}{2}\,e v_F^{2}\,
\big(n_i n_j\,\partial_i A_j(\mathbf r,t)\big)\,\check\Theta_{\mathbf n\epsilon}(\mathbf r,t)\\&+\;\frac{ie}{2m}\,
\big[\check\tau_3,\,(\mathbf A\!\cdot\!{\mbox{\boldmath $\nabla$}_\br})\,\check g\big]_{+}-\frac{ie}{2m}\big((\hat\phi_\bn\!\cdot\!{\mbox{\boldmath $\nabla$}_\br})(\bn{\mathbf A}(\br,t))\big)\big[\check\tau_3,\,\partial_{\phi_\bn}\check g\big]_{+}.
\end{aligned}
\en
\end{widetext}
Here convolutions are taken with respect to temporal variables only, and we introduced the functions
\beg\label{Theta}
\begin{aligned}
&\check\Theta_{\mathbf n\epsilon}(\mathbf r,t)
=\frac{i}{\pi}\int\limits_{-\veps_F}^\infty d\xi_\bp\left(1+\frac{\xi_\bp}{\veps_F}\right)\left[\check{\tau}_3,\frac{\partial \check{G}_{\bp\eps}}{\partial \xi_\bp}\right]_{+}, \\
&\check{\Lambda}_{\bn\eps}(\br,t)=\frac{i}{\pi}\int\limits_{-\veps_F}^\infty d\xi_\bp\sqrt{1+\frac{\xi_\bp}{\veps_F}}\frac{\partial}{\partial \xi_\bp}\check{G}_{\bp\eps}(\br,t),
\end{aligned}
\en
which enter the second-order sources listed in Appendix \ref{AppendixSource}.

The last term on the left hand side of Eq.~\eqref{EilenMain} accounts for the elastic scattering on point-like nonmagnetic impurities. The corresponding self-energy has the same matrix structure as $\check g$ and is given by
\beg\label{sigma-dis}
\check\sigma_{\rm dis}=-\frac{i}{2\tau}\langle\check g\rangle_\bn ,
\en
 which is valid within the Born approximation, with the elastic scattering rate $1/\tau$ given by Eq.~\eqref{ElasticRate}. For isotropic pairing, solving the Eilenberger equation with this self-energy in the presence of the field was shown to be equivalent to the exact summation of the impurity-ladder diagrams with current vertices \cite{Silaev2019-Disorder}. Since $\check\sigma_{\rm dis}$ does not depend on the momentum, and at linear order in the field $\langle\check g\rangle_\bn$ changes only at order $q$, the gradient corrections to the convolution of $\check\sigma_{\rm dis}$ with $\check G_{\bp\eps}$ in Eq.~\eqref{EilenMainG} are of second order in $q$ and have been omitted. 

We will work in the clean limit, $1/\tau\ll\min(\Delta,\omega)$, and keep all the terms to the leading order in $1/(\tau\Delta)$. It is important to realize that this limit has different consequences for the linear and for the second-order response. The linear response functions of Sec.~III B remain finite for $\tau\to\infty$, and in them the impurity term can be dropped. In their numerical evaluation the infinitesimal $i0$ in the retarded and advanced functions is replaced by $i\gamma$ with $\gamma=10^{-3}\Delta$. In the second-order dc response the impurity term must be retained, however small $1/\tau$ is. This is because the dc current necessarily requires the absorption of the radiation, and a quasiparticle can absorb a photon only if its momentum changes. For $v_Fq\ll\omega$ this change is supplied by the impurities, so that the absorption is proportional to $1/\tau$. The current-carrying part of the resulting nonequilibrium distribution relaxes by the same elastic scattering over the time $\tau$, which enters through the denominators $\eta^R_{\bn\eps}+\eta^A_{\bn\eps}$ of order $1/\tau$ (see discussion in Sec.~III A below). The two factors cancel, and the contribution of the impurity scattering to the dc current remains finite for $\tau\to\infty$ and does not depend on $\tau$.    It is absent only when there are no impurities at all, $n_{\rm imp}=0$. A weakly disordered superconductor is therefore not described by a theory without impurities, however weak the disorder is. Such a theory misses the order parameter contribution to the dc current altogether, and a phenomenological damping of the quasiparticles does not reproduce it either, as we explicitly demonstrate in Appendix~\ref{AppendixC}.
   
The terms on the right hand side of Eq.~\eqref{EilenMain} are of the order $\eps/\veps_F$ and represent the extension of the usual Eilenberger equation: the terms with $\check\Lambda$ and $\check\Theta$ contain the radial part of the momentum gradient \eqref{NablapGpeps}, the terms with $\partial_{\phi_\bn}\check g$ its angular part, and the remaining term is the spatial gradient of \eqref{ExpandLinearA}. The last term on the right hand side must be kept together with the term containing $({\mathbf A}\cdot{\mbox{\boldmath $\nabla$}}_\br)\check g$. Both come from Eq.~\eqref{ExpandLinearA}. The term with $({\mathbf A}\cdot{\mbox{\boldmath $\nabla$}}_\br)\check g$ accounts for the change of the quasiparticle velocity by $-e{\mathbf A}/m$. The last term accounts for the rotation of the quasiparticle momentum by the component of the force $(e/m){\mbox{\boldmath $\nabla$}}_\br(\bp{\mathbf A})$ perpendicular to $\bp$. At second order in the field, the dc current produced by either term alone contains a part that flows perpendicular to $\bq$, reverses its direction when the helicity of the light is reversed, and is proportional to $\tau$. As we show in Sec.~IV, these two parts cancel each other. Lastly, the convolution with $\mu(\br,t)$, the second term on the left hand side of Eq.~\eqref{EilenMain}, contains only the derivatives with respect to time and energy. It originates from the last two terms in Eq.~\eqref{PhiConvo}. We keep this term for completeness. It does not contribute to the dc component of the second-order current density.
 
\subsection{Ground state}
    In the ground state we set $\Delta'=0$ in Eq.~\eqref{SigmaPairing}, so that
 \beg\label{DLTgs}
 \hat{\Delta}_\bn=i\hat{\tau}_2\Delta_\bn,
 \en
 where $\Delta_{\bn}={\cal Y}_\bn\Delta$. 
 We represent the retarded and advanced parts of the ground state correlator $\hat{\cal G}_{\bn\eps}^{R(A)}$ as follows
\beg\label{ggsRA}
\hat{\cal G}_{\bn\eps}^{R(A)}=\hat{\tau}_3 g_{\bn\eps}^{R(A)}+i\hat{\tau}_2f_{\bn\eps}^{R(A)},
\en
while in equilibrium the Keldysh component is
\beg\label{ggsK}
\hat{\cal G}_{\bn\eps}^{K}=\left(\hat{\cal G}_{\bn\eps}^{R}-\hat{\cal G}_{\bn\eps}^{A}\right)\tanh\left(\frac{\eps}{2T}\right),
\en
where $T$ is the temperature.

In equilibrium $\check{\cal G}_{\bn\eps}$ satisfies a much simplified Eilenberger equation
\beg\label{EilenbergerGS}
 \left[\eps\check{\tau}_3+\check{\Delta}_{\bn}-\check\sigma_{\rm dis},\check{\cal G}_{\bn\eps}\right]_{-}=0.
\en
The impurity self-energy is determined by the two Fermi-surface averages
\beg\label{avgG}
\langle\hat{\cal G}^R_{\bn\eps}\rangle_\bn=\hat\tau_3{\cal G}_\eps+i\hat\tau_2{\cal F}_\eps ,
\en
and Eq.~\eqref{EilenbergerGS} takes the form of the Eilenberger equation of a clean superconductor in which $\eps$ and $\Delta_\bn$ are replaced by
\beg\label{xitilde}
\tilde\eps^R_\eps=\eps+\frac{i}{2\tau}{\cal G}_\eps ,\qquad \tilde\Delta^R_{\bn\eps}=\Delta_\bn+\frac{i}{2\tau}{\cal F}_\eps ,
\en
and $\tilde\eps^A_\eps=[\tilde\eps^R_\eps]^*$, $\tilde\Delta^A_{\bn\eps}=[\tilde\Delta^R_{\bn\eps}]^*$ for the advanced part. As a result, the propagators take the BCS form
\beg\label{gfeta}
\begin{aligned}
&g^{R(A)}_{\bn\eps}=\frac{\tilde\eps^{R(A)}_\eps}{\eta^{R(A)}_{\bn\eps}},\qquad
f^{R(A)}_{\bn\eps}=\frac{\tilde\Delta^{R(A)}_{\bn\eps}}{\eta^{R(A)}_{\bn\eps}},\\
&\eta^{R(A)}_{\bn\eps}=\sqrt{\big(\tilde\eps^{R(A)}_\eps\big)^2-\big(\tilde\Delta^{R(A)}_{\bn\eps}\big)^2},
\end{aligned}
\en
with $\eta^A_{\bn\eps}=-[\eta^R_{\bn\eps}]^*$ and the analytic branch fixed by ${\rm Im}\,\eta^R_{\bn\eps}>0$, so that $[g^{R(A)}_{\bn\eps}]^2-[f^{R(A)}_{\bn\eps}]^2=1$ in accordance with the normalization condition \eqref{norm}. 
In the clean limit, $\tau\to\infty$, these expressions reduce to $g_{\bn\eps}^R={\eps}/{\eta_{\bn\eps}^R}$ and $f_{\bn\eps}^R={\Delta_{\bn}}/{\eta_{\bn\eps}^R}$ with
\beg\label{etaRA}
\eta_{\bn\eps}^{R(A)}=\left\{\begin{aligned} &\pm\mathrm{sgn}(\eps)\sqrt{(\eps{\pm i0})^2-|\Delta_{\bn}|^2}, ~|\eps|\geq|\Delta_\bn|, \\
&i\sqrt{|\Delta_\bn|^2-\eps^2}, ~ |\eps|<|\Delta_\bn|.
\end{aligned}
\right.
\en
The advanced component of $\check{\cal G}$ can also be found using the rule $\hat{\cal G}^A=-\hat{\tau}_3[\hat{\cal G}^R]^\dagger\hat{\tau}_3$.

For $s$-wave pairing ${\cal G}_\eps=g^R_{\bn\eps}$, ${\cal F}_\eps=f^R_{\bn\eps}$, and Eqs.~\eqref{xitilde} and \eqref{gfeta} give $\tilde\eps^R_\eps/\tilde\Delta^R_{\bn\eps}=\eps/\Delta$, so that $g$ and $f$ coincide with their values in the clean limit (Anderson theorem), while $\eta^R_{\bn\eps}$ is shifted from its clean-limit value \eqref{etaRA} by $i/2\tau$. 
For $d$-wave pairing ${\cal F}_\eps\equiv0$. Indeed, Eqs.~\eqref{sigma-dis} and \eqref{EilenbergerGS} do not change under $\Delta_\bn\to-\Delta_\bn$, $\check{\cal G}\to\check\tau_3\check{\cal G}\check\tau_3$, and the rotation $\phi_\bn\to\phi_\bn+\pi/2$ changes the sign of ${\cal Y}_\bn$. Therefore $f_{\bn\eps}$ changes sign under this rotation and its average over the Fermi surface vanishes. Then $\tilde\Delta_{\bn\eps}=\Delta_\bn$, and inserting Eq.~\eqref{gfeta} into ${\cal G}_\eps=\langle g^R_{\bn\eps}\rangle_\bn$ reduces the impurity problem to a closed equation for a single complex function \cite{SteinemanDzero-Eliashberg},
\beg\label{Gfixedpoint}
{\cal G}_\eps=\frac{2}{\pi}K\!\left(\frac{\sqrt2\Delta}{\tilde\eps^R_\eps}\right),
\en
where $K(k)=\int_0^{\pi/2}d\vartheta/\sqrt{1-k^2\sin^2\vartheta}$ is the complete elliptic integral, continued to complex argument with the branch fixed by ${\cal G}_\eps\to1$ for $|\eps|\to\infty$.    The denominators $\eta^R_{\bn\eps}+\eta^A_{\bn\eps}=2i\,{\rm Im}\,\eta^R_{\bn\eps}$ of the second-order response are of order $1/\tau$ at energies where quasiparticles exist, so that the ground state has to be determined self-consistently to this order. In the subsequent numerical analysis of the nonlinear transport Eqs.~\eqref{xitilde} are solved at each energy. Apart from these denominators, the renormalizations can be neglected in the clean limit.

\paragraph{Functions $\check{\Lambda}_{\bn\eps}$ and $\check{\Theta}_{\bn\eps}$.} In the ground state the retarded part of function $\check{G}_{\bp\eps}$ is given by
\beg\label{hatGp}
\hat{G}_{\bp\eps}^R=\frac{\eps\hat{\tau}_3+i\hat{\tau}_2\Delta_\bn+\xi_{\bp}\hat{\tau}_0}{(\eps+i0)^2-\xi_\bp^2-|\Delta_\bn|^2}.
\en
Inserting these expressions into \eqref{Theta} and evaluating the integrals by parts where needed yields 
\beg\label{LambdaRAFinal}
\hat{\Lambda}_{\bn\eps}^{R(A)}\approx-\frac{\hat{\cal G}_{\bn\eps}^{R(A)}}{2\veps_F}, ~\hat\Theta_{\mathbf n\epsilon}^{R(A)}
\approx -\,\frac{[\hat\tau_3,\hat{\cal G}_{\bn\eps}^{R(A)}]_{+}}{\varepsilon_F}.
\en
Their Keldysh components follow from Eq.~\eqref{ggsK}. At first order in $q$ only $\check\Theta$ contributes to the dc component of the second-order current density (Sec.~IV), since $\check\Lambda$ appears in Eq.~\eqref{EilenMain} multiplied by the gradient of $\mu_1$, which is itself proportional to $(\bq{\mathbf E})$ (Sec.~III B), and therefore contributes only at second order in $q$.

\subsection{Linear analysis}
\paragraph{Retarded and advanced components.}
Using expressions above we can now compute the corrections to the quasi-classical function $\check{g}_{\bn\eps}$ in powers of electric field ${\mathbf E}$. Retaining terms up to the first order in electric field we write
\beg\label{Expand}
\check{g}(\bn\eps;\br t)=\check{g}_{0}+\check{g}_1(\bn\eps;\br t).
\en
Here $\check{g}_0=\check{\cal G}_{\bn\eps}$ is the quasiclassical propagator in the ground state. From the normalization condition (\ref{norm}) it follows that $\check{g}_1$  must satisfy $\check{g}_0\circ\check{g}_1+\check{g}_1\circ\check{g}_0=0$.

We start with the first order corrections to the retarded and advanced parts of $\check{g}_1$. 
Given \eqref{VectorPotential} we will look for $\hat{g}_1^{R(A)}(\bn\eps;\br t)$ in the form
\beg\label{g1four}
\hat{g}_1^{R(A)}(\bn\eps;\br,t)=\sum\limits_{s=\pm}\hat{g}_{1s}^{R(A)}(\bn\eps;q)e^{is(\bq\br-\omega t)}
\en
and we introduced the compact notation 
$q=(\bq,\omega)$.
Similarly, for the electrochemical potential we write
\beg\label{Scalar}
\mu_1(\br,t)=\sum\limits_{s=\pm}\varphi_{1s}(q)e^{is(\bq\br-\omega t)}.
\en
Electrochemical potential is a real quantity, which means that
$\varphi_{1-}(q)=\varphi_{1+}^*(q)$. Similar expression can be written for the linear correction for the order parameter
\beg\label{LineardeltaDelta1}
\delta\hat{\Delta}_1(\br,t)=i\hat{\tau}_2\sum\limits_{s=\pm}\delta\Delta_{1s}(q){\cal Y}_\bn e^{is(\bq\br-\omega t)}.
\en
Here we assumed that the phase of the order parameter can be absorbed into the chemical potential. 

Equation which determines the first order correction to the retarded and advanced components of $\hat{g}_1(\bn\eps;q)$ reads
\beg\label{Eq4g1}
\begin{aligned}
&\left[\eps\hat{\tau}_3+\hat\Delta_\bn,\hat{g}_{1+}\right]_{-}+\frac{1}{2}\left[{\omega}\hat{\tau}_3-{v}_F(\bn\bq)\hat{\tau}_0,\hat{g}_{1+}\right]_+\\&=
\left(\frac{ev_F}{i\omega}\right)(\bn{\mathbf E})\left(\hat{\cal G}_{\bn\eps+\frac{\omega}{2}}\hat{\tau}_3-
\hat{\tau}_3\hat{\cal G}_{\bn\eps-\frac{\omega}{2}}\right)+\hat{\cal R}_{\bn\eps}(q),
\end{aligned}
\en
where
\beg\label{calR}
\begin{aligned}
&\hat{\cal R}_{\bn\eps}=\varphi_{1+}\left\{\hat{\cal G}_{\bn\eps+\frac{\omega}{2}}-
\hat{\cal G}_{\bn\eps-\frac{\omega}{2}}\right\}\\&+
\hat{\cal G}_{\bn\eps+\frac{\omega}{2}}\,\delta\hat\Delta_{1+}
-\delta\hat\Delta_{1+}\,\hat{\cal G}_{\bn\eps-\frac{\omega}{2}}.
\end{aligned}
\en
and we have omitted the superscript $R,A$ for brevity.     Note that Eq.~\eqref{Eq4g1} does not contain the particle-hole asymmetric terms of Eq.~\eqref{EilenMain}, those with $\check\Lambda$, $\check\Theta$ and $\partial_{\phi_\bn}\check g$. The term with $\check\Lambda$ is proportional to $\varphi_1(\bn\bq)$, and since $\varphi_1\propto(\bq{\mathbf E})$, Eq.~\eqref{phi1pmMain}, it is of second order in $q$. The terms with $\check\Theta$ and $\partial_{\phi_\bn}\check g$ are of first order in $q$ and of order $v_Fq/\veps_F$. Their effect on the order parameter is included in $\delta\Delta_{1+}$, which we obtain in Appendix~\ref{AppendixA} from the exact solution of the Dyson equation. Their remaining effect on $\check g_1$ gives a contribution to the dc current that does not pass through the order parameter. Like the other contributions of this kind discussed in Sec.~IV A, it is not evaluated in this work.

From the normalization condition it follows
\beg\label{Norm1}
\hat{\cal G}_{\bn\eps+\frac{\omega}{2}}^{R(A)}\hat{g}_{1+}^{R(A)}+\hat{g}_{1+}^{R(A)}\hat{\cal G}_{\bn\eps-\frac{\omega}{2}}^{R(A)}=0.
\en
To solve Eq.~\eqref{Eq4g1} we introduce the projectors
\beg\label{Projectors}
\hat{\cal P}_{\lambda,\pm}=\frac{1}{2}\left(\hat{\tau}_0+\lambda\hat{\cal G}_{\bn\eps\pm\frac{\omega}{2}}\right), \quad \lambda=\pm.
\en
Given \eqref{Projectors} it follows that 
\beg\label{ProjectRel}
\hat{\cal G}_{\bn\eps\pm\frac{\omega}{2}}\hat{\cal P}_{\lambda,\pm}={\lambda}\hat{\cal P}_{\lambda,\pm}.
\en
Equation \eqref{Norm1} is equivalent to $\hat{\cal P}_{\lambda,+}\hat{g}_{1,+}\hat{\cal P}_{\lambda,-}=0$ for $\lambda=\pm$. For every source in Eqs.~\eqref{Eq4g1} and \eqref{calR} the projections $\hat{\cal P}_{\lambda,+}(\dots)\hat{\cal P}_{\lambda,-}$ vanish, so that the solution satisfies Eq.~\eqref{Norm1}. For $\hat{g}_{1,+}^{R(A)}=\hat{\tilde{g}}_{1,+}^{R(A)}+\delta\hat{g}_{1,+}^{R(A)}$ we then find
\beg\label{Solution4g1p}
\begin{aligned}
\hat{\tilde{g}}_{1,+}^{R(A)}&=\left(\frac{ev_F}{i\omega}\right)(\bn{\mathbf E})\\&\times\sum\limits_{\alpha,\beta=\pm}\frac{\hat{\cal P}_{\alpha,+}\left(\hat{\cal G}_{\bn\eps+\frac{\omega}{2}}^{R(A)}\hat{\tau}_3-
\hat{\tau}_3\hat{\cal G}_{\bn\eps-\frac{\omega}{2}}^{R(A)}\right)\hat{\cal P}_{\beta,-}}{\alpha\eta_{\bn\eps+\frac{\omega}{2}}^{R(A)}-\beta\eta_{\bn\eps-\frac{\omega}{2}}^{R(A)}-v_F(\bn\bq)}, \\
\delta\hat{g}_{1,+}^{R(A)}&=\sum\limits_{\alpha,\beta=\pm}\frac{\hat{\cal P}_{\alpha,+}\hat{\cal R}_{\bn\eps}^{R(A)}(\bq,\omega)\hat{\cal P}_{\beta,-}}{\alpha\eta_{\bn\eps+\frac{\omega}{2}}^{R(A)}-\beta\eta_{\bn\eps-\frac{\omega}{2}}^{R(A)}-v_F(\bn\bq)}.
\end{aligned}
\en
The second term contains the response to the imbalance potential and to the order parameter oscillation. 
Projection operators entering here must be evaluated for the retarded and advanced part correspondingly. 
The expression for the function $\hat{g}_{1-}^{R(A)}(\bn\eps;q)$ is immediately obtained from \eqref{Solution4g1p} by replacing ${\mathbf E}\to{\mathbf E}^*$, $\omega\to-\omega$ and $\bq\to-\bq$.
\paragraph{Keldysh component.}
Formally, the Keldysh component of $\check{g}_1$ satisfies the same equation as its retarded and advanced components, Eq. \eqref{Eq4g1}.
However the solution for the  function $\hat{g}_{1+}^K$ is different from \eqref{Solution4g1p}, for example, since it satisfies the different normalization condition
\beg\label{norm4g1K2}
\begin{aligned}
&\hat{\cal G}_{\bn\eps+\frac{\omega}{2}}^R\hat{g}_{1+}^K(\bn\eps;q)+\hat{\cal G}_{\bn\eps+\frac{\omega}{2}}^K\hat{g}_{1+}^A(\bn\eps;q)
\\&+\hat{g}_{1+}^R(\bn\eps;q)\hat{\cal G}_{\bn\eps-\frac{\omega}{2}}^K+\hat{g}_{1+}^K(\bn\eps;q)\hat{\cal G}_{\bn\eps-\frac{\omega}{2}}^A=0.
\end{aligned}
\en
To solve the equations above we will use the following ansatz 
\beg\label{g1kp}
\hat{g}_{1+}^K=\hat{g}_{1+}^Rt_{\eps-\frac{\omega}{2}}-\hat{g}_{1+}^At_{\eps+\frac{\omega}{2}}+\delta \hat{g}_{1+}^K, 
\en
where $t_\eps=\tanh({\eps}/{2T})$. Then \eqref{norm4g1K2} becomes
\beg\label{newg1Knorm}
\hat{\cal G}_{\bn\eps+\frac{\omega}{2}}^R\delta\hat{g}_{1+}^K+\delta\hat{g}_{1+}^K\hat{\cal G}_{\bn\eps-\frac{\omega}{2}}^A=0.
\en
Consequently, equation for $\delta\hat{g}_{1+}^K$ reads
\begin{widetext}
\beg\label{Eq4dektag1K}
\begin{aligned}
&\eta_{\bn\eps+\frac{\omega}{2}}^R\hat{\cal G}_{\bn\eps+\frac{\omega}{2}}^R\delta\hat{g}_{1+}^K-\eta_{\bn\eps-\frac{\omega}{2}}^A\delta\hat{g}_{1+}^K\hat{\cal G}_{\bn\eps-\frac{\omega}{2}}^A-{v}_F(\bn\bq)\hat{\tau}_0\delta\hat{g}_{1+}^K=
\left\{\left(\frac{ev_F}{i\omega}\right)(\bn{\mathbf E})\left(\hat{\cal G}_{\bn\eps+\frac{\omega}{2}}^R\hat{\tau}_3-
\hat{\tau}_3\hat{\cal G}_{\bn\eps-\frac{\omega}{2}}^A\right)\right.\\&\left.+\varphi_{1+}(q)\left(\hat{\cal G}_{\bn\eps+\frac{\omega}{2}}^{R}-
\hat{\cal G}_{\bn\eps-\frac{\omega}{2}}^{A}\right)\right\}\left(t_{\eps+\frac{\omega}{2}}-t_{\eps-\frac{\omega}{2}}\right)
+\left(t_{\eps+\frac{\omega}{2}}-t_{\eps-\frac{\omega}{2}}\right)
\left(\hat{\cal G}^{R}_{\bn\eps+\frac{\omega}{2}}\,\delta\hat\Delta_{1+}
-\delta\hat\Delta_{1+}\,\hat{\cal G}^{A}_{\bn\eps-\frac{\omega}{2}}\right).
\end{aligned}
\en
\end{widetext}
This equation can now be easily solved by using the same method as above for the functions $\hat{g}_{1+}^{R(A)}$:
\beg\label{Solution4dg1K}
\delta\hat{g}_{1+}^K=\sum\limits_{\mathrm{a}=1}^2\sum\limits_{\alpha,\beta=\pm}\frac{\hat{\cal P}_{\alpha,+}^R\hat{\cal Q}_{\bn\eps}^{(\mathrm{a})}\hat{\cal P}_{\beta,-}^A}{\alpha\eta_{\bn\eps+\frac{\omega}{2}}^{R}-\beta\eta_{\bn\eps-\frac{\omega}{2}}^{A}-v_F(\bn\bq)}
\en
and functions $\hat{\cal Q}_{\bn\eps}^{(\mathrm{a})}$ correspond to each of the two terms in the right hand side of \eqref{Eq4dektag1K}.

\paragraph{First order correction to electrochemical potential.} Expressions \eqref{g1kp} and \eqref{Solution4dg1K} above allow us to compute the Fourier component of the potential $\varphi_{1\pm}(\bq,\omega)$ which accounts for a shift in the electrochemical potential due to the population imbalance of the particle-hole branches. The linear correction to the electrochemical potential needs to be computed self-consistently. 
It can be found from the expression for the total particle density which also takes into account polarizability of the electron gas.  

At the linear order there is no hybridization between $\varphi_{1\pm}(\bq,\omega)$ and $\delta\hat{\Delta}_{1s}$ to first order in the photon momentum: the particle-hole-symmetric part of the corresponding kernel vanishes by the parity in $\xi_\bp$, and its $O(\eps/\veps_F)$ part is proportional to ${\cal Y}_\bn(\bn\bq)$ and vanishes upon the integration over $\bn$ (Appendix \ref{AppendixA}). As a result we have \cite{LO-Vortex,Eliashberg-Dynamics}:
\beg\label{varphi1pm}
\varphi_{1\pm}(q)=-\frac{1}{8}\int\limits_0^{2\pi}\frac{d\phi_\bn}{2\pi}\int\limits_{-\infty}^\infty\textrm{Tr}\left\{\hat{\tau}_0\hat{g}_{1\pm}^K(\bn\eps;q)\right\}{d\eps}.
\en
We would like to emphasize here that expression \eqref{varphi1pm} was derived under  assumption that the particle density remains constant under the action of external radiation \cite{Eliashberg-Dynamics}. This assumption is certainly justified for a broad range of temperatures not too close to a critical temperature when the frequency $\omega$ is much smaller than the superconducting plasma frequency. 
After we insert the expressions \eqref{g1kp} and \eqref{Solution4dg1K} into Eq. \eqref{varphi1pm} we find:
\beg\label{phi1pmMain}
\begin{aligned}
&\varphi_{1+}(\bq,\omega)\approx\left(\frac{ie\zeta_\omega v_F^2}{4\omega}\right)\left(\bq{\mathbf E}\right)\Pi_\mu(\mathbf{q},\omega),
\end{aligned}
\en
where both $\zeta_\omega$ and $\Pi_\mu(\mathbf{q},\omega)$ have been defined in Appendix \ref{AppendixA} while from requiring that $\mu_1(\br,t)$ is real we have $\varphi_{1-}(q)=\varphi_{1+}^*(q)$. Function $\Pi_\mu(\mathbf{q},\omega)$ has a clear physical meaning of the dielectric function in the charge imbalance channel computed with the equilibrium order parameter. In the limit $\omega\to 0$ in the superconductor $\Pi_\mu^{-1}\to 1$, while in the normal state as the $q\to 0$ we have $\Pi_\mu^{-1}=2$, i.e. metallic compressibility-type screening cuts the potential in half. This serves as a check that the anomalous term $\delta\hat{g}_1^K$ carries the entire normal-metal physics the remaining regular kernel misses.

\begin{figure}
\includegraphics[width=0.950\linewidth]{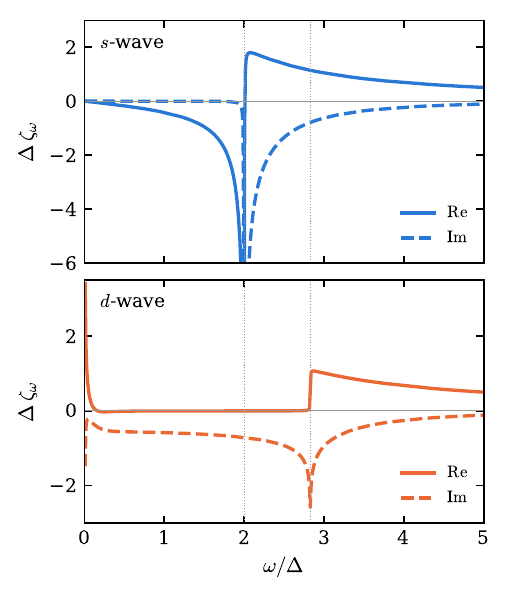}
\caption{Frequency dependence of the real and imaginary parts of the
dimensionless function $\zeta_\omega\Delta$, Eq.~\eqref{zetaom},
which determines the linear correction to the electrochemical potential,
Eq.~\eqref{phi1pmMain}, for the $s$-wave and $d$-wave symmetries of the order parameter.
For $s$-wave pairing $\zeta_\omega$ is purely real below the pair-breaking
threshold --- the imaginary parts of the regular and anomalous contributions
to the kernel cancel exactly for $\omega<2\Delta$ --- and develops a sharp
feature at $\omega=2\Delta$, whose depth is limited by the broadening
$\gamma$ that regularizes the clean limit. For $d$-wave pairing the nodal quasiparticles make the
response absorptive at all frequencies: $\mathrm{Im}\,\zeta_\omega$ dominates,
with a signature at the maximal-gap threshold $\omega=2\sqrt2\,\Delta$, while
$\mathrm{Re}\,\zeta_\omega$ nearly vanishes due to the cancellation between
the nodal and antinodal regions of the Fermi surface --- the same cancellation
that pins $\mathrm{Re}\,\Pi^{-1}_\omega$ near unity in Fig.~\ref{Fig-PiInvw}. For this plot
we fixed $T=0.005\,\omega_D$, $\Delta=0.125\,\omega_D$, and the broadening
$\gamma=10^{-3}\Delta$. The dotted lines mark $\omega=2\Delta$ and $2\sqrt2\Delta$.}
\label{Fig-zetaw}
\end{figure}

We show the frequency dependence of the function $\zeta_\omega$  in Fig. \ref{Fig-zetaw} for both $s$-wave and $d$-wave pairings. 
We observe that this function exhibits strong frequency dependence for $\omega\sim\Delta$ while it becomes weakly dependent on frequency for $\omega\gg\Delta$. Notably, in the case of a $d$-wave superconductor the real part of $\zeta_\omega$ nearly vanishes over a broad range of frequencies, reflecting the cancellation between the nodal and antinodal regions of the Fermi surface, while in the $s$-wave case it changes sign when $\omega=2\Delta$. The imaginary part of $\zeta_\omega$ has a pronounced signature at the pair-breaking threshold: $\omega=2\Delta$ in the $s$-wave case and $\omega=2\sqrt{2}\Delta$, set by the maximal gap, for $d$-wave pairing — frequencies coinciding with the Schmid–Higgs scale \cite{Kazi2026}. These features are beyond phenomenological treatments \cite{Mironov2021-IFESC}. They would control the frequency dependence of the imbalance channel of the rectified current, which we identify but do not evaluate (Sec.~IV A). 

\begin{figure}
\includegraphics[width=0.950\linewidth]{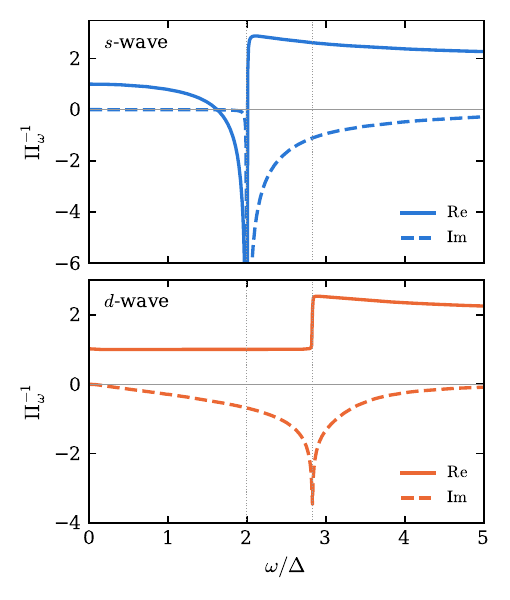}
\caption{Frequency dependence of the real and imaginary parts of the inverse
dielectric response function of the charge-imbalance channel,
$\Pi_\omega^{-1}=\Pi_\mu^{-1}(\mathbf{q}=0,\omega)$, Eq.~\eqref{Pikw}, evaluated in the limit
$v_F q \ll \Delta$ for the $s$-wave and $d$-wave symmetries of the order
parameter. In the static limit $\Pi_\omega^{-1} \to 1$: the condensate maintains
charge neutrality and the electrochemical potential is unscreened, while for
$\omega \gg \Delta$ both curves approach the normal-metal value
$\Pi_\omega^{-1} = 2$, which is carried entirely by the anomalous contribution $\delta\hat{g}_1^K$ to
$\hat{g}^K_{1}$. In the $s$-wave case $\mathrm{Re}\,\Pi_\omega^{-1}$ crosses zero
at $\omega \approx 1.63\Delta$, below the pair-breaking threshold
$\omega = 2\Delta$: the corresponding pole of $\Pi_\omega$ signals a collective
resonance of the branch-imbalance channel. Its location is protected at leading order by the parity selection rule that decouples the imbalance and order parameter channels (Sec. III B). 
In the $d$-wave case $\mathrm{Re}\,\Pi_\omega^{-1}$ remains close to unity up to $\omega \sim 2\sqrt{2}\Delta$ due to the near cancellation of the nodal and antinodal contributions, while the nodal quasiparticles produce an imaginary part whose magnitude grows with frequency up to $\omega=2\sqrt2\Delta$. For this plot we fixed $T = 0.005\,\omega_D$,
$\Delta = 0.125\,\omega_D$, and the broadening $\gamma = 10^{-3}\Delta$. The dotted lines mark $\omega=2\Delta$ and $2\sqrt2\Delta$.}
\label{Fig-PiInvw}
\end{figure}

\paragraph{First order correction to the order parameter.}\label{sec:dDelta}
The linear-in-field correction to the order parameter amplitude follows from the self-consistency equation
\beg\label{Self-consistency}
\delta\Delta_{1+}=\frac{\lambda}{8}\int\limits_0^{2\pi}{\cal Y}_\bn\frac{d\phi_\bn}{2\pi}\int\limits_{-\infty}^\infty{d\eps}\textrm{Tr}\left\{(-i\hat{\tau}_2)\hat{g}_{1+}^K\right\},
\en
where $\lambda=\nu_Fg$ is the dimensionless pairing strength. Since $\hat g^K_{1+}$ contains the response to $\delta\hat\Delta_{1+}$ itself, Eq. \eqref{Self-consistency} is a self-consistency condition rather than a formula. Eliminating the cutoff-dependent coupling $\lambda$ with the help of the equilibrium gap equation resums this feedback into the longitudinal (Schmid-Higgs) part of the pair susceptibility $\chi_{\mathrm{SH}}(\omega)$ \cite{Kazi2026}, defined in Eq. \eqref{eq:chiSH} below, which multiplies a drive term of first order in the particle-hole asymmetry.

Before writing the result it is useful to note what its form must be. In the single-channel model \eqref{Eq1} every fluctuation of the order parameter has the form $\delta\Delta_\bn=\delta\Delta\,{\cal Y}_\bn$, i.e. the form factor is fixed and only the amplitude fluctuates. The $d_{x^2-y^2}$ ground state is not invariant under the point group $D_4$ acting on $\bn$ alone, but it is invariant under $D_4$ supplemented by a gauge rotation: the quarter turn and the diagonal mirrors send ${\cal Y}_\bn\to-{\cal Y}_\bn$, which is undone by $\Delta\to e^{i\pi}\Delta$, while the axial mirrors leave ${\cal Y}_\bn$ invariant. Under every element of this combined group the amplitude $\delta\Delta$ is invariant, since $\delta\Delta_\bn$ transforms exactly as $\Delta_\bn$, whereas gauge rotations do not act on $\bq$ and ${\mathbf E}$. The coefficient tensor in the linear response $\delta\Delta_{1+}=T_{ij}(\omega)q_iE_j$ is therefore invariant under the geometric $D_4$, which leaves $T_{ij}\propto\delta_{ij}$ ($\epsilon_{ij}$ is excluded by the mirrors). Hence
\begin{equation}
\delta\Delta_{1+}\propto(\bq{\mathbf E})
\label{eq:symm}
\end{equation}
to the first order in $\eps/\veps_F$ and in the gradient expansion, for any Fermi surface of square symmetry and for either pairing symmetry: the linear order parameter response is fully symmetric ($A_{1g}$), and the coefficients of the remaining bilinears, $q_xE_x-q_yE_y$ ($B_{1g}$), $q_xE_y+q_yE_x$ ($B_{2g}$) and $[\bq\times{\mathbf E}]_z$ ($A_{2g}$), vanish identically. In particular, a response proportional to $q_xE_x-q_yE_y$, which one might expect for $d$-wave pairing, is forbidden: a rotation by $\pi/2$ changes the sign of $q_xE_x-q_yE_y$ but leaves the amplitude $\delta\Delta$ unchanged.

The explicit calculation, performed in Appendix \ref{AppendixA} from the exact solution of the Dyson equation to first order in the field, confirms \eqref{eq:symm} and gives
\begin{equation}
\delta\Delta_{1+}=2i\,\chi_{\mathrm{SH}}(\omega)\left(\frac{ev_F^2}{\veps_F}\right)(\bq{\mathbf E})\,{\cal A}(\omega),
\label{eq:dDelta1}
\end{equation}
where we introduced
\begin{equation}
\begin{aligned}
&{\cal A}(\omega)=\frac{\Delta}{8}\int\limits_{-\infty}^{\infty}d\eps\,\eps^2\Big\langle{\cal Y}_\bn^2\Big[\frac{t_{\eps-\frac{\omega}{2}}}{\eta^R_{+}\eta^R_{-}(\eta^R_{+}+\eta^R_{-})^3}\\&-\frac{t_{\eps+\frac{\omega}{2}}}{\eta^A_{+}\eta^A_{-}(\eta^A_{+}+\eta^A_{-})^3}+\frac{t_{\eps+\frac{\omega}{2}}-t_{\eps-\frac{\omega}{2}}}{\eta^R_{+}\eta^A_{-}(\eta^R_{+}+\eta^A_{-})^3}\Big]\Big\rangle_\bn,
\end{aligned}
\label{eq:Akernel}
\end{equation}
and $\eta^{R(A)}_\pm\equiv\eta^{R(A)}_{\bn,\eps\pm\omega/2}$, the vector potential is ${\mathbf A}=({\mathbf E}/i\omega)e^{i(\bq\br-\omega t)}+{\rm c.c.}$, Eq. \eqref{VectorPotential}, and $\chi_{\mathrm{SH}}(\omega)$ is the dimensionless amplitude (Schmid--Higgs) susceptibility,
\begin{equation}
\begin{aligned}
\chi_{\mathrm{SH}}^{-1}(\omega)=\frac12\int\limits_{-\infty}^{\infty} d\xi \Big\langle\frac{{\cal Y}_\bn^2\left(4\Delta_\bn^2-\omega^2\right)\tanh\left(\frac{E}{2T}\right)}{E\,[4E^2-(\omega+i0)^2]}\Big\rangle_\bn,
\end{aligned}
\label{eq:chiSH}
\end{equation}
where $E=\sqrt{\xi^2+\Delta_\bn^2}$. The angular part of the momentum gradient in Eq.~\eqref{NablapGpeps} does not contribute to Eq.~\eqref{eq:dDelta1}. Its contribution contains the product $(\bq\hat\phi_\bn)(\bn{\mathbf A})$, whose average over the Fermi surface has no part proportional to $(\bq{\mathbf A})$, since $\hat\phi_\bn\perp\bn$. By Eq.~\eqref{eq:symm} no other combination can appear, so this contribution vanishes (Appendix~\ref{AppendixA}).

Equation~\eqref{eq:chiSH} follows from the self-consistency condition \eqref{Self-consistency} after the coupling constant is eliminated with the help of the gap equation (Appendix \ref{AppendixA}): the amplitude response to any source is $g/(1-gP_{11})=\chi_{\mathrm{SH}}/\nu_F$, where $P_{11}$ is the amplitude--amplitude pair bubble, so that $\chi_{\mathrm{SH}}$ is independent of the ultraviolet cutoff and $\chi_{\mathrm{SH}}(0)=1$ at $T=0$ for either pairing symmetry. At $T=0$ the energy integral gives 
\beg\label{chiSHT0}
\chi_{\mathrm{SH}}^{-1}(\omega)=\langle{\cal Y}_\bn^2\sqrt{1-x_\bn^2}\,\arcsin(x_\bn)/x_\bn\rangle_\bn 
\en
with $x_\bn=\omega/2|\Delta_\bn|$. For an isotropic gap $\chi_{\mathrm{SH}}^{-1}$ vanishes as a square root at $\omega=2\Delta$, the Schmid--Higgs resonance. For $d$-wave pairing $\mathrm{Re}\,\chi_{\mathrm{SH}}^{-1}=1-\omega^2/12\Delta^2$ at $T=0$ for all $\omega<2\sqrt2\Delta$ and reaches its minimum value $1/3$ at the maximal-gap threshold, while the nodal quasiparticles keep $\mathrm{Im}\,\chi_{\mathrm{SH}}^{-1}$ finite at all frequencies, so that $\chi_{\mathrm{SH}}$ has a broad maximum rather than a pole ($|\chi_{\mathrm{SH}}|\simeq1.7$ at $\omega\simeq2.5\Delta$, and $\mathrm{Im}\,\chi_{\mathrm{SH}}$ peaks at $\omega\simeq2.75\Delta$ for $T=0.04\Delta$).

The kernel \eqref{eq:Akernel} is directly related to the susceptibility.    Using the identity
\beg\label{AuxRelation}
\frac{4\Delta_\bn^2-\omega^2}{4E^2-\omega^2}-\frac{\Delta_\bn^2}{E^2}=-\frac{\omega^2\xi^2}{E^2(4E^2-\omega^2)}, 
\en
one finds (Appendix \ref{AppendixA})
\begin{equation}
\begin{aligned}
{\cal A}(\omega)&=\frac{\Delta}{4\omega^2}\left[\chi_{\mathrm{SH}}^{-1}(\omega)-\chi_{\mathrm{SH}}^{-1}(0)-N_T\right],
\end{aligned}
\label{eq:Aidentity}
\end{equation}
where $\chi_{\mathrm{SH}}^{-1}(0)$ is the $\omega\to0$ limit of \eqref{eq:chiSH} and 
\beg\label{NT}
N_T=\left\langle{\cal Y}_\bn^2\int\limits_{-\infty}^{\infty}d\xi\,\frac{\xi^2}{E^2}\left(-\frac{\partial n_F}{\partial E}\right)\right\rangle_\bn
\en
is the weight of the thermally excited quasiparticles, exponentially small for $s$-wave and $\propto(T/\Delta)^3$ for $d$-wave pairing. Consequently
\begin{equation}\label{eq:dDelta1closed}
\delta\Delta_{1+}=\frac{i\chi_{\mathrm{SH}}(\omega)\Delta}{2\omega^2}\left(\frac{ev_F^2}{\veps_F}\right)(\bq{\mathbf E})\left[\delta\chi_{\mathrm{SH}}^{-1}(\omega)-N_T\right].
\end{equation}
Here $\delta\chi_{\mathrm{SH}}^{-1}(\omega)=\chi_{\mathrm{SH}}^{-1}(\omega)-\chi_{\mathrm{SH}}^{-1}(0)$.
Equation \eqref{eq:dDelta1closed} shows explicitly how the amplitude resonance enters the order parameter response already at the linear order in the electric field. In particular, $\mathrm{Im}\,{\cal A}=\Delta\,\mathrm{Im}\,\chi_{\mathrm{SH}}^{-1}/4\omega^2$, i.e. the drive and the susceptibility share the same pair-breaking continuum. 

\begin{figure}
\includegraphics[width=0.950\linewidth]{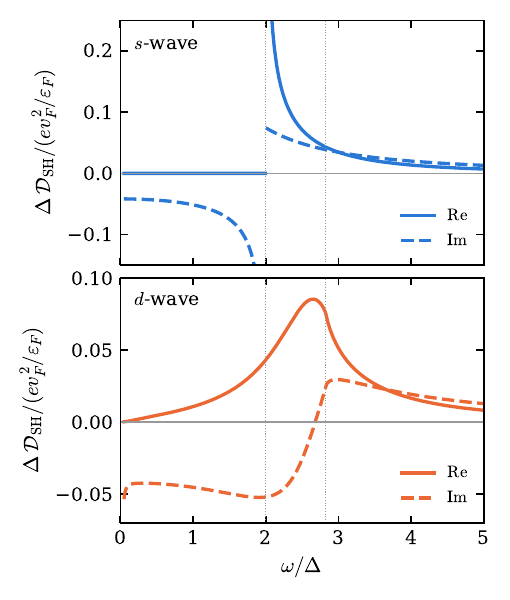}
\caption{Order parameter response function ${\cal D}_{\rm SH}(\omega)=\delta\Delta_{1+}/(\bq{\mathbf E})$,
Eqs.~\eqref{eq:dDelta1} and \eqref{eq:dDelta1closed}, in units of $(ev_F^2/\veps_F)/\Delta$, for
$s$-wave (top) and $d$-wave (bottom) pairing at $T=0.04\Delta$. For an isotropic gap
${\rm Re}\,{\cal D}_{\rm SH}\propto{\rm Im}\,\chi_{\rm SH}$ vanishes below $2\Delta$ and diverges at
the threshold together with $\chi_{\rm SH}$, while ${\rm Im}\,{\cal D}_{\rm SH}$ changes sign there.
For $d$-wave pairing both parts are finite at all frequencies, and ${\rm Re}\,{\cal D}_{\rm SH}$ has a
maximum near $2.65\Delta$ and a kink at the maximal-gap threshold. The dotted lines mark
$\omega=2\Delta$ and $2\sqrt2\Delta$.}
\label{Fig-DSH}
\end{figure}

Two properties of this result deserve a comment. (i) Our calculation shows that the response is proportional to $1/\veps_F$: in the canonical, particle-hole-symmetric quasiclassical theory the $i\hat\tau_2$ projection of every linear source vanishes on the Fermi-surface average, and the amplitude channel opens only through the particle-hole asymmetric terms. (ii) The response contains no static piece at $T=0$. Written in terms of the vector potential, ${\mathbf E}=i\omega{\mathbf A}$, it is proportional to $(\bq{\mathbf A})\omega$ and vanishes linearly as $\omega\to0$ at fixed ${\mathbf A}$, as it must by symmetry: a static supercurrent texture is odd under time reversal while the amplitude of the order parameter is even. This statement is in fact exact to all orders in the photon momentum, since the static current--amplitude correlator is odd in the Matsubara frequency (Appendix \ref{AppendixA}). For $s$-wave pairing (${\cal Y}_\bn=1$) $\mathrm{Im}\,{\cal A}$ vanishes below $\omega=2\Delta$ and sets in as a square root, with a cusp of $\mathrm{Re}\,{\cal A}$ at $2\Delta$. For $d$-wave pairing $\mathrm{Im}\,{\cal A}$ is finite at all frequencies owing to the nodes and has a kink at $2\sqrt2\Delta$, while by \eqref{eq:Aidentity} $\mathrm{Re}\,{\cal A}=-1/48\Delta$ at $T=0$ for all $\omega<2\sqrt2\Delta$. The resulting order parameter response, shown in Fig.~\ref{Fig-DSH}, does not depend on the scattering rate in the clean limit. In either case the smallness of the $\veps_F^{-1}$ prefactor in \eqref{eq:dDelta1} is partially compensated by the large value of $|\chi_{\mathrm{SH}}|$ near the pair-breaking threshold, where $\mathrm{Re}\,\chi_{\mathrm{SH}}^{-1}$ vanishes ($s$-wave) or reaches its minimum ($d$-wave).

It is instructive to summarize the hierarchy of scalar responses that emerges from our analysis so far. For strictly homogeneous radiation ($\mathbf{q}=0$) no scalar quantity can be driven at linear order in the field: both the electrochemical potential $\varphi_1$ and the order parameter correction $\delta\Delta_1$ vanish identically, in accordance with Eliashberg's observation that monochromatic homogeneous radiation does not disturb the branch populations \cite{Eliashberg-Dynamics}. At finite momentum both scalar fields are driven by the same longitudinal bilinear $(\bq{\mathbf E})$, but they open at \emph{different orders} in the particle-hole asymmetry: the charge-imbalance potential $\varphi_1$ appears already at order $(\varepsilon/\varepsilon_F)^{0}$, Eq.~\eqref{phi1pmMain}, whereas the amplitude of the order parameter opens only at order $(\varepsilon/\varepsilon_F)^1$, through the particle-hole asymmetric vertices: it is the finite-momentum, linear-in-field counterpart of the Eliashberg effect \cite{Eliashberg1970,Ivlev1973,Klapwijk1977}, resonantly enhanced at the pair-breaking threshold. The two channels do not hybridize at this order. The imbalance-channel physics described by $\zeta_\omega$, the dielectric function $\Pi_\mu$ and the collective resonance of the $s$-wave imbalance channel is therefore of order $(\varepsilon/\varepsilon_F)^0$. In the rectified current both channels appear at the same order, since the imbalance potential contributes to the current only through particle-hole asymmetric vertices (Sec.~IV A). The order parameter channel carries a qualitatively distinct fingerprint: the pair susceptibility $\chi_{\mathrm{SH}}(\omega)$, with its $2\Delta$ resonance for an isotropic gap and its nodal-broadened $2\sqrt2\Delta$ threshold for $d$-wave pairing.

\subsection{Second order corrections}
Having determined linear corrections to $\check{g}$ we proceed with the calculation of the second order corrections. Due to the presence of the gradient term \eqref{NablapGpeps} we first discuss the second order correction to $\check{G}_{\bp\eps}(\br,t)$. We have
\beg\label{NonLinearPart}
\check{G}_{\bp\eps}^{(2)}(\br,t)=\check{G}_{\bp\eps}^{(\mathrm{dc})}(q)+\sum\limits_{s=\pm}\check{G}_{\bp\eps}^{(2,s)}(q)e^{2is(\bq \br-\omega t)}.
\en
Since we are primarily interested in $\mathrm{dc}$-transport, we will only need to evaluate function  $\check{G}_{\bp\eps}^{(\mathrm{dc})}$. Formally, 
$\check{G}_{\bp\eps}^{(\mathrm{dc})}$ satisfies \eqref{EilenMainG} where we have to keep only terms which are of the second order in electric field. 

A few simplifications are in order. In the equation for $\check{G}_{\bp\eps}^{(\mathrm{dc})}$ we neglect the second-order corrections to the amplitude and to the phase of the order parameter. For a single plane wave they do not depend on $\br$, and they depend on $\bn$ only through ${\cal Y}_\bn$, so that they do not produce a dc current. For a beam they vary on the scale of its envelope and produce only a current along the wave vector of the envelope, which does not contribute to the magnetization (Sec.~IV). The dc correction to the amplitude is the one responsible for the Eliashberg effect \cite{Eliashberg1970,Ivlev1973,Klapwijk1977}. The first-order corrections $\delta\hat{\Delta}_{1\pm}$, on the other hand, enter the second-order sources at the order considered here and must be kept. Furthermore, the dc part of $\check g_2$ does not depend on time, and its spatial gradients produce corrections of higher order in the photon momentum. Thus, in Eq.~\eqref{EilenMain} we drop all the terms which contain the temporal and spatial gradients of $\check{g}_2$, while the gradient terms acting on the first-order functions are retained in the source function. Then the dc part of the second-order correction to the quasiclassical function \eqref{Quasi} is given by the solution of the equation
\beg\label{MainEq4g2}
[\eps\check{\tau}_3+\check\Delta_\bn,\check{g}_{2}(\bn\eps;q)]=\check{\Gamma}_{\bn\eps}(q),
\en
where the source function on the right hand side consists of the seven contributions listed in Appendix~\ref{AppendixSource}. Two of them, $\check S^{(\mu)}$ and $\check S^{(\Lambda)}$, contain the imbalance potential and vanish when the polarization is transverse to $\bq$. The source $\check S^{(\Lambda)}$ contains the additional factor $(\bn\bq)$ and is therefore of second order in $q$, since $\varphi_1\propto(\bq{\mathbf E})$. Like the term with $\check\Lambda$ at linear order (Sec.~III B), it does not contribute at the order considered here. With impurity scattering, $\eps\check\tau_3+\check\Delta_\bn$ on the left hand side is replaced by $\tilde\eps_\eps\check\tau_3+i\check\tau_2\tilde\Delta_{\bn\eps}$, Eq.~\eqref{xitilde}, for the part of $\check g_2$ odd in $\bn$, which carries the current. 

In order to determine the components of $\check{g}_2$ we also need to take into account the normalization condition \eqref{norm} which in this case becomes
\beg\label{Norm4G2dc}
\check{\cal G}_{\bn\eps}\check{g}_{2}+\check{g}_{2}\check{\cal G}_{\bn\eps}=\check{\cal N}_{\bn\eps}(q),
\en
where we introduced
\beg\label{checkN}
\begin{aligned}
\check{\cal N}_{\bn\eps}(q)=-\sum\limits_{s=\pm}\check{g}_{1s}(\bn\eps_{\overline{s}};q)\check{g}_{1\overline{s}}(\bn\eps_{\overline{s}};q)
\end{aligned}
\en
for brevity and $\overline{s}=-s$, $\eps_{\pm}=\eps\pm\omega/2$. 

In order to obtain the expressions for the retarded and advanced components of $\check{g}_2$ we repeat the same steps as in the linear analysis above using the projector operators $\hat{\cal P}_\lambda^{R(A)}=(\hat{\tau}_0+\lambda\hat{\cal G}_{\bn\eps}^{R(A)})/2$. This gives
\beg\label{Sol4g2RA}
\begin{aligned}
\hat{g}_2^{R(A)}&=\frac{\hat{\cal P}_{+}^{R(A)}\,\hat\Gamma_{\bn\eps}^{R(A)}\,\hat{\cal P}_{-}^{R(A)}-\hat{\cal P}_{-}^{R(A)}\,\hat\Gamma_{\bn\eps}^{R(A)}\,\hat{\cal P}_{+}^{R(A)}}{2\eta_{\bn\eps}^{R(A)}}\\&
\;+\;\frac{1}{2}\sum_{\lambda=\pm}{\lambda}\,\hat{\cal P}_{\lambda}^{R(A)}\,\hat {\cal N}_{\bn\eps}^{R(A)}\,\hat{\cal P}_\lambda^{R(A)}.
\end{aligned}
\en

It remains for us to find an expression for $\hat{g}_2^{K}(\bn\eps;q)$. We will look for $\hat{g}_2^{K}$ in the following form
\beg\label{Ansatz4gK}
\hat{g}_2^K=\left(\hat{g}_{2}^R-\hat{g}_{2}^A\right)t_{\eps}+\delta\hat{g}_2^K.
\en
After we insert ansatz \eqref{Ansatz4gK} into \eqref{MainEq4g2} we find the following equation for the function $\delta\hat{g}_2^K$
\beg\label{MainEq4dg2K}
\begin{aligned}
[\eps\hat{\tau}_3+\hat\Delta_\bn,\delta\hat{g}_{2}^K]&=\hat{\Gamma}_{\bn\eps}^K-\left(\hat{\Gamma}_{\bn\eps}^R-\hat{\Gamma}_{\bn\eps}^A\right)t_\eps.
\end{aligned}
\en
The normalization condition for the function $\delta\hat{g}_{2}^K(\bn\eps;q)$ reads
\beg\label{Norm4dg2K}
\begin{aligned}
&\hat{\cal G}_{\bn\eps}^R\delta\hat{g}_{2}^K+\delta\hat{g}_{2}^K\hat{\cal G}_{\bn\eps}^A=\hat{\cal N}_{\bn\eps}^K-\left(\hat{\cal N}_{\bn\eps}^R-\hat{\cal N}_{\bn\eps}^A\right)t_\eps.
\end{aligned}
\en
Given that Eqs.~\eqref{MainEq4dg2K} and \eqref{Norm4dg2K} are fully analogous to the equations above, 
for the function $\delta\hat{g}_2^K(\bn\eps;q)$ we have 
\beg\label{Sol4dg2K}
\begin{aligned}
\delta\hat{g}_2^K&=\frac{\hat{\cal P}_{+}^{R}\hat{\cal Q}_{\bn\eps}^K\hat{\cal P}_{-}^{A}-\hat{\cal P}_{-}^{R}\hat{\cal Q}_{\bn\eps}^K\hat{\cal P}_{+}^{A}}{\eta_{\bn\eps}^R+\eta_{\bn\eps}^A}
\\&+\sum\limits_{\alpha=\pm}\frac{\alpha}{2}\,\hat{\cal P}_{\alpha}^{R}\,\delta\hat {\cal N}_{\bn\eps}^K\,\hat{\cal P}_\alpha^{A}.
\end{aligned}
\en
Here $\delta\hat {\cal N}_{\bn\eps}^K=\hat{\cal N}_{\bn\eps}^K-(\hat{\cal N}_{\bn\eps}^R-\hat{\cal N}_{\bn\eps}^A)t_\eps$ and $\hat{\cal Q}_{\bn\eps}^K=\hat{\Gamma}_{\bn\eps}^K-(\hat{\Gamma}_{\bn\eps}^R-\hat{\Gamma}_{\bn\eps}^A)t_\eps$.
We will use \eqref{Sol4g2RA} and \eqref{Sol4dg2K} to compute the dc-component of the nonlinear current.   In the first term of Eq.~\eqref{Sol4dg2K} the denominator $\eta^R_{\bn\eps}+\eta^A_{\bn\eps}=2i\,{\rm Im}\,\eta^R_{\bn\eps}$ is of order $1/\tau$ at energies where quasiparticles exist, as we have mentioned in Section~III A. It supplies the quasiparticle lifetime $\tau$ that compensates the impurity term of order $1/\tau$ in the source, Eq.~\eqref{Simp}. Terms in which this factor is not compensated are proportional to $\tau$ and will cancel out at the end of the calculation.

\section{Nonlinear transport}
Second-order response of a metal to an electromagnetic field contains, in
general, three distinct contributions. The first one is the second-harmonic
generation (i.e. an ac current oscillating at twice the frequency of light) which
in a medium with inversion symmetry is necessarily nonlocal, i.e.\ it appears
only at a finite photon momentum \cite{Edel1988,Dzero2024Nice}. The second one is the photogalvanic
effect -- a local rectified current proportional to $\mathbf{E}\times\mathbf{E}^{*}$ --
which requires broken inversion symmetry and therefore vanishes in the bulk of
a centrosymmetric crystal. The third contribution is a nonlocal rectified
response which is a dc current of first (and higher odd) order in the photon momentum,
and is allowed even when the inversion center is present. It is this last
contribution that we analyze below. 

For an isotropic system the most general form of such a current to the lowest order in the photon momentum is
\begin{equation}
\begin{aligned}
\mathbf{j}_{\mathrm{dc}}(\mathbf{q})
&=\alpha(q)\mathbf{q}|\mathbf{E}|^{2}
+i\beta_{\textrm{a}}(q)\left[\mathbf{E}\,(\mathbf{q}\mathbf{E}^{*})
-\mathbf{E}^{*}(\mathbf{q}\mathbf{E})\right]\\&+
{\beta}_{\textrm{s}}(q)\left[\mathbf{E}\,(\mathbf{q}\mathbf{E}^{*})
+\mathbf{E}^{*}(\mathbf{q}\mathbf{E})\right]+...
\end{aligned}
\label{54}
\end{equation}
The three terms here describe different physics. The first one is a photon-drag
current flowing along $\mathbf{q}$ regardless of the polarization of light.
The second one is odd in the helicity of the light, ${\mathbf Z}=i[{\mathbf E}\times{\mathbf E}^*]$, and equals $\beta_{\rm a}[\bq\times{\mathbf Z}]$. This is a uniform current flowing perpendicular to $\bq$ whose direction is reversed together with the helicity (i.e.\ circular photon drag). The third term is the polarization-dependent part of the photon
drag. For $d$-wave pairing the $D_4$ symmetry of $|\Delta_\bn|$ allows one more term, proportional to $\sum_{i=x,y}\hat{\mathbf e}_iq_i|E_i|^2$ with $\hat{\mathbf e}_{x,y}$ along the antinodal directions. It is even in the helicity, so that $\beta_{\rm a}$ remains the only helicity-odd coefficient.

The helicity-odd current $\beta_{\rm a}[\bq\times{\mathbf Z}]$ must be
distinguished from the light-induced magnetization. A static magnetization
${\mathbf M}(\br)$ is accompanied by the current $\nabla\times{\mathbf M}$,
which flows where ${\mathbf M}$ varies, i.e.\ at the edges of the illuminated
spot, whereas Eq.~\eqref{54} is the uniform current of a single plane wave. For
a beam, which is a superposition of plane waves, the dc current at the wave
vector $\bQ$ of the envelope is produced by pairs of field components
${\mathbf E}(\bq_1)$ and ${\mathbf E}^*(\bq_2)$ with $\bq_1-\bq_2=\bQ$, and it
depends on $\bq_1$ and $\bq_2$ separately. To first order in the wave vectors
its helicity-odd part is the sum of a term proportional to the mean wave vector
$(\bq_1+\bq_2)/2$, which is the current $\beta_{\rm a}[\bq\times{\mathbf Z}]$ of
Eq.~\eqref{54}, and of a term proportional to $\bQ$, which has the form
$i\bQ\times{\mathbf M}$ and defines the magnetization. The two coefficients are
independent: in the normal state of the model \eqref{Eq1} the first one
vanishes (see below), while the second one does not \cite{Pit1961,Hertel2006}. For the order parameter channel
both are obtained in Sec.~IV A.

In a normal metal without spin-orbit coupling, such as our model, a circularly polarized beam induces an orbital magnetization, carried by currents that circulate around the beam \cite{Pit1961,Hertel2006}. It comes from the terms proportional to $\bQ$, which are not contained in Eq.~\eqref{54}. In the particle-hole symmetric approximation and for $\varphi_{1\pm}=0$ the dc current vanishes at first order in $\bq$, in the normal as well as in the superconducting state (Appendix~\ref{AppendixB}). The imbalance potential $\varphi_{1\pm}$ contributes to the current only through particle-hole asymmetric vertices (Sec.~IV A), so that at this order the entire dc response originates from the particle-hole asymmetric terms of Section III.

The normal state provides an exact benchmark for these vertices. For a two-dimensional parabolic band with elastic impurity scattering the dc current at second order in the field is fixed by momentum balance. It equals ${\mathbf j}=(e/m){\mathbf P}$, where the kinetic momentum density ${\mathbf P}$ relaxes on the impurities, $d{\mathbf P}/dt={\mathbf F}-{\mathbf P}/\tau$, and ${\mathbf F}$ is the dc force density exerted by the electromagnetic wave on the electrons. For any local conductivity $\sigma(\omega)$ this force is directed along the photon momentum $\bq$, ${\mathbf F}=(2/\omega)\,{\rm Re}\,\sigma(\omega)\,\bq|{\mathbf E}|^2$, so that $\beta_{\rm a}=\beta_{\rm s}=0$ in the normal state. In particular, the helicity-odd uniform current vanishes. With the Drude conductivity of the model the drag coefficient is
\beg\label{alphaq}
\alpha_{\Delta=0}(q)=\left(\frac{2e^2n_c\tau}{m}\right)\frac{e\tau}{m\omega(1+\omega^2\tau^2)},
\en
where $n_c=2\nu_F\veps_F$ is the electron density.

This argument does not apply to the
orbital magnetization, which is determined by the envelope of the beam and is
finite in the normal state \cite{Pit1961,Hertel2006}. Note that for $\omega\tau\gg1$ the
drag coefficient, $\alpha\to2e^3n_c/(m^2\omega^3)$, does not depend on $\tau$. Indeed, the
momentum absorbed from the wave, $\propto{\rm Re}\,\sigma(\omega)\propto1/\tau$, is
relaxed over the time $\tau$. The same compensation operates in the
order parameter channel of the superconducting state discussed below. The solution of the kinetic equation
with all particle-hole asymmetric terms (Appendix \ref{AppendixB}) reproduces
this result and shows how it comes about. The spatial-gradient term $(\mathbf A\cdot\nabla_{\mathbf r})\check g$ of Eq.~\eqref{EilenMain} generates a helicity-odd current proportional to $\tau$, which is cancelled exactly by the angular term with $\partial_{\phi_\bn}\check g$. In addition, the helicity-odd current of the force term with the energy-dependent velocity, which acts on the $O(q)$ part of the first-order distribution, is cancelled by that of the particle-hole asymmetric correction to the current vertex. The first of these two cancellations holds in the superconducting state as well (see Sec.~IV A).
The helicity-odd uniform current obtained below is therefore of superconducting origin.

Expression for the current density in terms of the Keldysh component of $\check{g}$ reads
\beg\label{j2dc}
\begin{aligned}
{\mathbf j}_{\mathrm{dc}}(q)&=-\left(\frac{e\nu_F}{4}\right)\int\limits_{-\infty}^\infty\left\langle{\mathbf v}_F\textrm{Tr}\left[\hat{\tau}_3\hat{g}_2^K(\bn\eps;q)\right]\right\rangle_\bn{d\eps},
\end{aligned}
\en
where the function $\hat{g}_2^K(\bn\eps;q)$ is defined by Eq.~\eqref{Ansatz4gK}. The prefactor follows from the identity $\langle\Psi^\dagger\hat M\Psi\rangle=\frac12{\rm Tr}\hat M-\frac i2{\rm Tr}[\hat M\hat G^K(t,t)]$ for the Nambu spinor $\Psi$ and any matrix $\hat M$ in Nambu space, together with Eq.~\eqref{Quasi}. It is fixed unambiguously by the clean London limit because at $\bq=0$ the same expression evaluated with the linear solution gives ${\mathbf j}=-(n_ce^2/m){\mathbf A}$ at all frequencies. Here the velocity is taken at the Fermi surface. The particle-hole asymmetric corrections to the current vertex, of relative order $\eps/\veps_F$ [the energy dependence of the velocity and its dependence on the field, cf. Eq.~\eqref{eq:kinNS}], contribute to the dc current at first order in $q$ only together with the longitudinal $O(q)$ part of $\check g_1$. They do not affect the order parameter and diamagnetic channels considered below, where they produce corrections of order $(\eps/\veps_F)^2$. 

\subsection{Contributions from the source terms}
\begin{figure}
\includegraphics[width=0.950\linewidth]{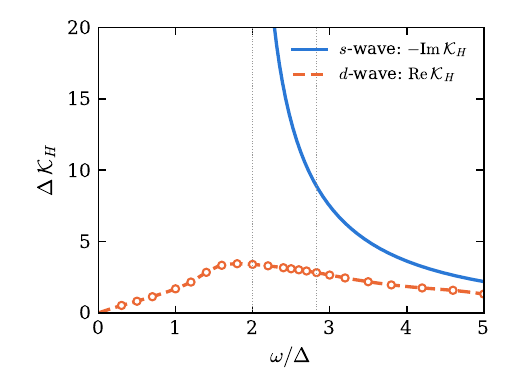}
\caption{Kernel ${\cal K}_H(\omega)$ of the order parameter channel, Eq.~\eqref{jdcDelta}, with Born impurity
scattering, in the clean limit, at $T=0.04\Delta$. Solid line: $s$-wave pairing,
$-{\rm Im}\,{\cal K}_H$ of Eq.~\eqref{eq:KH}, while ${\rm Re}\,{\cal K}_H$ is of order $1/(\tau\Delta^2)$. The
kernel vanishes below the pair-breaking threshold $2\Delta$ and diverges as
$(\omega-2\Delta)^{-1/2}$ above it. Symbols: $d$-wave pairing ($\Delta_\bn={\cal Y}_\bn\Delta$),
${\rm Re}\,{\cal K}_H$ from the numerical solution of the second-order equations with
$1/\tau=0.05\Delta$, while ${\rm Im}\,{\cal K}_H\lesssim0.03/\Delta$ is of order $1/(\tau\Delta^2)$. The dashed line is a
guide to the eye. The dotted lines mark $\omega=2\Delta$ and $2\sqrt2\Delta$.}
\label{Fig-K01w}
\end{figure}

The source function is defined by the terms listed in Eq.~\eqref{SumGamma} of Appendix \ref{AppendixSource}. We consider their contributions separately, and the details of the calculation can be found in Appendix \ref{AppendixC}. Four contributions are linear in the order parameter oscillation and are of the same order: $\check S^{(\Delta)}$ acting on the field-induced part $\check g_1^{(E)}$ of the linear solution, $\check S^{(E)}$ acting on the part $\check g_1^{(\Delta)}$ induced by $\delta\hat\Delta_1$ itself [the last line of \eqref{calR}], the cross terms $\check g_1^{(E)}\check g_1^{(\Delta)}$ of the normalization function \eqref{checkN}, and the impurity source $\check S^{(\rm imp)}$, Eq.~\eqref{Simp}, which couples $\check g_1^{(E)}$ to the impurity average of $\check g_1^{(\Delta)}$. Adding all four (Appendix \ref{AppendixC}) we find for the contribution of the Schmid--Higgs channel
\begin{equation}\label{jdcDelta}
j^{(\Delta)}_{{\rm dc},l}(\bq)=-\frac{e^2\nu_Fv_F^2}{4\omega}
\left\{i{E}_l^*(\bq{\mathbf E})\,{\cal D}_{\mathrm{SH}}(\omega)\,
{\mathcal K}_H(\omega)+\textrm{c.c.}\right\},
\end{equation}
where ${\cal D}_{\mathrm{SH}}(\omega)\equiv\delta\Delta_{1+}/(\bq{\mathbf E})=2i\chi_{\mathrm{SH}}(\omega)(ev_F^2/\veps_F){\cal A}(\omega)$ is the order parameter response of Eq.~\eqref{eq:dDelta1} per unit $(\bq{\mathbf E})$, Fig.~\ref{Fig-DSH}. To leading order in $1/(\tau\Delta)$ the kernel ${\cal K}_H$ does not depend on the scattering rate. For $s$-wave pairing it is given by
\begin{equation}\label{eq:KH}
{\mathcal K}_H(\omega)=-\frac{16\pi i\,\Delta}{\omega\sqrt{\omega^2-4\Delta^2}}\,\tanh\left(\frac{\omega}{4T}\right)\theta(\omega-2\Delta),
\end{equation}
which reproduces the numerical solution of the second-order equations within $0.1\%$ (Appendix \ref{AppendixC}). For $d$-wave pairing ${\cal K}_H$ is real and is shown in Fig.~\ref{Fig-K01w}. The term with $\delta\Delta_{1+}$ is directed along ${\mathbf E}^*$, the only vector available at $\bq\to0$. In terms of the coefficients of Eq.~\eqref{54},
\begin{equation}\label{eq:betaDelta}
\begin{aligned}
\beta_{\rm a}^{(\Delta)}&=\frac{e^2\nu_Fv_F^2}{4\omega}\,{\rm Re}\big[{\cal D}_{\mathrm{SH}}{\cal K}_H\big],\\
\beta_{\rm s}^{(\Delta)}&=\frac{e^2\nu_Fv_F^2}{4\omega}\,{\rm Im}\big[{\cal D}_{\mathrm{SH}}{\cal K}_H\big],
\end{aligned}
\end{equation}
and by Eq.~\eqref{eq:Aidentity} we have
\beg\label{ReImDSH}
\begin{aligned}
{\rm Re}\,{\cal D}_{\mathrm{SH}}&=\left(\frac{ev_F^2}{\veps_F}\right)\left(\frac{\Delta}{2\omega^2}\right)\,c_T\,{\rm Im}\,\chi_{\mathrm{SH}}, \\
{\rm Im}\,{\cal D}_{\mathrm{SH}}&=\left(\frac{ev_F^2}{\veps_F}\right)\left(\frac{\Delta}{2\omega^2}\right)\left(1-c_T\,{\rm Re}\,\chi_{\mathrm{SH}}\right)
\end{aligned}
\en
with $c_T=\chi^{-1}_{\mathrm{SH}}(0)+N_T$. The two pairing symmetries therefore select different quadratures of the order parameter response. For $s$-wave pairing ${\cal K}_H$ is imaginary and $\beta_{\rm a}^{(\Delta)}\propto|{\cal K}_H|\,{\rm Im}\,{\cal D}_{\rm SH}\propto|{\cal K}_H|\,[1-c_T{\rm Re}\,\chi_{\rm SH}]$, so that the helicity-odd current vanishes below $2\Delta$ at $T\to0$ and diverges as $(\omega-2\Delta)^{-1/2}$ above the threshold, where $1-c_T{\rm Re}\,\chi_{\rm SH}$ remains finite. For $d$-wave pairing ${\cal K}_H$ is real, and $\beta_{\rm a}^{(\Delta)}\propto{\cal K}_H\,{\rm Re}\,{\cal D}_{\rm SH}\propto{\cal K}_H\,{\rm Im}\,\chi_{\rm SH}$ follows the absorptive part of the amplitude susceptibility, with a broad maximum near $2.5\Delta$. The helicity-even coefficient $\beta_{\rm s}^{(\Delta)}$ is proportional to the other part of ${\cal D}_{\rm SH}$, that is, to ${\rm Re}\,{\cal D}_{\rm SH}$ for $s$-wave pairing and to ${\rm Im}\,{\cal D}_{\rm SH}$ for $d$-wave pairing.

The magnetization of the order parameter channel follows from the same result. The dependence of the current \eqref{jdcDelta} on the wave vector enters only through $\delta\Delta_1$, which is driven by the field component at its own wave vector. For two field components ${\mathbf E}e^{i\bq_1\br}$ and ${\mathbf E}^*e^{-i\bq_2\br}$ the dc current at $\bQ=\bq_1-\bq_2$ is therefore
\begin{equation}\label{eq:jQ}
\begin{aligned}
j^{(\Delta)}_{{\rm dc},l}=-\frac{e^2\nu_Fv_F^2}{4\omega}\Big\{&iE_l^*(\bq_1{\mathbf E})\,X\\&-iE_l(\bq_2{\mathbf E}^*)\,X^*\Big\},
\end{aligned}
\end{equation}
where $X={\cal D}_{\rm SH}{\cal K}_H$. This equation reduces to Eq.~\eqref{jdcDelta} for $\bq_1=\bq_2=\bq$. Its helicity-odd part proportional to $\bQ$ has the form $i\bQ\times{\mathbf M}^{(\Delta)}$ with
\begin{equation}\label{eq:MDelta}
{\mathbf M}^{(\Delta)}=\frac{e^2\nu_Fv_F^2}{8\omega}\,{\rm Im}\big[{\cal D}_{\rm SH}{\cal K}_H\big]\,{\mathbf Z}=\tfrac12\beta_{\rm s}^{(\Delta)}{\mathbf Z}.
\end{equation}
The helicity-even coefficient $\beta_{\rm s}^{(\Delta)}$ is proportional to the other part of ${\cal D}_{\rm SH}$, that is, to ${\rm Re}\,{\cal D}_{\rm SH}$ for $s$-wave pairing and to ${\rm Im}\,{\cal D}_{\rm SH}$ for $d$-wave pairing.
For $s$-wave pairing ${\mathbf M}^{(\Delta)}\propto|{\cal K}_H|\,{\rm Re}\,{\cal D}_{\rm SH}\propto|{\cal K}_H|\,{\rm Im}\,\chi_{\rm SH}$ and so it vanishes below $2\Delta$ at $T\to0$ and diverges as $(\omega-2\Delta)^{-1}$ above the threshold, since both factors diverge as $(\omega-2\Delta)^{-1/2}$. For $d$-wave pairing ${\mathbf M}^{(\Delta)}\propto{\cal K}_H\,{\rm Im}\,{\cal D}_{\rm SH}\propto{\cal K}_H\,[1-c_T{\rm Re}\,\chi_{\rm SH}]$, which is finite at all frequencies and changes sign near $2.7\Delta$, where ${\rm Im}\,{\cal D}_{\rm SH}$ vanishes (Fig.~\ref{Fig-DSH}).

Three features of this result deserve emphasis. First, the kernel ${\cal K}_H$ is generated entirely by impurity scattering, and yet it does not depend on the scattering rate. At $\bq\to0$ inside the second-order solution the vector potential enters every trajectory as a Doppler shift $ev_F(\bn{\mathbf A})\hat\tau_3$, which is removed by the trajectory-dependent gauge transformation $\check g(t,t')\to e^{i\Theta_\bn(t)}\check g(t,t')e^{-i\Theta_\bn(t')}$, $\dot\Theta_\bn=ev_F\bn{\mathbf A}(t)$. The current depends on the equal-time function and is invariant. The transformation does not commute with the impurity average $\langle\check g\rangle_\bn$, which connects trajectories with different phases. The field therefore survives only through the term which accounts for  scattering on impurities and therefore its amplitude must be proportional to $1/\tau$. Physically, a quasiparticle cannot absorb a photon of vanishing momentum without transferring momentum to the impurities, and the finite photon momentum does not help, since $v_Fq\ll\omega$. This small amplitude is compensated by the quasiparticle lifetime $\tau$, which enters through the factor $1/(\eta_{\bn\eps}^R+\eta_{\bn\eps}^A)$ in $\delta\hat g_2^K$ at energies where quasiparticles exist. The result is finite in the clean limit, as for the photon drag in the normal state [Eq.~\eqref{eq:NSexact} at $\omega\tau\gg1$]. The individual contributions contain this factor without the compensating amplitude and are therefore proportional to $\tau$. These terms cancel in the sum. Second, for $s$-wave pairing the impurity-assisted absorption requires breaking a Cooper pair, which sets the threshold at $\omega=2\Delta$. Equation \eqref{eq:KH} can be written as ${\cal K}_H=-(8\pi i/\omega)f(\omega/2)\tanh(\omega/4T)$ with $f(\eps)=\Delta/\sqrt{\eps^2-\Delta^2}$, so that its inverse square-root edge is that of the anomalous spectral function at the pair-breaking energy $\eps=\omega/2$. For $d$-wave pairing the nodal quasiparticles provide final states at all energies and ${\cal K}_H$ has no threshold. Third, one needs to appreciate that a phenomenological damping term $i\gamma\hat\tau_3$ does not capture this physics. With such a term the field then couples to the bath of the damping term, which supplies final states without pair breaking, and the resulting kernel has its threshold at $\omega=|\Delta_\bn|$ instead of $2\Delta$ (Appendix \ref{AppendixC}).

Let us now turn to the diamagnetic source $\check S^{(A)}$, which originates from the particle-hole asymmetric gradient terms of the Doppler vertex, Eq.~\eqref{ExpandLinearA}. It contains two parts, Eq.~\eqref{Sources}: the spatial-gradient (streaming) part $\propto(\bq{\mathbf A}_s)$ and the angular part $\propto(\bq\hat\phi_\bn)(\bn{\mathbf A}_s)\partial_{\phi_\bn}$, where ${\mathbf A}_\pm$ are the amplitudes of the two terms in Eq.~\eqref{VectorPotential}, ${\mathbf A}_+={\mathbf E}/i\omega$ and ${\mathbf A}_-={\mathbf A}_+^*$. The streaming part alone produces a helicity-odd current whose coefficient is proportional to the quasiparticle lifetime. In the normal state it equals $\beta_{\rm a}=-e^3n_c\tau^3/[m^2(1+\omega^2\tau^2)]$. This current is cancelled identically by the angular part. Indeed, at $\bq\to0$ the linear solution inside the source has the form $\check g_{1\bar s}=(\bn{\mathbf A}_{\bar s})\check Y_{\bar s}(\eps;\Delta_\bn)$, where $\check Y$ depends on $\bn$ only through $\Delta_\bn$. With $\partial_{\phi}(\bn{\mathbf A})=\hat\phi_\bn{\mathbf A}$ the source becomes
\begin{equation}\label{eq:jA}
\begin{aligned}
&\check S^{(A)}=\sum_{s=\pm}\frac{se}{2m}\big\langle\{\check\tau_3,\check X_s\}_+\big\rangle_s,
\end{aligned}
\end{equation}
where
\begin{equation}\label{eq:Xs}
\begin{aligned}
&\check X_s=\big[(\bq{\mathbf A}_s)(\bn{\mathbf A}_{\bar s})+(\bn{\mathbf A}_s)(\bq{\mathbf A}_{\bar s})\\
&\quad-(\bn\bq)(\bn{\mathbf A}_s)(\bn{\mathbf A}_{\bar s})\big]\check Y_{\bar s}\\
&\quad+(\bq\hat\phi_\bn)(\bn{\mathbf A}_s)(\bn{\mathbf A}_{\bar s})\,(\partial_{\phi}\Delta_\bn)\,\partial_{\Delta_\bn}\check Y_{\bar s}.
\end{aligned}
\end{equation}
Both angular factors are symmetric under ${\mathbf A}_s\leftrightarrow{\mathbf A}_{\bar s}$, i.e. under ${\mathbf E}\leftrightarrow{\mathbf E}^*$, and after the average over $\bn$ the polarization enters the current only through combinations symmetric in ${\mathbf E}$ and ${\mathbf E}^*$. Hence
\begin{equation}\label{eq:betaA}
\beta_{\rm a}^{(A)}=0
\end{equation}
for either pairing symmetry, at any temperature and scattering rate since the diamagnetic channel contributes to the photon drag, but not to $\beta_{\rm a}$. Its contribution to the magnetization, which is determined by the dependence on the envelope wave vector, has not been evaluated. The cancellation is confirmed by the numerical solution of the second-order equations (Appendix \ref{AppendixC}).

Two more source terms, $\check S^{(\Theta)}$ and $\check S^{(E)}$, do not contribute to $\beta_{\rm a}$ at the order considered here. The one given by $\check S^{(\Theta)}$ is 
\beg\label{SThetaDrag}
{\mathbf j}_{\rm dc}^{(\Theta)}=
-\frac{e^3\nu_Fv_F^4}{4\,\varepsilon_F\omega^2}\,
\Big\langle \bn\,(\bn\bq)(\bn{\mathbf E})(\bn{\mathbf E}^*)\,{\cal K}_4(\bn;\omega)\Big\rangle_{\bn},
\en
where ${\cal K}_4(\bn;\omega)$ is a scalar function of $\omega$ and of the direction $\bn$. The angular factor is symmetric under ${\mathbf E}\leftrightarrow{\mathbf E}^*$, so that after the average over $\bn$ this term contributes only to the photon drag. We do not discuss it further. The source $\check S^{(E)}$, evaluated with the field-induced part $\check g_1^{(E)}$ of the linear solution, contributes to the current only at second order in $q$ (its part containing $\check g_1^{(\Delta)}$ is of first order and is included in Eq.~\eqref{jdcDelta}):
\beg\label{jdcE}
\mathbf j^{(E)}_{\rm dc}(\bq)=O(q^2).
\en
This result is a re-statement of the fact that in the particle-hole symmetric case neither the helicity-odd nor the photon-drag current is generated. It holds as long as the velocity in the force term and in the current vertex is taken at the Fermi surface. The particle-hole asymmetric corrections to these vertices are the energy dependence of the velocity, $v(\eps)=v_F\sqrt{1+\eps/\veps_F}$, and the field-dependent part of the current vertex. They act on the longitudinal $O(q)$ part of $\check g_1$ and give contributions of first order in $q$ and in $\eps/\veps_F$. In the normal state their helicity-odd parts cancel each other (Appendix \ref{AppendixB}). In the superconducting state they belong to the longitudinal class discussed next. 

Finally, we comment on the two remaining source terms, $\check S^{(\mu)}$ and $\check S^{(\Lambda)}$ in Eq.~\eqref{SumGamma}, which contain the imbalance potential $\varphi_{1\pm}$. Since $\varphi_{1}\propto\zeta_\omega\Pi_\mu(q,\omega)(\mathbf{q}\mathbf{E})$, Eq.~\eqref{phi1pmMain}, both terms are proportional to the longitudinal projection of the polarization vector, exactly as in the order parameter channel. The source $\check S^{(\Lambda)}$ contains the additional factor $(\bn\bq)$ and is therefore of second order in $q$ (Sec.~III C), so that only $\check S^{(\mu)}$ contributes at the order considered here. The trace identities
\eqref{eq:id1}--\eqref{eq:id2} apply, and the resulting contribution to the current is of the
same order in $\epsilon/\varepsilon_F$ as \eqref{jdcDelta}, but with the
overall factor $\zeta_\omega\Pi_\mu(q,\omega)$ replacing the pair
susceptibility. This channel imports the collective physics of
the charge-imbalance sector into the rectified response for its frequency
dependence is controlled by the dielectric function of Fig.~\ref{Fig-PiInvw} and, for
$s$-wave pairing, it may be resonantly enhanced at the frequency where
$\mathrm{Re}\,\Pi^{-1}_\mu$ crosses zero, a resonance that has no counterpart
for $d$-wave pairing. Together with the particle-hole asymmetric force and
current-vertex terms mentioned above, the imbalance channel forms a
longitudinal class. All these terms act through the charge and imbalance
densities induced at first order, which are proportional to $(\bq{\mathbf E})$.
Its evaluation requires the self-consistent longitudinal (imbalance and phase)
response of the superconductor with impurity scattering, including the
corresponding vertex corrections, and the particle-hole asymmetric velocity in
the Nambu kinetic equation. We do not carry it out in the present work. The
longitudinal class is of the same order, $v_Fq/\veps_F$, as the
order parameter channel, and no symmetry argument forces its helicity-odd part
to vanish. In the normal state it does vanish, but only through a cancellation
between its pieces (Appendix \ref{AppendixB}), which has no reason to persist in
the superconducting state. The results for the order parameter channel must therefore be regarded as
estimates of the order of magnitude of the total response, away from the
imbalance resonance, and not of its sign or detailed lineshape. For the
magnetization this applies all the more, since it also contains the orbital
contribution present already in the normal state.

\subsection{Contributions from the $\check{\cal N}$ terms}
In addition, one also needs to consider the contributions from the terms which contain the 'normalization' functions $\check{\cal N}_{\bn\eps}$.
The cross terms $\check g_1^{(E)}\check g_1^{(\Delta)}$ are of the same order as the Higgs-channel source terms and are included in Eq.~\eqref{jdcDelta}. Without them the terms proportional to $\tau$ in the contributions of $\check S^{(\Delta)}$ and $\check S^{(E)}$ do not cancel. The terms bilinear in $\check g_1^{(E)}$ do not contribute in the particle-hole symmetric approximation (Appendix \ref{AppendixB}). Their particle-hole asymmetric part involves the longitudinal $O(q)$ part of $\check g_1$ and belongs to the longitudinal class.

\section{Discussion}
Let us now discuss how the light-induced magnetization of a superconducting film
can be observed and estimate its magnitude. 
In order to produce a static magnetization, the order parameter channel needs two properties
of the field in the plane of the film. The field must have a longitudinal
component, $\bq\cdot{\mathbf E}\neq0$, which drives the order parameter,
Eq.~\eqref{eq:dDelta1}, and it must carry a finite helicity
${\mathbf Z}=i[{\mathbf E}\times{\mathbf E}^*]$. A plane wave propagating in the
plane of the film has a single in-plane field component, perpendicular to its
wave vector, and both quantities vanish, so that the light has to fall on the
film from outside. At normal incidence the in-plane wave vectors of a beam are
those of its envelope, and Eq.~\eqref{eq:jQ} gives the magnetization
\eqref{eq:MDelta} directed along the film normal. The uniform current
$\beta_{\rm a}[\bq\times{\mathbf Z}]$ needs, in addition, a finite in-plane wave
vector of the incident wave and appears only at oblique incidence, as we illustrate in Fig.~\ref{Fig-geometry}. We take the $xz$-plane as the plane of incidence and
denote by $\theta_i$ the angle of incidence and by $E_p$ and $E_s$ the
amplitudes of the $p$- and $s$-polarized components of the incident wave, with
$E_s=\pm iE_p$ for circular polarization. The in-plane projections are
\begin{equation}
\bq_\parallel=\frac{\omega}{c}\sin\theta_i\,\hat x,\qquad
{\mathbf E}_\parallel=\left(t_pE_p\cos\theta_i,\;t_sE_s,\;0\right),
\end{equation}
where $t_{s,p}(\theta_i,\omega)$ are the transmission amplitudes of the film,
so that, with $q=\omega/c$,
\begin{equation}
\begin{aligned}
\bq_\parallel\cdot{\mathbf E}_\parallel&=q\sin\theta_i\cos\theta_i\,t_pE_p,\\
Z_z=i[{\mathbf E}_\parallel\times{\mathbf E}^*_\parallel]_z&=2\cos\theta_i\,\mathrm{Im}\left(t_p^*t_sE_p^*E_s\right).
\end{aligned}
\end{equation}
The helicity-odd part of the current \eqref{54},
$\bj_{\rm a}=\beta_{\rm a}[\bq_\parallel\times{\mathbf Z}]=-\beta_{\rm a}Z_zq\sin\theta_i\,\hat y$,
flows in the film perpendicular to the plane of incidence and is reversed
together with the helicity. For slowly varying $t_{s,p}$ it is proportional to
$\sin 2\theta_i$, while the magnetization
${\mathbf M}^{(\Delta)}=\tfrac12\beta_{\rm s}^{(\Delta)}Z_z\hat z$ is proportional
to $\cos\theta_i$. Neither magnitude depends on the orientation of
$\bq_\parallel$ with respect to the crystal axes. For $d$-wave pairing the
anisotropic term of Sec.~IV enters only the helicity-even part of the current.
 
\begin{figure}
\begin{center}
\begin{tikzpicture}[scale=0.85,>=stealth]
\draw[thick] (-3,0)--(3,0) node[right]{\small film};
\draw[thick,->] (0,0)--(0,2.3) node[above]{$\hat z$};
\draw[thick,->] (-2.4,2.0)--(-0.3,0.25) node[midway,above right]{$\bq$};
\draw[dashed] (0,0)--(-2.4,2.0);
\draw (0,0.8) arc (90:140:0.8); \node at (-0.45,1.15) {$\theta_i$};
\draw[->,red,thick] (-1.6,1.33)--(-1.0,2.05) node[above]{${\mathbf E}_p$};
\draw[->,blue,thick] (-1.6,1.33)--(-2.2,1.0) node[left]{${\mathbf E}_s\parallel\hat y$};
\draw[->] (0,0)--(2.2,0) node[below]{$\hat x$}; \draw[->,thick,gray] (0,0)--(1.6,-0.1);
\node[black] at (1.75,0.75) {\small $\boldsymbol{\bq_\parallel=q\sin\theta_i\,\hat x}$};
\end{tikzpicture}
\end{center}
\caption{Oblique-incidence geometry. The in-plane projection of the photon momentum, $\bq_\parallel$, and the in-plane components of the $p$- and $s$-polarized fields drive the rectified response of the film. For elliptical polarization the induced moment points along the film normal $\hat z$.}
\label{Fig-geometry}
\end{figure} 
 
The field ${\mathbf E}$ in these formulas is the field inside the film. For a
film of thickness $d$ much smaller than the penetration depth, on a substrate
with refractive index $n_s$, it is related to the incident field at normal
incidence by the Tinkham factor $t=2/(1+n_s+Z_0\sigma d)$, where
$Z_0=377\,\Omega$ and $\sigma(\omega)$ is the complex conductivity of the film.
At oblique incidence $t$ is replaced by $t_s$ and $t_p$. The longitudinal
in-plane component is also screened by the two-dimensional plasmon of the
film. For cuprate-like parameters and $q_\parallel\sim3\times10^4$~m$^{-1}$ the
plasmon frequency lies in the terahertz range, and the screening factor is of
order unity and depends on frequency. We leave these factors out of the
estimates below. They change the magnitude of the response and add a frequency
dependence of their own, which has to be taken into account when the theory is
compared with experiment.
 
For circular polarization, $|{\mathbf Z}|=E_0^2$, Eq.~\eqref{eq:MDelta} divided
by the electron density $n_c=2\nu_F\veps_F$ gives the moment per carrier of the
order parameter channel. In units of the Bohr magneton $\mu_B=e/2m$ it equals
\begin{equation}
\frac{{\mathbf M}^{(\Delta)}}{n_c\mu_B}=\frac{\rho^{2}}{4}\left(\frac{\varepsilon_F}{\Delta}\right)\frac{\omega}{\Delta}\;
\frac{{\rm Im}\big[\Delta{\cal K}_H(\omega)\,\Delta{\cal D}_{\rm SH}(\omega)\big]}{ev_F^2/\veps_F}\,\frac{{\mathbf Z}}{E_0^2},
\label{eq:moment}
\end{equation}
where $\rho={eE_0v_F}/({\omega\,\varepsilon_F})$. For parameters representative
of a $d$-wave cuprate ($\varepsilon_F\approx0.3$~eV, $\vF\approx2\times10^{5}$~m/s,
$\Delta\approx17$~meV) and for $E_0\approx10$~kV/cm one finds
$\rho\approx2.6\times10^{-2}\,\Delta/\omega$. 

\begin{figure}
\includegraphics[width=0.950\linewidth]{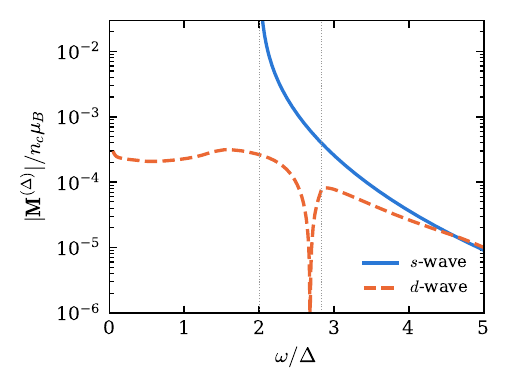}
\caption{Absolute value of the moment per carrier of the order parameter
channel, Eq.~\eqref{eq:moment}, induced by circularly polarized light, for
$s$-wave (solid) and $d$-wave (dashed) pairing, with
$\rho=2.6\times10^{-2}\,\Delta/\omega$, $\veps_F/\Delta=17.6$ and $T=0.04\Delta$.
The $d$-wave curve uses the kernel ${\cal K}_H$ of Fig.~\ref{Fig-K01w}, and its
dip at $\omega\approx2.7\Delta$ corresponds to the change of sign. The dotted
lines mark $\omega=2\Delta$ and $2\sqrt2\Delta$.}
\label{Fig-moment}
\end{figure}

The moment is plotted in
Fig.~\ref{Fig-moment}. For $d$-wave pairing it reaches about
$3\times10^{-4}\,\mu_B$ per carrier at $\omega\approx1.6\Delta$. Near $2.7\Delta$
it changes sign, and at higher frequencies it stays below $10^{-4}\,\mu_B$. An
isotropic ($s$-wave) gap with the same parameters gives no moment below $2\Delta$ at
$T\ll\Delta$. Above the threshold the moment diverges as $(\omega-2\Delta)^{-1}$
and falls to about $2.7\times10^{-3}\,\mu_B$ per carrier at $\omega=2.3\Delta$ and
to $8\times10^{-4}\,\mu_B$ at $2.6\Delta$. In a real material the divergence is
cut off by the anisotropy of the gap and by inelastic scattering, which our
model of an isotropic gap does not contain. Since
$\rho^2\veps_F/\Delta=(eE_0v_F/\omega\Delta)^2(\Delta/\veps_F)$, the moment is of
first order in the particle-hole asymmetry, in agreement with the naive
estimate of the Introduction, and in the clean limit it does not depend on the
scattering rate. The scale $1/(\tau_uT_c)$ considered in the Introduction, with
$\tau_u=\tau$ given by Eq.~\eqref{ElasticRate}, therefore does not appear. The
longitudinal class, which is of the same order, and the other contributions to
the magnetization discussed in Sec.~IV are not included, so the numbers above
are estimates of the order of magnitude.
 
Our results rest on several approximations. The particle-hole asymmetric
terms were kept to leading order in $\epsilon/\varepsilon_F$. Within this order
the order parameter channel is complete, and the symmetry statements, among
them $\beta_{\rm a}^{(A)}=0$, and the positions of the thresholds are exact.
Impurity scattering was treated in the Born approximation, in the clean limit
of Sec.~II and to leading order in $1/(\tau\Delta)$. The dirty and gapless
regimes, where the magnitude of the effect is expected to be set by the
scattering rates, are beyond the present treatment. The response
was computed to first order in the photon momentum and without spin-orbit
coupling. The expansion in powers of the field requires the Doppler shift
$ev_F(\bn{\mathbf A})$, whose amplitude is of the order of $\rho\varepsilon_F$,
to be small compared with $\omega$ and $\Delta$. For the parameters used above
$\rho\varepsilon_F\approx0.46\Delta$ at $\omega=\Delta$ and $0.18\Delta$ at
$\omega=2.5\Delta$. At $\omega\lesssim\Delta$ the field $E_0\approx10$~kV/cm is
at the edge of the perturbative regime or beyond it, and in this range the
curves of Fig.~\ref{Fig-moment}, which scale as $E_0^2$, apply to weaker fields
after rescaling. Stronger fields lead to the regime where Eliashberg-type
renormalizations of the gap become important. 

The existing theories of the
inverse Faraday effect in conventional superconductors
\cite{Mironov2021-IFESC,Putilov2023-IFESC} are based on the time-dependent
Ginzburg-Landau equations, which, as discussed in the Introduction, apply only
in the gapless regime where pair breaking dominates. They have no common range
of validity with the clean-limit theory developed here, and a quantitative
comparison with them is not possible.
 
The measurement we have in mind is a THz-pump, static-probe experiment, in
which elliptically polarized radiation with $\w\sim(1\text{--}3)\Delta$ falls on
a thin film and the static moment along the film normal is detected by polar
magneto-optical Kerr rotation or by SQUID magnetometry. Reversing the helicity
reverses the moment, and linear polarization gives none. Since ${\mathbf Z}$ is
the only axial vector that can be formed from ${\mathbf E}$ and ${\mathbf E}^*$,
every contribution to the magnetization is proportional to it, and the
contributions cannot be separated by changing the polarization or the angle of
incidence. They differ in their dependence on frequency and temperature. Only
the order parameter channel contains $\chi_{\rm SH}(\omega)$. For $s$-wave pairing
its moment sets in at the threshold $2\Delta$, while the imbalance channel may
add a resonance at $\omega\approx1.63\Delta$, where ${\rm Re}\,\Pi_\mu^{-1}$
crosses zero (Fig.~\ref{Fig-PiInvw}). For $d$-wave pairing the change of sign of
the order parameter channel comes from the factor $1-c_T{\rm Re}\,\chi_{\rm SH}$
in Eq.~\eqref{ReImDSH}, which vanishes just below the maximal-gap threshold
$2\sqrt2\Delta$. A measurement of the static moment is thus a probe of the
amplitude mode that complements the third-harmonic experiments
\cite{Katsumi2018,Katsumi2020}.
 
The same approach applies to noncentrosymmetric superconductors, where a local
photogalvanic current is allowed by symmetry, and to superconductors with
chiral order parameters, where the broken time-reversal symmetry allows a
light-induced magnetization even for linear polarization. For beams with
orbital angular momentum Eq.~\eqref{eq:jQ} determines the pattern of the
induced magnetization. The formalism also covers the second-harmonic response.

\section{Conclusions}
We have developed a microscopic theory of the inverse Faraday effect in conventional and unconventional ($d$-wave)
spin-singlet superconductors, based on an extension of the Keldysh--Nambu
quasiclassical formalism that retains the particle-hole asymmetric vertices
responsible for the branch population imbalance, with the terms of order
$\eps/\veps_F$ obtained from the exact finite-momentum expansion. The
rectified response was shown to decompose into additive channels with
distinct physical content, all driven by the longitudinal projection of the
field and by its helicity. Our central result is that at finite photon
momentum the radiation induces a linear-in-field oscillation of the amplitude
(Schmid--Higgs) mode of the order parameter which contributes directly to the
static magnetization: the corresponding current, Eq.~\eqref{jdcDelta}, is
proportional to the pair susceptibility and is resonantly enhanced at the
pair-breaking threshold. This response is fully symmetric under the point
group, exists for any singlet pairing, opens only at finite photon momentum
and at first order in the particle-hole asymmetry, and is exactly free of a
static piece. Its contribution to the current is an impurity-assisted
process which, in the clean limit, does not depend on the scattering rate, so
that the magnetization is of first order in the particle-hole asymmetry,
without a relaxation enhancement. The diamagnetic channel contributes only to
the photon drag, its helicity-odd part being cancelled identically by the
angular part of the momentum gradient, and in the normal state the
magnetization vanishes, as required by momentum balance. The charge-imbalance
channel, which is of the same order as the order parameter channel, remains to
be evaluated. In a thin film at oblique
incidence with elliptically polarized light the moment points along the film
normal, and its frequency dependence distinguishes the pairing symmetries
through the lineshape of the Higgs feature and, possibly, through the
collective resonance of the imbalance channel. Our results establish the rectified
nonlinear response as a dc probe of collective modes in superconductors, and
provide the framework for its extension to disordered, chiral, and multiband
systems.

\begin{acknowledgments}
We would like to thank Andrey Chubukov, Peter Gordon, Dima Pesin  and Emil Yuzbashyan for useful discussions related to various aspects of this study. MD acknowledges the financial support by the National Science Foundation grant No. DMR-2400484. The authors acknowledge the use of Anthropic's
Claude for cross-checking the analytical derivations, for independent
numerical verification of the results, and for figure preparation and
manuscript checking. All outputs of Claude were verified and validated by
the authors.
\end{acknowledgments}

\begin{appendix}
\begin{widetext}

\section{Expressions for the branch population imbalance and order parameter}\label{AppendixA}
It is sufficient to consider $\varphi_{1+}(\bq,\omega)$, since $\varphi_{1-}=\varphi_{1+}^*$. Inserting the linear solutions \eqref{g1kp} and \eqref{Solution4dg1K} into Eq.~\eqref{varphi1pm} we find
\beg\label{varphi1plus}
\begin{aligned}
&\varphi_{1+}(\mathbf{q},\w) = -\left(\frac{e v_F^2}{4i\omega}\right)\Pi_\mu(\mathbf{q},\omega)
\int_0^{2\pi}\frac{d\phi_{\mathbf n}}{2\pi}\,(\bn{\mathbf E})(\bn\bq)
\int_{-\infty}^{\infty}d\eps\,
\mathcal K_{\bn\eps}(\mathbf q,\omega),
\end{aligned}
\en
where we introduced the functions
\beg\label{Pikw}
\begin{aligned}
&\Pi_\mu^{-1}(\bq,\omega)=1+\frac{1}{4}\int\limits_0^{2\pi}\frac{d\phi_\bn}{2\pi}\int\limits_{-\infty}^\infty \left\{\frac{\left(\eta_{\bn\eps+\frac{\omega}{2}}^{R}+\eta_{\bn\eps-\frac{\omega}{2}}^{R}\right)\left(1-g_{\bn\eps+\frac{\omega}{2}}^Rg_{\bn\eps-\frac{\omega}{2}}^R+f_{\bn\eps+\frac{\omega}{2}}^Rf_{\bn\eps-\frac{\omega}{2}}^R\right)t_{\eps-\frac{\omega}{2}}}{\left(\eta_{\bn\eps+\frac{\omega}{2}}^{R}
+\eta_{\bn\eps-\frac{\omega}{2}}^{R}\right)^2-v_F^2(\bn\bq)^2}\right.\\&\left.-\frac{\left(\eta_{\bn\eps+\frac{\omega}{2}}^{A}+\eta_{\bn\eps-\frac{\omega}{2}}^{A}\right)\left(1-g_{\bn\eps+\frac{\omega}{2}}^Ag_{\bn\eps-\frac{\omega}{2}}^A+f_{\bn\eps+\frac{\omega}{2}}^Af_{\bn\eps-\frac{\omega}{2}}^A\right)t_{\eps+\frac{\omega}{2}}}{\left(\eta_{\bn\eps+\frac{\omega}{2}}^{A}
+\eta_{\bn\eps-\frac{\omega}{2}}^{A}\right)^2-v_F^2(\bn\bq)^2}\right.\\&\left.+\frac{\left(\eta_{\bn\eps+\frac{\omega}{2}}^{R}+\eta_{\bn\eps-\frac{\omega}{2}}^{A}\right)\left(1-g_{\bn\eps+\frac{\omega}{2}}^Rg_{\bn\eps-\frac{\omega}{2}}^A+f_{\bn\eps+\frac{\omega}{2}}^Rf_{\bn\eps-\frac{\omega}{2}}^A\right)(t_{\eps+\frac{\omega}{2}}-t_{\eps-\frac{\omega}{2}})}{\left(\eta_{\bn\eps+\frac{\omega}{2}}^{R}
+\eta_{\bn\eps-\frac{\omega}{2}}^{A}\right)^2-v_F^2(\bn\bq)^2}\right\}d\eps
\end{aligned}
\en
and 
\beg\label{Kernel}
{\mathcal K}_{\bn\eps}=\frac{\left(g_{\bn\eps+\frac{\omega}{2}}^R-g_{\bn\eps-\frac{\omega}{2}}^R\right)t_{\eps-\frac{\omega}{2}}}{\left(\eta_{\bn\eps+\frac{\omega}{2}}^{R}
+\eta_{\bn\eps-\frac{\omega}{2}}^{R}\right)^2-v_F^2(\bn\bq)^2}-\frac{\left(g_{\bn\eps+\frac{\omega}{2}}^A-g_{\bn\eps-\frac{\omega}{2}}^A\right)t_{\eps+\frac{\omega}{2}}}{\left(\eta_{\bn\eps+\frac{\omega}{2}}^{A}
+\eta_{\bn\eps-\frac{\omega}{2}}^{A}\right)^2-v_F^2(\bn\bq)^2}+\frac{\left(g_{\bn\eps+\frac{\omega}{2}}^R-g_{\bn\eps-\frac{\omega}{2}}^A\right)(t_{\eps+\frac{\omega}{2}}-t_{\eps-\frac{\omega}{2}})}{\left(\eta_{\bn\eps+\frac{\omega}{2}}^{R}
+\eta_{\bn\eps-\frac{\omega}{2}}^{A}\right)^2-v_F^2(\bn\bq)^2}.
\en
In realistic experiments $v_Fq\ll\Delta$ (Sec.~II), and in this limit Eq.~\eqref{varphi1plus} simplifies to
\beg\label{varphi1plusApp}
\begin{aligned}
&\varphi_{1+}(\bq,\omega) \approx (\bq{\mathbf E})\left(\frac{ie\zeta_\omega v_F^2}{4\omega}\right)\Pi_\mu(\mathbf{q},\omega),
\end{aligned}
\en
where
\beg\label{zetaom}
\zeta_\w = \frac{1}{2}\left\langle\int\limits_{-\infty}^{\infty}d\eps
\left[\frac{\left(g_{\bn\eps+\frac{\omega}{2}}^R-g_{\bn\eps-\frac{\omega}{2}}^R\right)t_{\eps-\frac{\omega}{2}}}{\left(\eta_{\bn\eps+\frac{\omega}{2}}^{R}
+\eta_{\bn\eps-\frac{\omega}{2}}^{R}\right)^2}-\frac{\left(g_{\bn\eps+\frac{\omega}{2}}^A-g_{\bn\eps-\frac{\omega}{2}}^A\right)t_{\eps+\frac{\omega}{2}}}{\left(\eta_{\bn\eps+\frac{\omega}{2}}^{A}
+\eta_{\bn\eps-\frac{\omega}{2}}^{A}\right)^2}+\frac{\left(g_{\bn\eps+\frac{\omega}{2}}^R-g_{\bn\eps-\frac{\omega}{2}}^A\right)(t_{\eps+\frac{\omega}{2}}-t_{\eps-\frac{\omega}{2}})}{\left(\eta_{\bn\eps+\frac{\omega}{2}}^{R}
+\eta_{\bn\eps-\frac{\omega}{2}}^{A}\right)^2}\right]\right\rangle_\bn.
\en
This is Eq.~\eqref{phi1pmMain} of the main text.

\subsection*{Linear response of the order parameter}
We now derive the drive of the order parameter amplitude, Eqs. \eqref{eq:dDelta1}--\eqref{eq:Akernel}. Since the drive is of first order in the particle-hole asymmetry, it is convenient to bypass the gradient expansion altogether. The response to a single Fourier component of the field is given exactly by the first-order solution of the Dyson equation, and the terms of order $\eps/\veps_F$ are the terms odd in $\xi_\bp$ under the energy integral. We use the standard Nambu spinor $\Psi_\bp=(c_{\bp\uparrow},c^\dagger_{-\bp\downarrow})^T$ and the Matsubara technique, so that
\begin{equation}
G_0(\bp,z)=\frac{z+\xi_\bp\tau_3+\Delta_\bp\tau_1}{z^2-\xi_\bp^2-\Delta_\bp^2},\qquad z=i\eps_n,\qquad \Delta_\bp=\Delta{\cal Y}(\phi_\bp)
\end{equation}
[in this representation $\hat{\cal G}=\tau_3G_0$, the amplitude vertex $i\hat\tau_2$ of the main text becomes $\tau_1$ and $\hat\tau_3$ becomes $\tau_0$]. A perturbation $\sum_\bp\Psi^\dagger_{\bp+\bq/2}V(\bp)\Psi_{\bp-\bq/2}e^{-i\omega t}$ with
\begin{equation}
V(\bp)=-\frac{e}{m}(\bp{\mathbf A})\tau_0+\varphi_1\tau_3+\delta\Delta'_1{\cal Y}_\bp\tau_1+\delta\Delta''_1{\cal Y}_\bp\tau_2,\qquad {\mathbf A}=\frac{{\mathbf E}}{i\omega},
\end{equation}
produces exactly $G_1(\bp;\bq,i\Omega)=G_0(\bp_+,z_+)V(\bp)G_0(\bp_-,z_-)$ with $\bp_\pm=\bp\pm\bq/2$, $z_\pm=z\pm i\Omega/2$. We write $G_0(\bp_\pm,z_\pm)=N_\pm/D_\pm$ with $N_\pm=z_\pm+\xi_\pm\tau_3+\Delta_\pm\tau_1$ and $D_\pm=z_\pm^2-\xi_\pm^2-\Delta_\pm^2$, where $\xi_\pm=\xi_{\bp_\pm}$ and $\Delta_\pm=\Delta_{\bp_\pm}$. The vertex $\bp/m$ and the form factor ${\cal Y}_\bp$ are evaluated at the average momentum. In this representation the gap equation reads $\Delta=-\frac{g}{2}\sum_\bp{\cal Y}_\bp T\sum_n{\rm Tr}[\tau_1G_0(\bp,z)]$, i.e. $1=g\nu_F\langle{\cal Y}_\bn^2\int d\xi\,\tanh(E/2T)/2E\rangle_\bn$, and the amplitude self-consistency reads $\delta\Delta'_1=-\frac{g}{2}\sum_\bp{\cal Y}_\bp T\sum_n{\rm Tr}[\tau_1G_1]$. Separating the term proportional to $\delta\Delta'_1$ itself, $\delta\Delta'_1=gP_{1A}/(1-gP_{11})$, with the amplitude--amplitude and amplitude--current bubbles
\begin{equation}
\begin{aligned}
P_{11}(i\Omega)&=-\frac12\sum_\bp{\cal Y}_\bp^2\,T\sum_n{\rm Tr}[\tau_1G_0(\bp,z_+)\tau_1G_0(\bp,z_-)],\\
P_{1A}(\bq,i\Omega)&=-\frac12\sum_\bp{\cal Y}_\bp\,T\sum_n{\rm Tr}[\tau_1G_0(\bp_+,z_+)V_AG_0(\bp_-,z_-)],
\end{aligned}
\end{equation}
where $V_A=-(e/m)(\bp{\mathbf A})\tau_0$ is the paramagnetic part of $V(\bp)$. At $\bq=0$, ${\rm Tr}[\tau_1N_+\tau_1N_-]=2(z_+z_--\xi^2+\Delta_\bp^2)$. Combining $P_{11}$ with the gap equation, written in the symmetric form $1=\frac{g}{2}\sum_\bp{\cal Y}_\bp^2T\sum_n[(\eps_+^2+E^2)^{-1}+(\eps_-^2+E^2)^{-1}]$, and using $T\sum_n[(\eps_+^2+E^2)(\eps_-^2+E^2)]^{-1}=\tanh(E/2T)/[E(\Omega^2+4E^2)]$ for $\Omega\neq0$, we obtain
\begin{equation}
1-gP_{11}(i\Omega)=\frac{g}{2}\sum_\bp{\cal Y}_\bp^2\left(\Omega^2+4\Delta_\bp^2\right)T\sum_n\frac{1}{(\eps_+^2+E^2)(\eps_-^2+E^2)}=g\nu_F\,\chi_{\mathrm{SH}}^{-1}(i\Omega),
\end{equation}
whose continuation $i\Omega\to\omega+i0$ is Eq. \eqref{eq:chiSH}. The coupling constant and the cutoff drop out, and $\delta\Delta'_1=\chi_{\mathrm{SH}}P_{1A}/\nu_F$. The traces with the current, imbalance and phase vertices are
\begin{equation}
\begin{aligned}
&{\rm Tr}[\tau_1N_+\tau_0N_-]=2(z_+\Delta_-+z_-\Delta_+),\qquad
{\rm Tr}[\tau_1N_+\tau_3N_-]=2(\xi_+\Delta_-+\xi_-\Delta_+),\\
&{\rm Tr}[\tau_1N_+\tau_2N_-]=-2iz(\xi_+-\xi_-)-\Omega(\xi_++\xi_-),
\end{aligned}
\label{eq:traces}
\end{equation}
so that
\begin{equation}
P_{1A}(\bq,i\Omega)=\frac{e}{2m}\,\nu_F\int d\xi\int\frac{d\phi}{2\pi}\,{\cal Y}_\bp(\bp{\mathbf A})\,T\sum_n\frac{2(z_+\Delta_-+z_-\Delta_+)}{D_+D_-}.
\label{eq:PiA}
\end{equation}
All sources of particle-hole asymmetry are now explicit in
\begin{equation}
\xi_\pm=\xi\pm a,\quad a=\tfrac12v(\xi)(\bn\bq),\qquad
\Delta_\pm=\Delta[{\cal Y}\pm b],\quad b=\frac{(\bq\hat\phi_\bn)}{2p(\xi)}{\cal Y}'(\phi),\qquad
\frac{\bp{\mathbf A}}{m}=v(\xi)(\bn{\mathbf A}),
\end{equation}
with $p(\xi)=p_F\sqrt{1+\xi/\veps_F}$ and $v(\xi)=v_F\sqrt{1+\xi/\veps_F}$. The density of states $\nu_F=m/2\pi$ of a two-dimensional parabolic band is energy independent. The term $b$ is the angular part of the momentum gradient in \eqref{NablapGpeps} and is of order $v_Fq/\veps_F$, since $1/p_F=v_F/2\veps_F$. Expanding \eqref{eq:PiA} to first order in $\bq$, using $D^0_--D^0_+=-2i\Omega z$ for $D^0_\pm=z_\pm^2-\xi^2-\Delta_\bn^2$, the integrand splits into three terms,
\begin{equation}
T_1=\frac{-8i\Omega z^2\Delta{\cal Y}^2(\bn{\mathbf A})(\bn\bq)\,\xi v(\xi)^2}{(D^0_+)^2(D^0_-)^2},\qquad
T_2=\frac{-8i\Omega z^2\Delta^3{\cal Y}^3{\cal Y}'(\bq\hat\phi_\bn)(\bn{\mathbf A})}{m(D^0_+)^2(D^0_-)^2},\qquad
T_3=\frac{-i\Omega\Delta{\cal Y}{\cal Y}'(\bq\hat\phi_\bn)(\bn{\mathbf A})}{mD^0_+D^0_-},
\end{equation}
of which $T_2$ and $T_3$ contain the angular vertex. Both carry the angular factor ${\cal Y}{\cal Y}'(\bq\hat\phi_\bn)(\bn{\mathbf A})$ times an even function of ${\cal Y}^2$, and since ${\cal Y}{\cal Y}'=-2\sin4\phi$ and $(\bq\hat\phi_\bn)(\bn{\mathbf A})=\frac12(q_yA_x-q_xA_y)+\frac12(q_yA_x+q_xA_y)\cos2\phi-\frac12(q_xA_x-q_yA_y)\sin2\phi$, they vanish identically upon averaging over the Fermi surface. The angular vertex therefore does not contribute to the order parameter response at this order. In $T_1$ the part odd in $\xi$ integrates to zero and $\xi v(\xi)^2\to\xi^2v_F^2/\veps_F$. Using $\langle n_in_j{\cal Y}^2R({\cal Y}^2)\rangle_\bn=\frac12\delta_{ij}\langle{\cal Y}^2R\rangle_\bn$ and $\int d\xi\,\xi^2/[(\xi^2+E_+^2)^2(\xi^2+E_-^2)^2]=\pi/[2E_+E_-(E_++E_-)^3]$ one finds
\begin{equation}
P_{1A}(\bq,i\Omega)=2ie\nu_F\frac{v_F^2}{\veps_F}\,\Delta\,(\bq{\mathbf A})\,T\sum_n\Omega\,\eps_n^2\left\langle{\cal Y}_\bn^2\frac{\pi}{2E_+E_-(E_++E_-)^3}\right\rangle_\bn,\qquad E_\pm=\sqrt{\left(\eps_n\pm\tfrac{\Omega}{2}\right)^2+\Delta_\bn^2}.
\label{eq:drive}
\end{equation}
The drive is manifestly $A_{1g}$, in agreement with \eqref{eq:symm}, proportional to $1/\veps_F$, and proportional to $\Omega(\bq{\mathbf A})\to-(\bq{\mathbf E})$, so that at $T=0$ it contains no static piece. At $\Omega=0$ the trace \eqref{eq:traces} for the current vertex is odd in $\eps_n$ for any $\bq$ and any dispersion, so that the static current--amplitude correlator vanishes identically at any temperature. The feedback of $\varphi_1$ and of the phase $\delta\Delta''_1$ into the amplitude vanishes at this order. The $\tau_3$ trace in \eqref{eq:traces} equals $4\Delta({\cal Y}\xi-ab)$, and the first term of the $\tau_2$ trace is odd in $\bn$. The remaining terms, $4\Delta{\cal Y}\xi$ and $-\Omega(\xi_++\xi_-)=-2\Omega\xi$, give contributions that are odd in $\xi$ or proportional to $(\bn\bq)$, and $ab$ is of second order in $q$.

The analytic continuation $i\Omega\to\omega+i0$ is performed with the two-cut formula
\begin{equation}
T\sum_nS\left(i\eps_n+\tfrac{i\Omega}{2},i\eps_n-\tfrac{i\Omega}{2}\right)\to\frac{1}{4\pi i}\int\limits_{-\infty}^{\infty}d\eps\left\{t_{\eps-\frac{\omega}{2}}S^{RR}-t_{\eps+\frac{\omega}{2}}S^{AA}+\left(t_{\eps+\frac{\omega}{2}}-t_{\eps-\frac{\omega}{2}}\right)S^{RA}\right\},
\end{equation}
with $S^{BB'}=S(\eps_++i0_B,\eps_-+i0_{B'})$, which reproduces the three-channel structure and the thermal factors of \eqref{Kernel}. On the real axis $\sqrt{\Delta_\bn^2-(\eps\pm i0)^2}=-i\eta^{R/A}_{\bn\eps}$ with $\eta^{R(A)}$ of \eqref{etaRA}, and \eqref{eq:drive} becomes $P_{1A}(\bq,\omega)=2ie\nu_F(v_F^2/\veps_F)(\bq{\mathbf E}){\cal A}(\omega)$ with ${\cal A}(\omega)$ of \eqref{eq:Akernel}, which together with $\delta\Delta'_1=\chi_{\mathrm{SH}}P_{1A}/\nu_F$ yields \eqref{eq:dDelta1}. Alternatively, the frequency sum in \eqref{eq:drive} can be done before the energy integral. By residues, $T\sum_n\eps_n^2[(\eps_+^2+E^2)(\eps_-^2+E^2)]^{-2}=\tanh(E/2T)/[32E^3(E^2+\Omega^2/4)]+n_F'(E)/(4E^2\Omega^2)$, and the continuation gives
\begin{equation}
{\cal A}(\omega)=-\frac{\Delta}{8}\left\langle{\cal Y}_\bn^2\int d\xi\,\frac{\xi^2\tanh(E/2T)}{E^3[4E^2-(\omega+i0)^2]}\right\rangle_\bn-\frac{\Delta\,N_T}{4\omega^2},
\end{equation}
identical to \eqref{eq:Akernel} and, by the identity \eqref{AuxRelation}, to \eqref{eq:Aidentity}. We have checked the equivalence of the two representations numerically to $10^{-5}$ and used the latter, which is free of large cancellations, for the numbers in the text.

To verify that no term has been lost we have also evaluated the full two-dimensional bubble \eqref{eq:PiA} numerically without any expansion, with the exact $\xi_{\bp\pm\bq/2}$, ${\cal Y}(\phi_{\bp\pm\bq/2})$ and $\bp{\mathbf A}/m$, for $m=\veps_F=e=1$, $\Delta=0.05$, $T=0.01$, $\Omega=8\pi T$ and 800 Matsubara frequencies, and extracted the part odd in $\bq$.
\begin{center}
\begin{tabular}{llll}
\hline
pairing & geometry & exact $O(q)$ bubble & Eq.~\eqref{eq:drive}\\
\hline
$d$ & $\bq\parallel{\mathbf A}\parallel x$, $q=0.01$ & $2.097\times10^{-4}\,i$ & $2.106\times10^{-4}\,i$\\
$d$ & $\bq\parallel{\mathbf A}\parallel x$, $q=0.005$ & $1.050\times10^{-4}\,i$ & $1.053\times10^{-4}\,i$\\
$d$ & $\bq\parallel{\mathbf A}\parallel[110]$, $q=0.01$ & $2.097\times10^{-4}\,i$ & $2.106\times10^{-4}\,i$\\
$d$ & $\bq\parallel[110]$, ${\mathbf A}\parallel[\bar110]$ & $<10^{-19}$ & $0$\\
$d$ & $\bq\parallel x$, ${\mathbf A}\parallel y$ & $<10^{-19}$ & $0$\\
$s$ & $\bq\parallel{\mathbf A}\parallel x$, $q=0.01$ & $2.431\times10^{-4}\,i$ & $2.441\times10^{-4}\,i$\\
$s$ & $\bq\parallel[110]$, ${\mathbf A}\parallel[\bar110]$ & $<10^{-19}$ & $0$\\
\hline
\end{tabular}
\end{center}
The residual difference ($0.4\%$ at $q=0.01$, $0.26\%$ at $q=0.005$) consists of the $O(q^3)$ terms and of the discretization error of the two-dimensional integral. The drive vanishes for ${\mathbf A}\perp\bq$ and is the same for $\bq\parallel[100]$ and $\bq\parallel[110]$, as required by \eqref{eq:symm}.


\section{Source terms defining the nonlinear response}\label{AppendixSource}
The source function on the right hand side of Eq.~\eqref{MainEq4g2} is the sum
\beg\label{SumGamma}
\check{\Gamma}_{\bn\eps}(q)=\check {S}_{\bn\eps}^{(\Delta)}+\check S_{\bn\eps}^{(A)}+\check S_{\bn\eps}^{(\Theta)}+\check{S}_{\bn\eps}^{(E)}
+\check S_{\bn\eps}^{(\mu)}+\check S_{\bn\eps}^{(\Lambda)}+\check S_{\bn\eps}^{(\rm imp)},
\en
where $\bar s=-s$, $\eps_\pm=\eps\pm\omega/2$, ${\mathbf A}_\pm$ are the amplitudes of the two terms in Eq.~\eqref{VectorPotential}, ${\mathbf A}_+={\mathbf E}/i\omega$ and ${\mathbf A}_-={\mathbf A}_+^*$, and
\beg\label{Sources}
\begin{aligned}
&\check{S}_{\bn\eps}^{(E)}=\left(\frac{ev_F}{i\omega}\right)\left[(\bn{\mathbf E})\left(\check{g}_{1-}(\bn\eps_{+};q)\check{\tau}_3-\check{\tau}_3\check{g}_{1-}(\bn\eps_{-};q)\right)
-(\bn{\mathbf E}^*)\left(\check{g}_{1+}(\bn\eps_{-};q)\check{\tau}_3-\check{\tau}_3\check{g}_{1+}(\bn\eps_{+};q)\right)\right], \\
&\check {S}_{\bn\eps}^{(\Delta)}=\sum_{s=\pm}
\Big[\check g_{1\bar s}\big(\epsilon+\tfrac{s\omega}{2}\big)\,\delta\hat\Delta_{1s}-\delta\hat\Delta_{1s}\,\check g_{1\bar s}\big(\epsilon-\tfrac{s\omega}{2}\big)
\Big],\\
&\check S_{\bn\eps}^{(\Theta)}=-\sum_{s=\pm}\frac{se v_F^{2}}{2\varepsilon_F}\,
(\mathbf{nq})(\mathbf n{\mathbf A}_{s})\;
\langle{\{\check\tau_3,\check g_{1\bar s}\}}_{+}\rangle_s,
\\&\check S_{\bn\eps}^{(A)}=+\sum_{s=\pm}\frac{se}{2m}\,
\Big\langle\Big\{\check\tau_3,\big[(\mathbf q{\mathbf A}_{s})+(\bq\hat\phi_\bn)(\bn{\mathbf A}_s)\,\partial_{\phi_\bn}\big]\check g_{1\bar s}\Big\}_{+}\Big\rangle_s.
\end{aligned}
\en
Here we used the shorthand notation 
\beg\label{BarNotation}
\langle X_{\mathbf n\epsilon}\rangle_s \equiv
\tfrac12\Big(X_{\mathbf n\epsilon-\frac{\omega}{2}}+X_{\mathbf n\epsilon+\frac{\omega}{2}}\Big).
\en
The second term of $\check S^{(A)}$ originates from the angular part of the momentum gradient, the last term of Eq.~\eqref{EilenMain}. The derivative $\partial_{\phi_\bn}$ acts on the explicit dependence of $\check g_{1\bar s}$ on $\bn$ and on its dependence on $\Delta_\bn$. The second-order part of the impurity scattering term gives the source
\beg\label{Simp}
\check S_{\bn\eps}^{(\rm imp)}=-\frac{i}{2\tau}\sum_{s=\pm}\Big[\check M_s(\eps_{\bar s})\,\check g^{(E)}_{1\bar s}(\bn\eps_{\bar s})-\check g^{(E)}_{1\bar s}(\bn\eps_{s})\,\check M_s(\eps_{s})\Big],\qquad \check M_s(\eps)=\langle\check g^{(\Delta)}_{1s}(\bn\eps)\rangle_\bn,
\en
where $\check g^{(E)}_1$ and $\check g^{(\Delta)}_1$ are the parts of the linear solution induced by the field and by $\delta\hat\Delta_1$. The impurity averages of $\check g^{(E)}_1$ at $\bq\to0$ and of the current-carrying part of $\check g_2$ vanish.
The remaining two source terms contain the imbalance potential $\varphi_{1\pm}$,
\beg\label{TwoMore}
\check S_{\bn\eps}^{(\mu)}=-\sum_{s=\pm}\varphi_{1s}
\Big[\check g_{1\bar s}\big(\epsilon-\tfrac{s\omega}{2}\big)
-\check g_{1\bar s}\big(\epsilon+\tfrac{s\omega}{2}\big)\Big], \quad 
\check S_{\bn\eps}^{(\Lambda)}=-\sum_{s=\pm}\,s\,\frac{v_F(\mathbf{nq})}{2\varepsilon_F}\,
\varphi_{1s}\,\langle\check g_{1\bar s}\rangle_s.
\en
Both are proportional to $(\bq{\mathbf E})$ and vanish when the polarization is transverse to $\bq$. The source $\check S^{(\mu)}$ is of the same order as $\check S^{(\Delta)}$, while $\check S^{(\Lambda)}$ contains the additional factor $(\bn\bq)$ and is of second order in $q$ (Secs.~III C and IV A). The source $\check S^{(\Delta)}$ connects the Schmid--Higgs resonance with the dc response.


\section{Nonlinear response in the particle-hole symmetric case}\label{AppendixB}
In this appendix we show that in the particle-hole symmetric approximation and for $\varphi_{1\pm}=0$ the dc current vanishes at first order in $\bq$, for $s$- and $d$-wave pairing alike. The physical field of the wave is
\begin{equation}
{\mathbf E}_{\rm phys}(\br,t)={\mathbf E}\,e^{i(\bq\br-\omega t)}+{\mathbf E}^{*}e^{-i(\bq\br-\omega t)},
\label{eq:field}
\end{equation}
so that linear polarization corresponds to ${\mathbf E}=e^{i\vartheta}{\mathbf E}_0$ with a real vector ${\mathbf E}_0$, and circular polarization in the plane of the orthogonal unit vectors $\hat{\mathbf e}_1$ and $\hat{\mathbf e}_2$ to ${\mathbf E}=\tfrac{E_0}{\sqrt2}(\hat{\mathbf e}_1\pm i\hat{\mathbf e}_2)$. In a medium with inversion symmetry the dc current at second order in the field is nonlocal. To first order in $\bq$, and after the average over $\bn$, it has the general form \eqref{54},
\beg\label{jqwphenom}
{\mathbf j}_{\rm dc}(\bq)=\alpha(q)\,\bq|{\mathbf E}|^2+i\beta_{\textrm{a}}(q)\left[{\mathbf E}(\bq {\mathbf E}^*)-{\mathbf E}^*(\bq {\mathbf E})\right]+
{\beta}_{\textrm{s}}(q)\left[{\mathbf E}(\bq {\mathbf E}^*)+{\mathbf E}^*(\bq {\mathbf E})\right]+\dots,
\en
where the dots stand for the helicity-even term allowed for $d$-wave pairing by the $D_4$ symmetry of $|\Delta_\bn|$ (Sec.~IV). With the identity
\begin{equation}\label{BecomeCurl}
i\left[{\mathbf E}(\bq{\mathbf E}^*)-{\mathbf E}^*(\bq{\mathbf E})\right]=[\bq\times{\mathbf Z}],\qquad {\mathbf Z}=i[{\mathbf E}\times{\mathbf E}^*],
\end{equation}
the helicity-odd term is the uniform current $\beta_{\rm a}[\bq\times{\mathbf Z}]$, which is reversed together with the helicity and vanishes for linear polarization. The magnetization is a different quantity. It is determined by the dependence of the current on the wave vector of the envelope of a beam (Sec.~IV).

The current \eqref{j2dc} contains the average of $\bn\,{\rm Tr}[\hat\tau_3\hat g_2^K]$ over the directions of $\bn$. The terms of $\hat g_2^K$ proportional to $(\bn{\mathbf E})(\bn{\mathbf E}^*)$ do not contribute, since their product with $\bn$ averages to zero, and at first order in $\bq$ only the terms with an additional factor $(\bn\bq)$ remain. For the normalization part of the solution one finds
\beg\label{TrPiRA}
\textrm{Tr}\left\{\hat{\tau}_3\hat{\cal G}_{\bn\eps}^{R(A)}\hat{\cal N}^{R(A)}_{\bn\eps}(\bq,\omega)\right\}\propto 1-(g_{\bn\eps}^{R(A)})^2+(f_{\bn\eps}^{R(A)})^2=0,
\en
so that the terms with $\check{\cal N}$ do not contribute to Eq.~\eqref{jqwphenom}. A direct calculation shows that the retarded and advanced components of the source vanish, $\Gamma_{\bn\eps}^{R(A)}(q)=0$, and that the traces of the terms with $\Gamma_{\bn\eps}^{K}(q)$ vanish as well,
\beg\label{TracesGammaK}
\begin{aligned}
&\textrm{Tr}\left\{\hat{\tau}_3{\hat{\cal G}_{\bn\eps}^{R(A)}}\left[\hat{\cal G}_{\bn\eps}^{R(A)}-
\hat{\tau}_3\hat{\cal G}_{\bn\eps\pm\omega}^{R(A)}\hat{\tau}_3\right]\right\}=\textrm{Tr}\left\{\hat{\tau}_3{\hat{\cal G}_{\bn\eps}^{R}}\left[\hat{\cal G}_{\bn\eps}^{A}-
\hat{\tau}_3\hat{\cal G}_{\bn\eps\pm\omega}^{A}\hat{\tau}_3\right]\right\}=0.
\end{aligned}
\en
The dc current at first order in $\bq$ therefore vanishes in the particle-hole symmetric approximation for $\varphi_{1\pm}=0$, in agreement with Eq.~\eqref{jdcE}.

\subsection*{Normal-state limit}
In the normal state the dc current at second order in the field and to first order in $\bq$ can be found exactly for the model \eqref{Eq1} with elastic impurity scattering, including all particle-hole asymmetric terms. In the gauge with zero scalar potential we describe the electrons by the distribution function $\Phi(\bn,\eps;\br,t)$ at fixed direction $\bn$ of the momentum and fixed energy $\eps=\xi_\bp-e(\bp{\mathbf A})/m$, the energy variable of the quasiclassical theory. To first order in ${\mathbf A}$ in the Hamiltonian $(\bp-e{\mathbf A})^2/2m$ the Boltzmann equation and the current read exactly
\begin{equation}
\begin{aligned}
&\partial_t\Phi+v(\eps)(\bn\nabla_\br)\Phi+ev(\eps)(\bn{\mathbf E})\,\partial_\eps\Phi-\frac em({\mathbf A}_\perp\nabla_\br)\Phi+\frac em\big((\hat\phi_\bn\nabla_\br)(\bn{\mathbf A})\big)\partial_{\phi_\bn}\Phi=-\frac{\Phi-\langle\Phi\rangle_\bn}{\tau},\\
&{\mathbf j}=2e\nu_F\int d\eps\,\Big\langle\Big[v(\eps)\bn+\frac{2e}{m}(\bn{\mathbf A})\bn-\frac em{\mathbf A}\Big]\Phi\Big\rangle_\bn,
\end{aligned}
\label{eq:kinNS}
\end{equation}
where $v(\eps)=v_F\sqrt{1+\eps/\veps_F}$ and ${\mathbf A}_\perp={\mathbf A}-\bn(\bn{\mathbf A})$. The radial momentum derivatives cancel identically against the gradient of the energy shift, and elastic scattering conserves $\eps$. The fourth and fifth terms are the normal-state counterparts of the spatial-gradient and angular terms of Eq.~\eqref{EilenMain}. The energy dependence of $v$ and the field-dependent part of the current vertex are the particle-hole asymmetric corrections discussed in Sec.~IV. Eliminating ${\mathbf A}$ in favor of ${\mathbf E}$ and ${\mathbf B}$ one finds ${\mathbf j}=(e/m){\mathbf P}$ for the dc current, where the kinetic momentum density obeys $d{\mathbf P}/dt={\mathbf F}-{\mathbf P}/\tau$ and ${\mathbf F}=\langle\rho_1{\mathbf E}_{\rm phys}+{\mathbf j}_1\times{\mathbf B}_1\rangle$ is the dc force density of the wave. Here $\rho_1$, ${\mathbf j}_1$ and ${\mathbf B}_1$ are the charge density, the current density and the magnetic field of the wave to first order in the field, and the brackets denote the average over the period $2\pi/\omega$. With ${\mathbf j}_1=\sigma(\omega){\mathbf E}$ and the continuity equation, ${\mathbf F}=(2/\omega){\rm Re}\,\sigma(\omega)\,\bq|{\mathbf E}|^2$ for any local conductivity, whence
\begin{equation}
\alpha=\frac{2e^3n_c\tau^2}{m^2\omega(1+\omega^2\tau^2)},\qquad\beta_{\rm a}=\beta_{\rm s}=0 .
\label{eq:NSexact}
\end{equation}
We have solved Eqs.~\eqref{eq:kinNS} numerically to second order in the field and to first order in $\bq$. In units $e=m=1$, with $\veps_F=50$, $T=0.2$ and $\omega=1$, the contributions of the individual terms are
\begin{center}
\begin{tabular}{lcccc}
\hline
term & $\alpha$ & $\beta_{\rm s}$ & $\beta_{\rm a}$ ($\tau=10$) & $\beta_{\rm a}$ ($\tau=100$)\\
\hline
spatial gradient $-(e/m)({\mathbf A}_\perp\nabla_\br)\Phi$ & $7.879$ & $-7.879$ & $-157.58$ & $-1591.4$\\
angular term $(e/m)((\hat\phi_\bn\nabla_\br)(\bn{\mathbf A}))\partial_{\phi_\bn}\Phi$ & $7.879$ & $-7.879$ & $+157.58$ & $+1591.4$\\
force term acting on the $O(q)$ part of the first-order $\Phi$ & $31.20$ & $15.76$ & $-3.120$ & $-0.318$\\
field-dependent current vertex & $-15.45$ & $0$ & $+3.120$ & $+0.318$\\
\hline
total & $31.516$ & $<10^{-9}$ & $<10^{-10}$ & $<10^{-11}$\\
Eq.~\eqref{eq:NSexact} & $31.516$ & $0$ & $0$ & $0$\\
\hline
\end{tabular}
\end{center}
(the $\alpha$ and $\beta_{\rm s}$ columns are for $\tau=10$, and the same agreement is found for $\tau=3$--$100$ and $\omega=0.5$--$3$). The table shows how the exact result \eqref{eq:NSexact} is realized. The helicity-odd currents of the spatial-gradient and angular terms, each proportional to $\tau$, cancel exactly. This cancellation holds in the superconducting state as well [Eq.~\eqref{eq:jA}]. The helicity-odd currents of the force term acting on the $O(q)$ part of the first-order distribution and of the field-dependent current vertex, each proportional to $1/\tau$, cancel each other.


\section{Nonlinear response in the presence of particle-hole asymmetry}\label{AppendixC}
\subsection{Order parameter channel}
The calculation rests on two algebraic identities that hold for an arbitrary
source. First, using $\hat{\cal P}_+\hat\Gamma\hat{\cal P}_- -\hat{\cal P}_-\hat\Gamma\hat{\cal P}_+
=\tfrac12[\hat{\mathcal G},\hat\Gamma]$ for the projectors $\hat{\cal P}_\pm$ of Eq.~\eqref{Sol4g2RA} and
$[\hat\tau_3,\hat{\mathcal G}^B]=2f^B\hat\tau_1$ for $B=R,A$, the off-diagonal
(commutator) part of Eq.~\eqref{Sol4g2RA} gives
\begin{equation}
\Tr\big[\hat\tau_3\,\hat g_2^{R(A)}\big]
=\frac{f^{R(A)}_{\bn\eps}}{2\eta_{\bn\eps}^{R(A)}}\,
\Tr\big[\hat\tau_1\hat\Gamma_{\bn\eps}^{R(A)}\big].
\label{eq:id1}
\end{equation}
Second, the $\alpha\neq\beta$ part of Eq.~\eqref{Sol4dg2K} collapses to
$(\hat{\mathcal G}_{\bn\eps}^R\hat{\mathcal Q}_{\bn\eps}^K-\hat{\mathcal Q}_{\bn\eps}^K\hat{\mathcal G}_{\bn\eps}^A)/
2(\eta_{\bn\eps}^R+\eta_{\bn\eps}^A)$, so that
\begin{equation}
\Tr\big[\hat\tau_3\,\delta\hat g_2^{K}\big]
=\frac{(g_{\bn\eps}^R-g_{\bn\eps}^A)\,\Tr\hat{\mathcal Q}_{\bn\eps}^K
+(f_{\bn\eps}^R+f_{\bn\eps}^A)\,\Tr[\hat\tau_1\hat{\mathcal Q}_{\bn\eps}^K]}{2(\eta_{\bn\eps}^R+\eta_{\bn\eps}^A)},
\label{eq:id2}
\end{equation}
where $\hat{\mathcal Q}_{\bn\eps}^K$ is defined below Eq.~\eqref{Sol4dg2K}. 
Only the $\hat\tau_0$ and $\hat\tau_1$ projections of the sources contribute to the current. The normalization parts of \eqref{Sol4g2RA} and \eqref{Sol4dg2K} give in the same way
\begin{equation}
\begin{aligned}
&\Tr\big[\hat\tau_3\,\tfrac14\{\hat{\cal G}^{R(A)}_{\bn\eps},\hat{\cal N}^{R(A)}_{\bn\eps}\}\big]=\tfrac12 g^{R(A)}_{\bn\eps}\Tr\hat{\cal N}^{R(A)}_{\bn\eps},\\
&\Tr\big[\hat\tau_3\,\tfrac14(\hat{\cal G}^{R}_{\bn\eps}\delta\hat{\cal N}^{K}_{\bn\eps}+\delta\hat{\cal N}^{K}_{\bn\eps}\hat{\cal G}^{A}_{\bn\eps})\big]
=\tfrac14\big[(g^R_{\bn\eps}+g^A_{\bn\eps})\Tr\delta\hat{\cal N}^K_{\bn\eps}+(f^R_{\bn\eps}-f^A_{\bn\eps})\Tr[\hat\tau_1\delta\hat{\cal N}^K_{\bn\eps}]\big].
\end{aligned}
\label{eq:id3}
\end{equation}
With impurity scattering these identities hold, with $\eta$, $g$ and $f$ given by Eq.~\eqref{gfeta} and $\hat h_{\bn\eps}=\tilde\eps_\eps\hat\tau_3+i\tilde\Delta_{\bn\eps}\hat\tau_2=\eta_{\bn\eps}\hat{\cal G}_{\bn\eps}$, for all quantities odd in $\bn$, whose impurity averages vanish. Such are the current-carrying part of $\check g_2$ and the field-induced part $\check g^{(E)}_1$ of the linear solution at $\bq\to0$. For $s$-wave pairing $g$ and $f$ coincide with their clean-limit values and ${\rm Im}\,\eta^R_{\bn\eps}=1/2\tau$ for $|\eps|>\Delta$. For $d$-wave pairing only $\eps$ is renormalized, Eqs.~\eqref{xitilde} and \eqref{Gfixedpoint}.

\paragraph*{Linear solutions.} At $\bq\to0$ the linear solution is the sum $\check g_1=\check g_1^{(E)}+\check g_1^{(\Delta)}$ of the responses to the Doppler vertex $a_s\hat\tau_3$, $a_s=ev_F(\bn{\mathbf A}_s)$, and to $\delta\hat\Delta_{1s}=i\delta\Delta_{1s}{\cal Y}_\bn\hat\tau_2$. Both solve $\hat h_L\hat X-\hat X\hat h_R=\hat{\cal G}_L\hat V-\hat V\hat{\cal G}_R$, where $L$ and $R$ refer to the energies $\eps\pm s\omega/2$. The retarded and advanced components follow from Eq.~\eqref{Solution4g1p}, and the Keldysh components from $\hat g_1^K=\hat g_1^Rt_{\eps_R}-t_{\eps_L}\hat g_1^A+(t_{\eps_L}-t_{\eps_R})\hat g_1^{RA}$, where $\hat g_1^{RA}$ is the solution with the retarded functions at $\eps_L$ and the advanced ones at $\eps_R$. For $\check g_1^{(E)}$, $\hat V=a_s\hat\tau_3$. The response to $\delta\hat\Delta_1$ is even in $\bn$ and carries the vertex correction $(i/2\tau)\check M_s$, so that $\hat V=\delta\hat\Delta_{1s}+(i/2\tau)\check M_s$, with $\check M_s=\langle\check g^{(\Delta)}_{1s}\rangle_\bn$ determined self-consistently. For $s$-wave pairing $\check g_1^{(\Delta)}$ coincides with its value in the clean limit (the Anderson theorem for a uniform variation of the gap). For $d$-wave pairing $\langle{\cal Y}_\bn\rangle_\bn=0$ and only the $\hat\tau_3$ component of $\check M_s$ survives.

\paragraph*{The four contributions.} The dc current linear in $\delta\Delta_1$ is
\begin{equation}
\int d\eps\,\Tr\big[\hat\tau_3\hat g_2^K\big]=c_\bn(\omega)\,a_-\delta\Delta_{1+}+{\rm c.c.},
\end{equation}
with contributions from $\check S^{(\Delta)}[\check g_1^{(E)}]$, $\check S^{(E)}[\check g_1^{(\Delta)}]$, the cross terms $\check{\cal N}=-\sum_s(\check g^{(E)}_{1s}\check g^{(\Delta)}_{1\bar s}+\check g^{(\Delta)}_{1s}\check g^{(E)}_{1\bar s})$ and the impurity source \eqref{Simp}, evaluated with \eqref{eq:id1}--\eqref{eq:id3}. With $\langle n_ln_jF\rangle_\bn=\delta_{lj}\langle F\rangle_\bn/2$ and the normalization \eqref{j2dc} this gives Eq.~\eqref{jdcDelta} with ${\cal K}_H=\langle c_\bn\rangle_\bn/2$. The second-order problem is overdetermined by Eqs.~\eqref{MainEq4g2} and \eqref{Norm4G2dc}, which provides a stringent check. Without $\check S^{(\rm imp)}$ the equation and the normalization condition give incompatible diagonal ($\alpha=\beta$) blocks of $\delta\hat g_2^K$, while with it they agree to $10^{-9}$.

\paragraph*{Gauge invariance and the clean limit.} At $\bq\to0$ the Doppler vertex is removed by $\check g(t,t')=e^{i\Theta_\bn(t)}\tilde{\check g}(t,t')e^{-i\Theta_\bn(t')}$ with $\dot\Theta_\bn=ev_F\bn{\mathbf A}(t)$. The current depends on the equal-time function and is unchanged. The impurity term $-[\check\sigma_{\rm dis}\stackrel{\circ},\check g]_-$ is not invariant, since the impurity average mixes trajectories with different $\Theta_\bn$. To first order in the field it produces a perturbation of order $a_s/\tau$. The current linear in $\delta\Delta_1$ is therefore the second-order response to $\delta\hat\Delta_1$ and to this perturbation. The factor $1/\tau$ is compensated by the on-shell factor $1/(\eta^R_{\bn\eps}+\eta^A_{\bn\eps})$, which equals $-i\tau$ for $|\eps|>\Delta$ in the $s$-wave case, so that $c_\bn$ is finite and independent of $\tau$ for $1/\tau\to0$. In the original gauge the same on-shell factor appears in each of the four contributions without the compensating $1/\tau$. These large terms cancel in the sum.

\paragraph*{Results.} We have solved the second-order equations numerically, with all four sources, for $1/\tau=0.01$--$0.1\,\Delta$ and an additional regularization $10^{-6}\Delta$ of the retarded and advanced functions. For $s$-wave pairing, at $\omega=2.5\Delta$ and $T=0.04\Delta$, $\Delta\,c_\bn=0.150-26.817i$, $0.288-26.838i$ and $0.707-26.876i$ for $1/\tau=0.01$, $0.02$ and $0.05\,\Delta$. The imaginary part does not depend on $\tau$, and the real part is proportional to $1/\tau$. The imaginary part is reproduced within $0.1\%$ by Eq.~\eqref{eq:KH} for $\omega$ up to $4.5\Delta$ and $T=0.04$--$0.4\,\Delta$, and it vanishes below $2\Delta$. We have not attempted an analytical derivation of this closed form. For $d$-wave pairing (${\cal Y}_\bn=\sqrt2\cos2\phi_\bn$) the kernel is real up to corrections of order $1/\tau$. At $\omega=2.5\Delta$, with an energy cutoff $10\Delta$, $\Delta{\cal K}_H=3.123+0.056i$, $3.173+0.029i$ and $3.191+0.015i$ for $1/\tau=0.1$, $0.05$ and $0.025\,\Delta$. The values of $\Delta\,{\rm Re}\,{\cal K}_H$ for $1/\tau=0.05\Delta$ and $T=0.04\Delta$, extrapolated to infinite cutoff (the correction is at most $0.2$, largest at $5\Delta$), are
\begin{center}
\begin{tabular}{lcccccccccccc}
\hline
$\omega/\Delta$ & 0.3 & 0.7 & 1.0 & 1.4 & 1.8 & 2.2 & 2.6 & 2.83 & 3.0 & 3.5 & 4.2 & 5.0\\
$\Delta\,{\rm Re}\,{\cal K}_H$ & 0.52 & 1.13 & 1.68 & 2.84 & 3.44 & 3.30 & 3.02 & 2.82 & 2.65 & 2.18 & 1.75 & 1.32\\
\hline
\end{tabular}
\end{center}
(Fig.~\ref{Fig-K01w} contains ten more points). In the clean limit they are larger by about $1\%$. In the limit $1/\tau\to0$ at fixed phenomenological damping (see below) the solver reproduces the damping-model kernel, which we have checked independently against the Gor'kov three-point function. The $d$-wave result was also obtained with both lobes of the gap computed explicitly (agreement to $10^{-6}$), and the general solver with vertex corrections reproduces a simplified $s$-wave solver.

\paragraph*{Phenomenological damping.} If the scattering term is replaced by a damping term $i\gamma\hat\tau_3$ in $\hat h^{R(A)}$, with the Keldysh component $2i\gamma t_\eps\hat\tau_3$, the same steps lead to a kernel that again does not depend on $\gamma$,
\begin{equation}
{\cal K}^{(\gamma)}_H(\omega)=\frac{4\Delta}{\omega^2}\left\langle{\cal Y}_\bn^2\int\limits_0^\infty du\,
\Big[2t_{\eps}-t_{\eps+\omega}-t_{\eps-\omega}\Big]_{\eps=|\Delta_\bn|\cosh u}\right\rangle_\bn .
\label{eq:cn}
\end{equation}
In this model the field couples to the Keldysh component of the damping term, i.e. to a bath which supplies final states at all energies, and the threshold is at $\omega=|\Delta_\bn|$ rather than $2\Delta$. For $s$-wave pairing Eq.~\eqref{eq:cn} is therefore an artifact of the damping model. For $d$-wave pairing, where the nodal quasiparticles provide final states at all energies, ${\cal K}^{(\gamma)}_H$ has the same shape as the Born result but is smaller by a factor $1.4$--$2.1$.

\subsection{Diamagnetic channel}
The streaming part of $\check S^{(A)}$ alone gives a current of the form \eqref{54} with a helicity-odd coefficient proportional to the quasiparticle lifetime. In the damping model it scales as $1/\gamma$, and in the normal state it reduces to $\beta_{\rm a}=-e^3n_c\tau^3/[m^2(1+\omega^2\tau^2)]$, in agreement with the solution of Eqs.~\eqref{eq:kinNS}. The angular part cancels it exactly, Eqs.~\eqref{eq:jA} and \eqref{eq:betaA}. We have verified the cancellation with the second-order solver used for the order parameter channel. For $s$-wave pairing at $\omega=2.5\Delta$ the helicity-odd coefficient of the streaming part alone grows tenfold when the damping is reduced from $10^{-3}\Delta$ to $10^{-4}\Delta$, while its sum with the angular part vanishes to a relative accuracy better than $10^{-13}$. The same holds for $d$-wave pairing at $\omega=1.5\Delta$, including the term with $\partial_\phi\Delta_\bn$, and for Born scattering. For $1/\tau=0.02\Delta$ and $0.04\Delta$ the streaming part alone is proportional to $\tau$, and the sum vanishes to a relative accuracy better than $10^{-13}$ for both pairing symmetries, with the self-consistent ground state \eqref{Gfixedpoint} in the $d$-wave case.
\end{widetext}

\end{appendix}

\bibliography{pgedwave}

\begin{thebibliography}{46}%
\makeatletter
\providecommand \@ifxundefined [1]{%
 \@ifx{#1\undefined}
}%
\providecommand \@ifnum [1]{%
 \ifnum #1\expandafter \@firstoftwo
 \else \expandafter \@secondoftwo
 \fi
}%
\providecommand \@ifx [1]{%
 \ifx #1\expandafter \@firstoftwo
 \else \expandafter \@secondoftwo
 \fi
}%
\providecommand \natexlab [1]{#1}%
\providecommand \enquote  [1]{``#1''}%
\providecommand \bibnamefont  [1]{#1}%
\providecommand \bibfnamefont [1]{#1}%
\providecommand \citenamefont [1]{#1}%
\providecommand \href@noop [0]{\@secondoftwo}%
\providecommand \href [0]{\begingroup \@sanitize@url \@href}%
\providecommand \@href[1]{\@@startlink{#1}\@@href}%
\providecommand \@@href[1]{\endgroup#1\@@endlink}%
\providecommand \@sanitize@url [0]{\catcode `\\12\catcode `\$12\catcode
  `\&12\catcode `\#12\catcode `\^12\catcode `\_12\catcode `\%12\relax}%
\providecommand \@@startlink[1]{}%
\providecommand \@@endlink[0]{}%
\providecommand \url  [0]{\begingroup\@sanitize@url \@url }%
\providecommand \@url [1]{\endgroup\@href {#1}{\urlprefix }}%
\providecommand \urlprefix  [0]{URL }%
\providecommand \Eprint [0]{\href }%
\providecommand \doibase [0]{https://doi.org/}%
\providecommand \selectlanguage [0]{\@gobble}%
\providecommand \bibinfo  [0]{\@secondoftwo}%
\providecommand \bibfield  [0]{\@secondoftwo}%
\providecommand \translation [1]{[#1]}%
\providecommand \BibitemOpen [0]{}%
\providecommand \bibitemStop [0]{}%
\providecommand \bibitemNoStop [0]{.\EOS\space}%
\providecommand \EOS [0]{\spacefactor3000\relax}%
\providecommand \BibitemShut  [1]{\csname bibitem#1\endcsname}%
\let\auto@bib@innerbib\@empty
\bibitem [{\citenamefont {Artemenko}\ and\ \citenamefont
  {Volkov}(1974)}]{Artemenko1974}%
  \BibitemOpen
  \bibfield  {author} {\bibinfo {author} {\bibfnamefont {S.~N.}\ \bibnamefont
  {Artemenko}}\ and\ \bibinfo {author} {\bibfnamefont {A.~F.}\ \bibnamefont
  {Volkov}},\ }\bibfield  {title} {\bibinfo {title} {Thermoelectric power in
  superconductors},\ }\href@noop {} {\bibfield  {journal} {\bibinfo  {journal}
  {Sov. Phys. - JETP Lett.}\ }\textbf {\bibinfo {volume} {21}},\ \bibinfo
  {pages} {313} (\bibinfo {year} {1974})}\BibitemShut {NoStop}%
\bibitem [{\citenamefont {Aronov}(1976)}]{Aronov1976}%
  \BibitemOpen
  \bibfield  {author} {\bibinfo {author} {\bibfnamefont {A.~G.}\ \bibnamefont
  {Aronov}},\ }\bibfield  {title} {\bibinfo {title} {Photoelectric and
  acousto-electric fields in superconductors},\ }\href@noop {} {\bibfield
  {journal} {\bibinfo  {journal} {Sov. Phys. - JETP}\ }\textbf {\bibinfo
  {volume} {43}},\ \bibinfo {pages} {770} (\bibinfo {year} {1976})}\BibitemShut
  {NoStop}%
\bibitem [{\citenamefont {Artemenko}\ and\ \citenamefont
  {Volkov}(1979)}]{Artemenko1979}%
  \BibitemOpen
  \bibfield  {author} {\bibinfo {author} {\bibfnamefont {S.~N.}\ \bibnamefont
  {Artemenko}}\ and\ \bibinfo {author} {\bibfnamefont {A.~F.}\ \bibnamefont
  {Volkov}},\ }\bibfield  {title} {\bibinfo {title} {Electric fields and
  collective excitations in superconductors},\ }\href@noop {} {\bibfield
  {journal} {\bibinfo  {journal} {Sov. Phys. - Uspekhi}\ }\textbf {\bibinfo
  {volume} {22}},\ \bibinfo {pages} {295} (\bibinfo {year} {1979})}\BibitemShut
  {NoStop}%
\bibitem [{\citenamefont {Zaitsev}(1986)}]{Zaitsev1986}%
  \BibitemOpen
  \bibfield  {author} {\bibinfo {author} {\bibfnamefont {A.~V.}\ \bibnamefont
  {Zaitsev}},\ }\bibfield  {title} {\bibinfo {title} {Photoelectric effect in
  superconductors},\ }\href@noop {} {\bibfield  {journal} {\bibinfo  {journal}
  {Sov. Phys. - JETP}\ }\textbf {\bibinfo {volume} {63}},\ \bibinfo {pages}
  {579} (\bibinfo {year} {1986})}\BibitemShut {NoStop}%
\bibitem [{\citenamefont {Zaitsev}\ and\ \citenamefont
  {Zaitsev}(1982)}]{ZaiZai1982}%
  \BibitemOpen
  \bibfield  {author} {\bibinfo {author} {\bibfnamefont {A.~V.}\ \bibnamefont
  {Zaitsev}}\ and\ \bibinfo {author} {\bibfnamefont {V.~V.}\ \bibnamefont
  {Zaitsev}},\ }\bibfield  {title} {\bibinfo {title} {Anomalous photoelectric
  effect in superconductors with paramagnetic impurities},\ }\href@noop {}
  {\bibfield  {journal} {\bibinfo  {journal} {Sov. Phys. - Tech. Phys.}\
  }\textbf {\bibinfo {volume} {27}},\ \bibinfo {pages} {773} (\bibinfo {year}
  {1982})}\BibitemShut {NoStop}%
\bibitem [{\citenamefont {Mineev}\ and\ \citenamefont
  {Samokhin}(1999)}]{MineevSamokhin1999}%
  \BibitemOpen
  \bibfield  {author} {\bibinfo {author} {\bibfnamefont {V.~P.}\ \bibnamefont
  {Mineev}}\ and\ \bibinfo {author} {\bibfnamefont {K.~V.}\ \bibnamefont
  {Samokhin}},\ }\href@noop {} {\emph {\bibinfo {title} {Introduction to
  Unconventional Superconductivity}}}\ (\bibinfo  {publisher} {CRC Press},\
  \bibinfo {year} {1999})\BibitemShut {NoStop}%
\bibitem [{\citenamefont {Elesin}\ and\ \citenamefont
  {Kopaev}(1981)}]{Elesin1981}%
  \BibitemOpen
  \bibfield  {author} {\bibinfo {author} {\bibfnamefont {V.~F.}\ \bibnamefont
  {Elesin}}\ and\ \bibinfo {author} {\bibfnamefont {Y.~V.}\ \bibnamefont
  {Kopaev}},\ }\bibfield  {title} {\bibinfo {title} {Superconductors with
  excess quasiparticles},\ }\href@noop {} {\bibfield  {journal} {\bibinfo
  {journal} {Sov. Phys. - Uspekhi}\ }\textbf {\bibinfo {volume} {24}},\
  \bibinfo {pages} {116} (\bibinfo {year} {1981})}\BibitemShut {NoStop}%
\bibitem [{\citenamefont {Kalenkov}\ and\ \citenamefont
  {Zaikin}(2015)}]{Zaikin2015}%
  \BibitemOpen
  \bibfield  {author} {\bibinfo {author} {\bibfnamefont {M.~S.}\ \bibnamefont
  {Kalenkov}}\ and\ \bibinfo {author} {\bibfnamefont {A.~D.}\ \bibnamefont
  {Zaikin}},\ }\bibfield  {title} {\bibinfo {title} {Diffusive superconductors
  beyond the usadel approximation: Electron-hole asymmetry and large
  photoelectric effect},\ }\href {https://doi.org/10.1103/PhysRevB.92.014507}
  {\bibfield  {journal} {\bibinfo  {journal} {Phys. Rev. B}\ }\textbf {\bibinfo
  {volume} {92}},\ \bibinfo {pages} {014507} (\bibinfo {year}
  {2015})}\BibitemShut {NoStop}%
\bibitem [{\citenamefont {Galperin}\ \emph {et~al.}(1981)\citenamefont
  {Galperin}, \citenamefont {Kozub},\ and\ \citenamefont
  {Spivak}}]{Galperin1981}%
  \BibitemOpen
  \bibfield  {author} {\bibinfo {author} {\bibfnamefont {Y.~M.}\ \bibnamefont
  {Galperin}}, \bibinfo {author} {\bibfnamefont {V.~I.}\ \bibnamefont
  {Kozub}},\ and\ \bibinfo {author} {\bibfnamefont {B.~Z.}\ \bibnamefont
  {Spivak}},\ }\bibfield  {title} {\bibinfo {title} {Dissipationless bcs
  dynamics with large branch imbalance},\ }\href@noop {} {\bibfield  {journal}
  {\bibinfo  {journal} {Sov. Phys. JETP}\ }\textbf {\bibinfo {volume} {54}},\
  \bibinfo {pages} {1126} (\bibinfo {year} {1981})}\BibitemShut {NoStop}%
\bibitem [{\citenamefont {Mironov}\ \emph {et~al.}(2021)\citenamefont
  {Mironov}, \citenamefont {Mel'nikov}, \citenamefont {Tokman}, \citenamefont
  {Vadimov}, \citenamefont {Lounis},\ and\ \citenamefont
  {Buzdin}}]{Mironov2021-IFESC}%
  \BibitemOpen
  \bibfield  {author} {\bibinfo {author} {\bibfnamefont {S.~V.}\ \bibnamefont
  {Mironov}}, \bibinfo {author} {\bibfnamefont {A.~S.}\ \bibnamefont
  {Mel'nikov}}, \bibinfo {author} {\bibfnamefont {I.~D.}\ \bibnamefont
  {Tokman}}, \bibinfo {author} {\bibfnamefont {V.}~\bibnamefont {Vadimov}},
  \bibinfo {author} {\bibfnamefont {B.}~\bibnamefont {Lounis}},\ and\ \bibinfo
  {author} {\bibfnamefont {A.~I.}\ \bibnamefont {Buzdin}},\ }\bibfield  {title}
  {\bibinfo {title} {Inverse faraday effect for superconducting condensates},\
  }\href {https://doi.org/10.1103/PhysRevLett.126.137002} {\bibfield  {journal}
  {\bibinfo  {journal} {Phys. Rev. Lett.}\ }\textbf {\bibinfo {volume} {126}},\
  \bibinfo {pages} {137002} (\bibinfo {year} {2021})}\BibitemShut {NoStop}%
\bibitem [{\citenamefont {Pershan}\ \emph {et~al.}(1966)\citenamefont
  {Pershan}, \citenamefont {van~der Ziel},\ and\ \citenamefont
  {Malmstrom}}]{Pershan1966}%
  \BibitemOpen
  \bibfield  {author} {\bibinfo {author} {\bibfnamefont {P.~S.}\ \bibnamefont
  {Pershan}}, \bibinfo {author} {\bibfnamefont {J.~P.}\ \bibnamefont {van~der
  Ziel}},\ and\ \bibinfo {author} {\bibfnamefont {L.~D.}\ \bibnamefont
  {Malmstrom}},\ }\bibfield  {title} {\bibinfo {title} {Theoretical discussion
  of the inverse faraday effect, raman scattering, and related phenomena},\
  }\href {https://doi.org/10.1103/PhysRev.143.574} {\bibfield  {journal}
  {\bibinfo  {journal} {Phys. Rev.}\ }\textbf {\bibinfo {volume} {143}},\
  \bibinfo {pages} {574} (\bibinfo {year} {1966})}\BibitemShut {NoStop}%
\bibitem [{\citenamefont {Battiato}\ \emph {et~al.}(2014)\citenamefont
  {Battiato}, \citenamefont {Barbalinardo},\ and\ \citenamefont
  {Oppeneer}}]{Battiato2014}%
  \BibitemOpen
  \bibfield  {author} {\bibinfo {author} {\bibfnamefont {M.}~\bibnamefont
  {Battiato}}, \bibinfo {author} {\bibfnamefont {G.}~\bibnamefont
  {Barbalinardo}},\ and\ \bibinfo {author} {\bibfnamefont {P.~M.}\ \bibnamefont
  {Oppeneer}},\ }\bibfield  {title} {\bibinfo {title} {Quantum theory of the
  inverse faraday effect},\ }\href {https://doi.org/10.1103/PhysRevB.89.014413}
  {\bibfield  {journal} {\bibinfo  {journal} {Phys. Rev. B}\ }\textbf {\bibinfo
  {volume} {89}},\ \bibinfo {pages} {014413} (\bibinfo {year}
  {2014})}\BibitemShut {NoStop}%
\bibitem [{\citenamefont {Yang}\ \emph {et~al.}(2022)\citenamefont {Yang},
  \citenamefont {Mou}, \citenamefont {Zapata}, \citenamefont {Reynier},
  \citenamefont {Gallas},\ and\ \citenamefont {Mivelle}}]{yang2022inverse}%
  \BibitemOpen
  \bibfield  {author} {\bibinfo {author} {\bibfnamefont {X.}~\bibnamefont
  {Yang}}, \bibinfo {author} {\bibfnamefont {Y.}~\bibnamefont {Mou}}, \bibinfo
  {author} {\bibfnamefont {H.}~\bibnamefont {Zapata}}, \bibinfo {author}
  {\bibfnamefont {B.}~\bibnamefont {Reynier}}, \bibinfo {author} {\bibfnamefont
  {B.}~\bibnamefont {Gallas}},\ and\ \bibinfo {author} {\bibfnamefont
  {M.}~\bibnamefont {Mivelle}},\ }\href@noop {} {\bibinfo {title} {An inverse
  faraday effect through linear polarized light}} (\bibinfo {year} {2022}),\
  \Eprint {https://arxiv.org/abs/2206.00954} {arXiv:2206.00954
  [physics.optics]} \BibitemShut {NoStop}%
\bibitem [{\citenamefont {Mou}\ \emph {et~al.}(2023{\natexlab{a}})\citenamefont
  {Mou}, \citenamefont {Yang}, \citenamefont {Gallas},\ and\ \citenamefont
  {Mivelle}}]{mou2023reversed}%
  \BibitemOpen
  \bibfield  {author} {\bibinfo {author} {\bibfnamefont {Y.}~\bibnamefont
  {Mou}}, \bibinfo {author} {\bibfnamefont {X.}~\bibnamefont {Yang}}, \bibinfo
  {author} {\bibfnamefont {B.}~\bibnamefont {Gallas}},\ and\ \bibinfo {author}
  {\bibfnamefont {M.}~\bibnamefont {Mivelle}},\ }\href@noop {} {\bibinfo
  {title} {A reversed inverse faraday effect}} (\bibinfo {year}
  {2023}{\natexlab{a}}),\ \Eprint {https://arxiv.org/abs/2305.14469}
  {arXiv:2305.14469 [physics.optics]} \BibitemShut {NoStop}%
\bibitem [{\citenamefont {Mou}\ \emph {et~al.}(2023{\natexlab{b}})\citenamefont
  {Mou}, \citenamefont {Yang}, \citenamefont {Gallas},\ and\ \citenamefont
  {Mivelle}}]{mou2023chiral}%
  \BibitemOpen
  \bibfield  {author} {\bibinfo {author} {\bibfnamefont {Y.}~\bibnamefont
  {Mou}}, \bibinfo {author} {\bibfnamefont {X.}~\bibnamefont {Yang}}, \bibinfo
  {author} {\bibfnamefont {B.}~\bibnamefont {Gallas}},\ and\ \bibinfo {author}
  {\bibfnamefont {M.}~\bibnamefont {Mivelle}},\ }\href@noop {} {\bibinfo
  {title} {A chiral inverse faraday effect mediated by an inversely designed
  plasmonic antenna}} (\bibinfo {year} {2023}{\natexlab{b}}),\ \Eprint
  {https://arxiv.org/abs/2301.05971} {arXiv:2301.05971 [physics.optics]}
  \BibitemShut {NoStop}%
\bibitem [{\citenamefont {Parafilo}\ \emph {et~al.}(2022)\citenamefont
  {Parafilo}, \citenamefont {Boev}, \citenamefont {Kovalev},\ and\
  \citenamefont {Savenko}}]{Parafilo2022Fl}%
  \BibitemOpen
  \bibfield  {author} {\bibinfo {author} {\bibfnamefont {A.~V.}\ \bibnamefont
  {Parafilo}}, \bibinfo {author} {\bibfnamefont {M.~V.}\ \bibnamefont {Boev}},
  \bibinfo {author} {\bibfnamefont {V.~M.}\ \bibnamefont {Kovalev}},\ and\
  \bibinfo {author} {\bibfnamefont {I.~G.}\ \bibnamefont {Savenko}},\
  }\bibfield  {title} {\bibinfo {title} {Photogalvanic transport in fluctuating
  ising superconductors},\ }\href {https://doi.org/10.1103/PhysRevB.106.144502}
  {\bibfield  {journal} {\bibinfo  {journal} {Phys. Rev. B}\ }\textbf {\bibinfo
  {volume} {106}},\ \bibinfo {pages} {144502} (\bibinfo {year}
  {2022})}\BibitemShut {NoStop}%
\bibitem [{\citenamefont {Croitoru}\ \emph {et~al.}(2022)\citenamefont
  {Croitoru}, \citenamefont {Mironov}, \citenamefont {Lounis},\ and\
  \citenamefont {Buzdin}}]{Croitoru2022}%
  \BibitemOpen
  \bibfield  {author} {\bibinfo {author} {\bibfnamefont {M.~D.}\ \bibnamefont
  {Croitoru}}, \bibinfo {author} {\bibfnamefont {S.~V.}\ \bibnamefont
  {Mironov}}, \bibinfo {author} {\bibfnamefont {B.}~\bibnamefont {Lounis}},\
  and\ \bibinfo {author} {\bibfnamefont {A.~I.}\ \bibnamefont {Buzdin}},\
  }\bibfield  {title} {\bibinfo {title} {Toward the light-operated
  superconducting devices: Circularly polarized radiation manipulates the
  current-carrying states in superconducting rings},\ }\href
  {https://doi.org/https://doi.org/10.1002/qute.202200054} {\bibfield
  {journal} {\bibinfo  {journal} {Advanced Quantum Technologies}\ }\textbf
  {\bibinfo {volume} {5}},\ \bibinfo {pages} {2200054} (\bibinfo {year}
  {2022})}\BibitemShut {NoStop}%
\bibitem [{\citenamefont {Plastovets}\ and\ \citenamefont
  {Buzdin}(2023)}]{Buzdin2023}%
  \BibitemOpen
  \bibfield  {author} {\bibinfo {author} {\bibfnamefont {V.}~\bibnamefont
  {Plastovets}}\ and\ \bibinfo {author} {\bibfnamefont {A.}~\bibnamefont
  {Buzdin}},\ }\bibfield  {title} {\bibinfo {title} {Fluctuation-mediated
  inverse faraday effect in superconducting rings},\ }\href
  {https://doi.org/https://doi.org/10.1016/j.physleta.2023.129001} {\bibfield
  {journal} {\bibinfo  {journal} {Physics Letters A}\ }\textbf {\bibinfo
  {volume} {481}},\ \bibinfo {pages} {129001} (\bibinfo {year}
  {2023})}\BibitemShut {NoStop}%
\bibitem [{\citenamefont {Putilov}\ \emph {et~al.}(2023)\citenamefont
  {Putilov}, \citenamefont {Mironov}, \citenamefont {Mel'nikov},\ and\
  \citenamefont {Bespalov}}]{Putilov2023-IFESC}%
  \BibitemOpen
  \bibfield  {author} {\bibinfo {author} {\bibfnamefont {A.~V.}\ \bibnamefont
  {Putilov}}, \bibinfo {author} {\bibfnamefont {S.~V.}\ \bibnamefont
  {Mironov}}, \bibinfo {author} {\bibfnamefont {A.~S.}\ \bibnamefont
  {Mel'nikov}},\ and\ \bibinfo {author} {\bibfnamefont {A.~A.}\ \bibnamefont
  {Bespalov}},\ }\bibfield  {title} {\bibinfo {title} {Inverse faraday effect
  in superconductors with a finite gap in the excitation spectrum},\ }\href
  {https://doi.org/10.1134/S0021364023601239} {\bibfield  {journal} {\bibinfo
  {journal} {JETP Letters}\ }\textbf {\bibinfo {volume} {117}},\ \bibinfo
  {pages} {827} (\bibinfo {year} {2023})}\BibitemShut {NoStop}%
\bibitem [{\citenamefont {Dzero}(2024)}]{Dzero-IFE2024}%
  \BibitemOpen
  \bibfield  {author} {\bibinfo {author} {\bibfnamefont {M.}~\bibnamefont
  {Dzero}},\ }\bibfield  {title} {\bibinfo {title} {Inverse faraday effect in
  superconductors with potential impurities},\ }\href
  {https://doi.org/10.1103/PhysRevB.110.054506} {\bibfield  {journal} {\bibinfo
   {journal} {Phys. Rev. B}\ }\textbf {\bibinfo {volume} {110}},\ \bibinfo
  {pages} {054506} (\bibinfo {year} {2024})}\BibitemShut {NoStop}%
\bibitem [{\citenamefont {Ginzburg}\ and\ \citenamefont {Landau}(2009)}]{GL}%
  \BibitemOpen
  \bibfield  {author} {\bibinfo {author} {\bibfnamefont {V.~L.}\ \bibnamefont
  {Ginzburg}}\ and\ \bibinfo {author} {\bibfnamefont {L.~D.}\ \bibnamefont
  {Landau}},\ }\bibinfo {title} {On the theory of superconductivity},\ in\
  \href {https://doi.org/10.1007/978-3-540-68008-6_4} {\emph {\bibinfo
  {booktitle} {On Superconductivity and Superfluidity: A Scientific
  Autobiography}}}\ (\bibinfo  {publisher} {Springer Berlin Heidelberg},\
  \bibinfo {address} {Berlin, Heidelberg},\ \bibinfo {year} {2009})\ pp.\
  \bibinfo {pages} {113--137}\BibitemShut {NoStop}%
\bibitem [{\citenamefont {Abrahams}\ and\ \citenamefont
  {Tsuneto}(1966)}]{Elihu1966}%
  \BibitemOpen
  \bibfield  {author} {\bibinfo {author} {\bibfnamefont {E.}~\bibnamefont
  {Abrahams}}\ and\ \bibinfo {author} {\bibfnamefont {T.}~\bibnamefont
  {Tsuneto}},\ }\bibfield  {title} {\bibinfo {title} {Time variation of the
  ginzburg-landau order parameter},\ }\href
  {https://doi.org/10.1103/PhysRev.152.416} {\bibfield  {journal} {\bibinfo
  {journal} {Phys. Rev.}\ }\textbf {\bibinfo {volume} {152}},\ \bibinfo {pages}
  {416} (\bibinfo {year} {1966})}\BibitemShut {NoStop}%
\bibitem [{\citenamefont {Gor'kov}\ and\ \citenamefont
  {Eliashberg}(1968{\natexlab{a}})}]{GorkovEliashberg1968a}%
  \BibitemOpen
  \bibfield  {author} {\bibinfo {author} {\bibfnamefont {L.~P.}\ \bibnamefont
  {Gor'kov}}\ and\ \bibinfo {author} {\bibfnamefont {G.~M.}\ \bibnamefont
  {Eliashberg}},\ }\bibfield  {title} {\bibinfo {title} {Generalization of the
  ginzburg-landau equations for non-stationary problems in the case of alloys
  with paramagnetic impurities},\ }\href@noop {} {\bibfield  {journal}
  {\bibinfo  {journal} {Sov. Phys. - JETP}\ }\textbf {\bibinfo {volume} {27}},\
  \bibinfo {pages} {328} (\bibinfo {year} {1968}{\natexlab{a}})}\BibitemShut
  {NoStop}%
\bibitem [{\citenamefont {Gor'kov}\ and\ \citenamefont
  {Eliashberg}(1968{\natexlab{b}})}]{GorkovEliashberg1968b}%
  \BibitemOpen
  \bibfield  {author} {\bibinfo {author} {\bibfnamefont {L.~P.}\ \bibnamefont
  {Gor'kov}}\ and\ \bibinfo {author} {\bibfnamefont {G.~M.}\ \bibnamefont
  {Eliashberg}},\ }\bibfield  {title} {\bibinfo {title} {The behavior of a
  superconductor in a variable field},\ }\href@noop {} {\bibfield  {journal}
  {\bibinfo  {journal} {Sov. Phys. - JETP}\ }\textbf {\bibinfo {volume} {38}},\
  \bibinfo {pages} {550} (\bibinfo {year} {1968}{\natexlab{b}})}\BibitemShut
  {NoStop}%
\bibitem [{\citenamefont {Kopnin}(2001)}]{Kopnin2001}%
  \BibitemOpen
  \bibfield  {author} {\bibinfo {author} {\bibfnamefont {N.}~\bibnamefont
  {Kopnin}},\ }\href@noop {} {\emph {\bibinfo {title} {Theory of Nonequilibrium
  Superconductivity}}}\ (\bibinfo  {publisher} {Oxford University Press},\
  \bibinfo {year} {2001})\BibitemShut {NoStop}%
\bibitem [{\citenamefont {Katsumi}\ \emph {et~al.}(2018)\citenamefont
  {Katsumi}, \citenamefont {Tsuji}, \citenamefont {Hamada}, \citenamefont
  {Matsunaga}, \citenamefont {Schneeloch}, \citenamefont {Zhong}, \citenamefont
  {Gu}, \citenamefont {Aoki}, \citenamefont {Gallais},\ and\ \citenamefont
  {Shimano}}]{Katsumi2018}%
  \BibitemOpen
  \bibfield  {author} {\bibinfo {author} {\bibfnamefont {K.}~\bibnamefont
  {Katsumi}}, \bibinfo {author} {\bibfnamefont {N.}~\bibnamefont {Tsuji}},
  \bibinfo {author} {\bibfnamefont {Y.~I.}\ \bibnamefont {Hamada}}, \bibinfo
  {author} {\bibfnamefont {R.}~\bibnamefont {Matsunaga}}, \bibinfo {author}
  {\bibfnamefont {J.}~\bibnamefont {Schneeloch}}, \bibinfo {author}
  {\bibfnamefont {R.~D.}\ \bibnamefont {Zhong}}, \bibinfo {author}
  {\bibfnamefont {G.~D.}\ \bibnamefont {Gu}}, \bibinfo {author} {\bibfnamefont
  {H.}~\bibnamefont {Aoki}}, \bibinfo {author} {\bibfnamefont {Y.}~\bibnamefont
  {Gallais}},\ and\ \bibinfo {author} {\bibfnamefont {R.}~\bibnamefont
  {Shimano}},\ }\bibfield  {title} {\bibinfo {title} {Higgs mode in the
  $d$-wave superconductor
  ${\mathrm{bi}}_{2}{\mathrm{sr}}_{2}{\mathrm{cacu}}_{2}{\mathrm{o}}_{8+x}$
  driven by an intense terahertz pulse},\ }\href
  {https://doi.org/10.1103/PhysRevLett.120.117001} {\bibfield  {journal}
  {\bibinfo  {journal} {Phys. Rev. Lett.}\ }\textbf {\bibinfo {volume} {120}},\
  \bibinfo {pages} {117001} (\bibinfo {year} {2018})}\BibitemShut {NoStop}%
\bibitem [{\citenamefont {Katsumi}\ \emph {et~al.}(2020)\citenamefont
  {Katsumi}, \citenamefont {Li}, \citenamefont {Raffy}, \citenamefont
  {Gallais},\ and\ \citenamefont {Shimano}}]{Katsumi2020}%
  \BibitemOpen
  \bibfield  {author} {\bibinfo {author} {\bibfnamefont {K.}~\bibnamefont
  {Katsumi}}, \bibinfo {author} {\bibfnamefont {Z.~Z.}\ \bibnamefont {Li}},
  \bibinfo {author} {\bibfnamefont {H.}~\bibnamefont {Raffy}}, \bibinfo
  {author} {\bibfnamefont {Y.}~\bibnamefont {Gallais}},\ and\ \bibinfo {author}
  {\bibfnamefont {R.}~\bibnamefont {Shimano}},\ }\bibfield  {title} {\bibinfo
  {title} {Superconducting fluctuations probed by the higgs mode in
  ${\mathrm{bi}}_{2}{\mathrm{sr}}_{2}\mathrm{Ca}{\mathrm{cu}}_{2}{\mathrm{o}}_{8+x}$
  thin films},\ }\href {https://doi.org/10.1103/PhysRevB.102.054510} {\bibfield
   {journal} {\bibinfo  {journal} {Phys. Rev. B}\ }\textbf {\bibinfo {volume}
  {102}},\ \bibinfo {pages} {054510} (\bibinfo {year} {2020})}\BibitemShut
  {NoStop}%
\bibitem [{\citenamefont {von Hoegen}\ \emph {et~al.}(2026)\citenamefont {von
  Hoegen}, \citenamefont {Tai}, \citenamefont {Allington}, \citenamefont
  {Yeung}, \citenamefont {Pettine}, \citenamefont {Michael}, \citenamefont
  {Vi{\~n}as~Bostr{\"o}m}, \citenamefont {Cui}, \citenamefont {Torres},
  \citenamefont {Kossak}, \citenamefont {Lee}, \citenamefont {Beach},
  \citenamefont {Gu}, \citenamefont {Rubio}, \citenamefont {Kim},\ and\
  \citenamefont {Gedik}}]{Gedik2026}%
  \BibitemOpen
  \bibfield  {author} {\bibinfo {author} {\bibfnamefont {A.}~\bibnamefont {von
  Hoegen}}, \bibinfo {author} {\bibfnamefont {T.}~\bibnamefont {Tai}}, \bibinfo
  {author} {\bibfnamefont {C.~J.}\ \bibnamefont {Allington}}, \bibinfo {author}
  {\bibfnamefont {M.}~\bibnamefont {Yeung}}, \bibinfo {author} {\bibfnamefont
  {J.}~\bibnamefont {Pettine}}, \bibinfo {author} {\bibfnamefont {M.~H.}\
  \bibnamefont {Michael}}, \bibinfo {author} {\bibfnamefont {E.}~\bibnamefont
  {Vi{\~n}as~Bostr{\"o}m}}, \bibinfo {author} {\bibfnamefont {X.}~\bibnamefont
  {Cui}}, \bibinfo {author} {\bibfnamefont {K.}~\bibnamefont {Torres}},
  \bibinfo {author} {\bibfnamefont {A.~E.}\ \bibnamefont {Kossak}}, \bibinfo
  {author} {\bibfnamefont {B.}~\bibnamefont {Lee}}, \bibinfo {author}
  {\bibfnamefont {G.~S.~D.}\ \bibnamefont {Beach}}, \bibinfo {author}
  {\bibfnamefont {G.~D.}\ \bibnamefont {Gu}}, \bibinfo {author} {\bibfnamefont
  {A.}~\bibnamefont {Rubio}}, \bibinfo {author} {\bibfnamefont
  {P.}~\bibnamefont {Kim}},\ and\ \bibinfo {author} {\bibfnamefont
  {N.}~\bibnamefont {Gedik}},\ }\bibfield  {title} {\bibinfo {title} {Imaging a
  terahertz superfluid plasmon in a two-dimensional superconductor},\
  }\bibfield  {journal} {\bibinfo  {journal} {Nature}\ }\href
  {https://doi.org/10.1038/s41586-025-10082-2} {10.1038/s41586-025-10082-2}
  (\bibinfo {year} {2026})\BibitemShut {NoStop}%
\bibitem [{\citenamefont {Awelewa}\ and\ \citenamefont
  {Dzero}(2025)}]{Awelewa2025}%
  \BibitemOpen
  \bibfield  {author} {\bibinfo {author} {\bibfnamefont {S.}~\bibnamefont
  {Awelewa}}\ and\ \bibinfo {author} {\bibfnamefont {M.}~\bibnamefont
  {Dzero}},\ }\bibfield  {title} {\bibinfo {title} {Dynamics of the
  schmid-higgs mode in $d$-wave superconductors},\ }\href@noop {} {\bibfield
  {journal} {\bibinfo  {journal} {pre-print}\ }\textbf {\bibinfo {volume}
  {arXiv:2511.03790}} (\bibinfo {year} {2025})}\BibitemShut {NoStop}%
\bibitem [{\citenamefont {Islam}\ \emph {et~al.}(2026)\citenamefont {Islam},
  \citenamefont {Awelewa}, \citenamefont {Chubukov},\ and\ \citenamefont
  {Dzero}}]{Kazi2026}%
  \BibitemOpen
  \bibfield  {author} {\bibinfo {author} {\bibfnamefont {K.~R.}\ \bibnamefont
  {Islam}}, \bibinfo {author} {\bibfnamefont {S.}~\bibnamefont {Awelewa}},
  \bibinfo {author} {\bibfnamefont {A.~V.}\ \bibnamefont {Chubukov}},\ and\
  \bibinfo {author} {\bibfnamefont {M.}~\bibnamefont {Dzero}},\ }\bibfield
  {title} {\bibinfo {title} {Spatially resolved collective modes in $d$-wave
  superconductors},\ }\href@noop {} {\bibfield  {journal} {\bibinfo  {journal}
  {pre-print}\ }\textbf {\bibinfo {volume} {arXiv:2601.09782}} (\bibinfo {year}
  {2026})}\BibitemShut {NoStop}%
\bibitem [{\citenamefont {Belzig}\ \emph {et~al.}(1999)\citenamefont {Belzig},
  \citenamefont {Wilhelm}, \citenamefont {Bruder}, \citenamefont {Schön},\
  and\ \citenamefont {Zaikin}}]{Belzig1999}%
  \BibitemOpen
  \bibfield  {author} {\bibinfo {author} {\bibfnamefont {W.}~\bibnamefont
  {Belzig}}, \bibinfo {author} {\bibfnamefont {F.~K.}\ \bibnamefont {Wilhelm}},
  \bibinfo {author} {\bibfnamefont {C.}~\bibnamefont {Bruder}}, \bibinfo
  {author} {\bibfnamefont {G.}~\bibnamefont {Schön}},\ and\ \bibinfo {author}
  {\bibfnamefont {A.~D.}\ \bibnamefont {Zaikin}},\ }\bibfield  {title}
  {\bibinfo {title} {Quasiclassical green’s function approach to mesoscopic
  superconductivity},\ }\href
  {https://doi.org/https://doi.org/10.1006/spmi.1999.0710} {\bibfield
  {journal} {\bibinfo  {journal} {Superlattices and Microstructures}\ }\textbf
  {\bibinfo {volume} {25}},\ \bibinfo {pages} {1251} (\bibinfo {year}
  {1999})}\BibitemShut {NoStop}%
\bibitem [{\citenamefont {Kalenkov}\ and\ \citenamefont
  {Zaikin}(2014)}]{Kalenkov2014}%
  \BibitemOpen
  \bibfield  {author} {\bibinfo {author} {\bibfnamefont {M.~S.}\ \bibnamefont
  {Kalenkov}}\ and\ \bibinfo {author} {\bibfnamefont {A.~D.}\ \bibnamefont
  {Zaikin}},\ }\bibfield  {title} {\bibinfo {title} {Electron-hole imbalance
  and large thermoelectric effect in superconducting hybrids with spin-active
  interfaces},\ }\href {https://doi.org/10.1103/PhysRevB.90.134502} {\bibfield
  {journal} {\bibinfo  {journal} {Phys. Rev. B}\ }\textbf {\bibinfo {volume}
  {90}},\ \bibinfo {pages} {134502} (\bibinfo {year} {2014})}\BibitemShut
  {NoStop}%
\bibitem [{\citenamefont {Masaki}\ and\ \citenamefont
  {Kato}(2018)}]{Masaki2018}%
  \BibitemOpen
  \bibfield  {author} {\bibinfo {author} {\bibfnamefont {Y.}~\bibnamefont
  {Masaki}}\ and\ \bibinfo {author} {\bibfnamefont {Y.}~\bibnamefont {Kato}},\
  }\bibfield  {title} {\bibinfo {title} {Charged and uncharged vortices in
  quasiclassical theory},\ }\href
  {https://doi.org/10.1088/1742-6596/969/1/012054} {\bibfield  {journal}
  {\bibinfo  {journal} {Journal of Physics: Conference Series}\ }\textbf
  {\bibinfo {volume} {969}},\ \bibinfo {pages} {012054} (\bibinfo {year}
  {2018})}\BibitemShut {NoStop}%
\bibitem [{\citenamefont {Silaev}(2019)}]{Silaev2019-Disorder}%
  \BibitemOpen
  \bibfield  {author} {\bibinfo {author} {\bibfnamefont {M.}~\bibnamefont
  {Silaev}},\ }\bibfield  {title} {\bibinfo {title} {Nonlinear electromagnetic
  response and higgs-mode excitation in bcs superconductors with impurities},\
  }\href {https://doi.org/10.1103/PhysRevB.99.224511} {\bibfield  {journal}
  {\bibinfo  {journal} {Phys. Rev. B}\ }\textbf {\bibinfo {volume} {99}},\
  \bibinfo {pages} {224511} (\bibinfo {year} {2019})}\BibitemShut {NoStop}%
\bibitem [{\citenamefont {Steineman}\ and\ \citenamefont
  {Dzero}(2026)}]{SteinemanDzero-Eliashberg}%
  \BibitemOpen
  \bibfield  {author} {\bibinfo {author} {\bibfnamefont {A.}~\bibnamefont
  {Steineman}}\ and\ \bibinfo {author} {\bibfnamefont {M.}~\bibnamefont
  {Dzero}},\ }\bibfield  {title} {\bibinfo {title} {Stimulation of
  superconductivity in $d$-wave superconductors}} (\bibinfo {year} {2026}),\
  \bibinfo {note} {unpublished}\BibitemShut {NoStop}%
\bibitem [{\citenamefont {Larkin}\ and\ \citenamefont
  {Ovchinnikov}(1969)}]{LO}%
  \BibitemOpen
  \bibfield  {author} {\bibinfo {author} {\bibfnamefont {A.~I.}\ \bibnamefont
  {Larkin}}\ and\ \bibinfo {author} {\bibfnamefont {Y.}~\bibnamefont
  {Ovchinnikov}},\ }\bibfield  {title} {\bibinfo {title} {Quasiclassical method
  in the theory of superconductivity},\ }\href@noop {} {\bibfield  {journal}
  {\bibinfo  {journal} {Sov. Phys. - JETP}\ }\textbf {\bibinfo {volume} {28}},\
  \bibinfo {pages} {1200} (\bibinfo {year} {1969})}\BibitemShut {NoStop}%
\bibitem [{\citenamefont {Eilenberger}(1968)}]{Eilenberger1968}%
  \BibitemOpen
  \bibfield  {author} {\bibinfo {author} {\bibfnamefont {G.}~\bibnamefont
  {Eilenberger}},\ }\bibfield  {title} {\bibinfo {title} {Transformation of
  gorkov's equation for type ii superconductors into transport-like
  equations},\ }\href {https://doi.org/10.1007/BF01379803} {\bibfield
  {journal} {\bibinfo  {journal} {Zeitschrift f{\"u}r Physik A Hadrons and
  nuclei}\ }\textbf {\bibinfo {volume} {214}},\ \bibinfo {pages} {195}
  (\bibinfo {year} {1968})}\BibitemShut {NoStop}%
\bibitem [{\citenamefont {Larkin}\ and\ \citenamefont
  {Ovchinnikov}(1977)}]{LO-Vortex}%
  \BibitemOpen
  \bibfield  {author} {\bibinfo {author} {\bibfnamefont {A.~I.}\ \bibnamefont
  {Larkin}}\ and\ \bibinfo {author} {\bibfnamefont {Y.}~\bibnamefont
  {Ovchinnikov}},\ }\bibfield  {title} {\bibinfo {title} {Nonlinear effects
  during the motion of vortices in superconductors},\ }\href@noop {} {\bibfield
   {journal} {\bibinfo  {journal} {Sov. Phys. - JETP}\ }\textbf {\bibinfo
  {volume} {46}},\ \bibinfo {pages} {155} (\bibinfo {year} {1977})}\BibitemShut
  {NoStop}%
\bibitem [{\citenamefont {Eliashberg}(1972)}]{Eliashberg-Dynamics}%
  \BibitemOpen
  \bibfield  {author} {\bibinfo {author} {\bibfnamefont {G.~M.}\ \bibnamefont
  {Eliashberg}},\ }\bibfield  {title} {\bibinfo {title} {Inelastic electron
  collisions and non-equilibrium stationary states in superconductors},\
  }\href@noop {} {\bibfield  {journal} {\bibinfo  {journal} {Sov. Phys. -
  JETP}\ }\textbf {\bibinfo {volume} {34}},\ \bibinfo {pages} {668} (\bibinfo
  {year} {1972})}\BibitemShut {NoStop}%
\bibitem [{\citenamefont {Eliashberg}(1970)}]{Eliashberg1970}%
  \BibitemOpen
  \bibfield  {author} {\bibinfo {author} {\bibfnamefont {G.}~\bibnamefont
  {Eliashberg}},\ }\bibfield  {title} {\bibinfo {title} {Film superconductivity
  stimulated by a high-frequency field},\ }\href@noop {} {\bibfield  {journal}
  {\bibinfo  {journal} {Sov. Phys. - JETP Lett.}\ }\textbf {\bibinfo {volume}
  {11}},\ \bibinfo {pages} {114} (\bibinfo {year} {1970})}\BibitemShut
  {NoStop}%
\bibitem [{\citenamefont {Ivlev}\ \emph {et~al.}(1973)\citenamefont {Ivlev},
  \citenamefont {Lisitsyn},\ and\ \citenamefont {Eliashberg}}]{Ivlev1973}%
  \BibitemOpen
  \bibfield  {author} {\bibinfo {author} {\bibfnamefont {B.~I.}\ \bibnamefont
  {Ivlev}}, \bibinfo {author} {\bibfnamefont {S.~G.}\ \bibnamefont
  {Lisitsyn}},\ and\ \bibinfo {author} {\bibfnamefont {G.~M.}\ \bibnamefont
  {Eliashberg}},\ }\bibfield  {title} {\bibinfo {title} {Nonequilibrium
  excitations in superconductors in high-frequency fields},\ }\href
  {https://doi.org/10.1007/BF00654920} {\bibfield  {journal} {\bibinfo
  {journal} {Journal of Low Temperature Physics}\ }\textbf {\bibinfo {volume}
  {10}},\ \bibinfo {pages} {449} (\bibinfo {year} {1973})}\BibitemShut
  {NoStop}%
\bibitem [{\citenamefont {Klapwijk}\ \emph {et~al.}(1977)\citenamefont
  {Klapwijk}, \citenamefont {van~den Bergh},\ and\ \citenamefont
  {Mooij}}]{Klapwijk1977}%
  \BibitemOpen
  \bibfield  {author} {\bibinfo {author} {\bibfnamefont {T.~M.}\ \bibnamefont
  {Klapwijk}}, \bibinfo {author} {\bibfnamefont {J.~N.}\ \bibnamefont {van~den
  Bergh}},\ and\ \bibinfo {author} {\bibfnamefont {J.~E.}\ \bibnamefont
  {Mooij}},\ }\bibfield  {title} {\bibinfo {title} {Radiation-stimulated
  superconductivity},\ }\href {https://doi.org/10.1007/BF00655418} {\bibfield
  {journal} {\bibinfo  {journal} {Journal of Low Temperature Physics}\ }\textbf
  {\bibinfo {volume} {26}},\ \bibinfo {pages} {385} (\bibinfo {year}
  {1977})}\BibitemShut {NoStop}%
\bibitem [{\citenamefont {Edel'shtein}(1988)}]{Edel1988}%
  \BibitemOpen
  \bibfield  {author} {\bibinfo {author} {\bibfnamefont {V.~M.}\ \bibnamefont
  {Edel'shtein}},\ }\bibfield  {title} {\bibinfo {title} {Second-harmonic
  generation in two-dimensional systems without inversion centers},\
  }\href@noop {} {\bibfield  {journal} {\bibinfo  {journal} {Sov. Phys. -
  JETP}\ }\textbf {\bibinfo {volume} {68}},\ \bibinfo {pages} {1446} (\bibinfo
  {year} {1988})}\BibitemShut {NoStop}%
\bibitem [{\citenamefont {Dzero}\ \emph {et~al.}(2025)\citenamefont {Dzero},
  \citenamefont {Hasan},\ and\ \citenamefont {Levchenko}}]{Dzero2024Nice}%
  \BibitemOpen
  \bibfield  {author} {\bibinfo {author} {\bibfnamefont {M.}~\bibnamefont
  {Dzero}}, \bibinfo {author} {\bibfnamefont {J.}~\bibnamefont {Hasan}},\ and\
  \bibinfo {author} {\bibfnamefont {A.}~\bibnamefont {Levchenko}},\ }\bibfield
  {title} {\bibinfo {title} {Resonant second harmonic generation in a
  two-dimensional electron system},\ }\href {https://doi.org/10.1103/y2k3-7fwh}
  {\bibfield  {journal} {\bibinfo  {journal} {Phys. Rev. B}\ }\textbf {\bibinfo
  {volume} {112}},\ \bibinfo {pages} {045416} (\bibinfo {year}
  {2025})}\BibitemShut {NoStop}%
\bibitem [{\citenamefont {Pitaevskii}(1961)}]{Pit1961}%
  \BibitemOpen
  \bibfield  {author} {\bibinfo {author} {\bibfnamefont {L.~P.}\ \bibnamefont
  {Pitaevskii}},\ }\bibfield  {title} {\bibinfo {title} {Electric forces in a
  transparent dispersive medium},\ }\href@noop {} {\bibfield  {journal}
  {\bibinfo  {journal} {Sov. Phys. - JETP}\ }\textbf {\bibinfo {volume} {12}},\
  \bibinfo {pages} {1008} (\bibinfo {year} {1961})}\BibitemShut {NoStop}%
\bibitem [{\citenamefont {Hertel}(2006)}]{Hertel2006}%
  \BibitemOpen
  \bibfield  {author} {\bibinfo {author} {\bibfnamefont {R.}~\bibnamefont
  {Hertel}},\ }\bibfield  {title} {\bibinfo {title} {Theory of the inverse
  faraday effect in metals},\ }\href@noop {} {\bibfield  {journal} {\bibinfo
  {journal} {J. Magn. Magn. Mater.}\ }\textbf {\bibinfo {volume} {303}},\
  \bibinfo {pages} {L1} (\bibinfo {year} {2006})}\BibitemShut {NoStop}%
\end{thebibliography}%

\end{document}